# Janus $MgAlB_2$ MBene: a dipole-engineered anode for ultrafast Li-ion transport and exceptional lithium storage

Pritam Samanta[1], Sashank Kumar Pandey[1], Amit Kumar Jana[1], Prakash Parida[1*],
[1]Department of Physics, Indian Institute of Technology Patna, Bihta, Patna, 801106, Bihar, India
*Corresponding authors: pparida@iitp.ac.in

## Abstract

In this work, we propose a group II/IIIA-based Janus MBene, $MgAlB_2$, and investigate its electrochemical properties using first-principles calculations. The substitution of one Mg layer in $Mg_2B_2$ MBene by an Al layer breaks the structural symmetry and generates a permanent out-of-plane polarization, giving rise to a distinct electronic environment compared with the parent $Mg_2B_2$ and $Al_2B_2$ monolayers. Electronic-structure analysis reveals enhanced orbital hybridization among B, Mg, and Al states near the Fermi level, resulting in improved electronic delocalization across the monolayer. The Janus $MgAlB_2$ monolayer is found to possess excellent dynamical, mechanical, and thermal stability. Owing to its polarization-modified energy landscape, Li ions migrate with an exceptionally low diffusion barrier of 17.1 meV, corresponding to a room-temperature diffusion coefficient of $3.43\times10^{-10}$ $cm^2s^{-1}$. Unlike the pristine $Mg_2B_2$ and $Al_2B_2$ monolayers, which support only a single stable adsorption layer, $MgAlB_2$ accommodates two complete Li layers. Detailed analysis shows that the residual polarization retained after first-layer lithiation continues to promote Li adsorption, whereas increasing Li-Li electrostatic interactions eventually limit further storage. As a result, the Janus monolayer delivers a high theoretical specific capacity of 1470.24 $mAhg^{-1}$ together with a small volume expansion of only 3.7% during maximum lithiation. The present study demonstrates that intrinsic polarization can be utilized to regulate both the thermodynamics and kinetics of Li storage, providing a design strategy for high-rate and high-capacity two-dimensional electrode materials.

## 1. Introduction

With the advent of the 21st century, the rising demand for energy has become a major challenge in the era of rapid industrial and technological development. [1, 2] But the limited availability of fossil-based natural resources and their attendant environmental concerns pose a tough challenge to human sustainable development. Therefore, there is a pressing need to explore efficient and environmentally benign alternative energy storage systems. [3-7] Under the circumstances, lithium-ion batteries (LIBs) have gained a significant attention for energy-storage applications for their superior energy density, long cycle life, and excellent reversibility. [8-13] LIBs play a crucial role in storing excess energy generated from intermittent sources such as solar energy, wind energy, etc., and releasing it whenever needed, thus playing a faithful role in a sustainable energy storage system. [14] Among portable batteries, LIBs have drawn rising attention for energy storage systems. They compensate for the daily increasing social needs. Moreover, LIBs find extensive use in portable electronics, electric vehicles, and various digital devices as they offer high energy density, reversible storage capacity, long cycle life, portability, and suitability for miniaturization and lightweight design. [15] The quality and the perpetuation of LIBs are predominantly decided by the characteristics of their three components, as cathode, anode, and electrolyte materials, particularly on the anode, which directly influences storage capacity, rate capability, cycling stability, etc. [16] A promising electrode material should possess a large specific capacity, a better electrical conductivity, faster ion diffusion kinetics, and structural integrity at the time of repeated charge/discharge cycles. Currently, graphite remains the most well-known commercial anode material as it offers high Coulombic efficiency, excellent cycling stability, etc. However, it's relatively low theoretical capacity (372 $mAhg^{-1}$) limits its applications in next-generation high-energy-density storage systems. [17, 18] Hence, there is an urgent need to design a novel electrode material with excellent electrochemical properties to improve the performance of LIBs. [19, 20] Of late, Two-dimensional (2D) materials have emerged as a favorable electrode material for metal ion batteries as they possess a large surface/volume ratio, a sufficient number of adsorption sites, reduced ion diffusion barrier compared to their bulk counterparts. A wide range of 2D materials such as graphene, transition metal oxides (TMOs) [21], transition metal dichalcogenides [22, 23], novel 2D materials such as $MoS_2$ [24, 25], $VS_2$ [26], silicene [27, 28], phosphorene [29, 30], borophene [31], transition metal carbides and nitrides (MXenes) [32, 33] have been studied extensively and reported as emerging anode materials for metal ion batteries. Despite with these advances, there are a lot of searches that need to be addressed in finding superior anode materials with superior electrochemical characteristics, specific capacity, and fast ion transport capability to tackle the challenges such as low storage capacity, poor diffusion rate, structural degradation, etc.

Recently, boron-rich 2D materials have become emerging electrodes for metal-ion batteries owing to their high theoretical capacities, excellent electronic conductivity, and tunable chemistry among the emerging 2D materials. [34] Among these, the MBenes family (transition metal borides with general formula $M_xB_y$, where M is transition metal and B is boron atom) possessing the similar structure to that of MXenes (transition metal carbides and borides) have emerged as a promising anode material due to their metallic nature and large surface area. [35, 36] Notably, Guo et al. in recent years reported $Fe_2B_2$ and $Mo_2B_2$, which showed storage capacities of 665 $mAhg^{-1}$ and 444 $mAhg^{-1}$, respectively, motivating further exploration of this class of materials. [37] Building upon the seminal work of Guo et al., who introduced several representative MBenes, including $Fe_2B_2$ and $Mo_2B_2$, a diverse family of MBene materials has since been extensively investigated and reported in the literature. Among different types of MBenes, the most reported were the orthogonal MBenes, including $Cr_2B_2$, $Mo_2B_2$, $W_2B_2$, and $Fe_2B_2$. [34] Whereas, Bo. et al. [38, 39] reported the hexagonal $M_2B_2$- type MBenes including $Ti_2B_2$, $Sc_2B_2$, $V_2B_2$, $Cr_2B_2$, $Y_2B_2$, $Mo_2B_2$, $Zr_2B_2$, etc. However, conventional symmetric MBenes often suffer from a low storage capacity. This uniform chemical environment in symmetric MBenes restricts adsorption tunability and further limits its progress.

Over the past decades, extensive research has focused on developing effective approaches to improve the storage capacity of MXenes and MBenes for energy-storage applications.[40-42] Recently, Janus materials [43-44] have drawn attention as an important 2D system and have been extensively studied for applications such as optoelectronics[43], thermoelectrics [44, 48] metal ion batteries[49], Solar cells[50], and water splitting[51]. Their structural asymmetry opens up distinct surface properties, making them particularly attractive for energy storage applications. [52] This dual functionality promotes improved interaction with metal ions, which can enhance both performance and stability in metal-ion batteries. In Janus monolayers, the replacement of atomic species on one side of the monolayer with different atomic species breaks inversion symmetry and generates a strong built-in out-of-plane electric field. [53] This polarization created from asymmetric charge redistribution, leading to energetically distinct adsorption sites and enhanced electrostatic interaction with alkali ions. Due to this built-in polarization, Janus monolayers exhibit moderate adsorption strength, better charge transfer, lower diffusion barriers, enabling faster ion transport and enhanced storage capacity compared to their symmetric counterparts. [53, 54] It was reported that the intrinsic dipole moment due to inherent asymmetry helped more adsorption of lithium atoms on the Janus monolayer, thus enhancing the storage capacity of MoSSe to 776.5 $mAhg^{-1}$ [53] from 669 $mAhg^{-1}$ in layered symmetric hexagonal $Ti_2B_2$ [39]. Also, this idea of the breaking of the out-of-plane structural symmetry was also applied to transition metal MBenes, such as $ScTiB_2$, where replacing one layer of $Ti_2B_2$ with Sc leads to enhanced storage capacity compared to its symmetric counterparts. [54] In the case of $ScTiB_2$ (937 $mAhg^{-1}$) [54], the storage capacity was nearly doubled compared to both hexagonal $Ti_2B_2$ (456 $mAhg^{-1}$) [39] and $Sc_2B_2$ (480 $mAhg^{-1}$) [89]. This enhanced storage capacity in $ScTiB_2$ resulted from the induced intrinsic dipole moment, which modifies the Coulomb interaction between atomic layers and creates site-dependent, distinct adsorption environments. [54] Such asymmetry not only strengthens ion adsorption on one surface while maintaining moderate binding on the other, but also gives faster ion diffusion.

While transition-metal MBenes benefit from excellent metallic conductivity, their localised electronic states originating from partially filled d-orbitals of transition metals often leads to excessively strong ion binding and localised electronic states, which may hinder ion diffusion and may lead to higher migration barriers. [55, 56] In contrast, non-transition metal-based MBenes composed of outermost s-p orbitals offer a more delocalized electronic environment, resulting in moderate adsorption strength and improved electronic conductivity. [57] This balance is particularly advantageous for achieving fast ion kinetics without compromising storage capacity. Moreover, the absence of strongly localized d-states in non-transition metal MBenes may provide a delocalised sheet of electrons that prevents ion trapping, enabling smoother diffusion pathways compared to transition-metal counterparts. The presence of dispersive bands in these non-transition metal MBene materials exhibiting metallic or semi-metallic, crossing the Fermi level, offers continuous electronic states and high carrier mobility, which are essential for efficient charge transport during lithiation/delithiation and faster electrochemical kinetics. [58, 59] Moreover, the relatively low atomic mass of light weight non-transition metal constituents provides enhanced gravimetric storage performance due to their reduced formula weight compared to transition-metal MBenes containing heavier transition-metal elements with higher formula weight. [57] As a result, metal MBenes with metallic or semi-metallic electronic structures and low atomic weight provide both enhanced storage capacity and fast ion/electron transport for designing high-performance anode materials.

Despite these advantages, the lithiation behavior of non-transition-metal MBenes remains constrained by the limited number of energetically favorable adsorption sites available on symmetric surfaces. In many cases, stable adsorption is restricted to a single Li layer, thereby limiting the achievable storage capacity. An appealing strategy to overcome this limitation is to introduce an intrinsic electric polarization capable of modifying the adsorption-energy landscape and promoting additional Li uptake. Janus architectures are particularly attractive in this regard because the chemical asymmetry between the two surfaces generates a built-in electric field that can influence charge redistribution, adsorption energetics, and ion-transport pathways.

Guided by this idea, we design a Janus $MgAlB_2$ MBene by replacing one Mg layer of the centrosymmetric $Mg_2B_2$ MBene with an Al layer. The choice of Al is motivated by its close chemical similarity and comparable atomic radius to Mg, which allows symmetry breaking to be introduced without significantly perturbing the parent lattice. More importantly, $Al_2B_2$ has been predicted to host highly dispersive Dirac nodal-line states near the Fermi level arising from strong Al-B orbital hybridization, indicative of enhanced electronic delocalization and metallic transport characteristics.[50] By integrating these electronic features with the intrinsic polarization generated by Janus asymmetry, $MgAlB_2$ provides an ideal platform for investigating whether symmetry breaking can simultaneously enhance Li adsorption, accelerate ion diffusion, and extend the lithiation limit beyond that of the parent $Mg_2B_2$ and $Al_2B_2$ monolayers. Through this study, we demonstrate that intrinsic polarization is not merely a structural consequence of Janus engineering but a powerful design parameter for regulating both the thermodynamics and kinetics of lithium storage in two-dimensional materials.

Using first principles calculations, we examined a Janus hexagonal $MgAlB_2$ monolayer as a lithium-ion battery anode. After geometry optimization, we evaluated thermodynamic stability, mechanical response, electronic structure, and dynamical stability. Electrochemical behavior was assessed through adsorption energies, theoretical capacity, and diffusion barriers. The results show multiple favorable adsorption sites with moderate binding, an ultralow Li migration barrier, and a high theoretical capacity. These advantages arise from the broken out of plane symmetry with built in polarization and from the low formula mass of the Janus framework. Together, they outline a design strategy that links symmetry breaking with band structure features to achieve high capacity and fast charge and discharge kinetics. In particular, Janus $MgAlB_2$ stably accommodates two Li layers, delivering a high storage capacity of 1470.24 $mAhg^{-1}$, with a low diffusion barrier of 17.1 meV. This group II and IIIA Janus, non-transition metal MBene therefore emerges as an alternative anode candidate for ultrafast charging and high energy storage.

## 2. Computational methods

The first-principles calculations were carried out using the Vienna Ab Initio Simulation Package (VASP) [61, 62] based on density functional theory (DFT). A high plane-wave energy cutoff of 750 eV was employed to ensure convergence accuracy. Both structural optimization and static calculations were performed using a Monkhorst-Pack method [63] with a k-mesh of 11×11×1, with strict convergence criteria of $1\times10^{-10}$ eV for total energy and $6\times10^{-4}$ eV/Å for atomic forces. The exchange-correlation functional was described within the generalized gradient approximation (GGA) using the Perdew-Burke-Ernzerhof (PBE) functional. [64] We have shown the convergence test results of the cutoff energy and Monkhorst-Pack k-grid in **Fig. S1** (see **SI**). We have also shown the Brillouin zone of the Janus hexagonal $MgAlB_2$ lattice in **Fig. S2** (see **SI**). We performed an electronic band structure, and density of states (DOS) of the unit cell of pristine Janus $MgAlB_2$. We considered DFT-D3 (Becke-Johnson damping) dispersion correction to include weak van der Waals (vdW) interactions during the optimization of both the pristine $MgAlB_2$ and the adsorption of lithium on both surfaces of $MgAlB_2$. [65] We set the vacuum layer distance at 20 Å along the z direction to avoid the interlayer interactions between periodic images. A supercell of dimensions $4\times4\times1$ was taken to determine the phonon dispersion spectra with the VASP-PHONOPY interface to check the dynamical stability based on the finite displacement (FD). [66] The converged charge density is subsequently used to compute the non-self-consistent electronic band structure and the projected density of states (PDOS). Band structures are plotted along high-symmetry paths using the GGA-PBE functional. [67] We have used the LOBSTER programme for calculating the integrated crystal orbital Hamiltonian population (ICOHP) calculation for bonding strength analysis. [68] Electrochemical properties were assessed on a $2\times2\times1$ supercell of $MgAlB_2$ monolayer. The climbing image nudged elastic band (CI-NEB) method was used to determine the diffusion barrier in the diffusion process. [69] We took 7 images for the CI-NEB calculations. We have performed ab initio molecular dynamics (AIMD) with a 10 ps simulation time with a time step of 1 fs to find the endurance of both pristine $MgAlB_2$ and lithium adsorbed $MgAlB_2$ monolayer. The NVT ensemble under the Nose-Hoover thermostat method was adopted at a finite temperature of T = 500K. [70] The atomic configurations and charge density distributions were visualized using VESTA, while the high-symmetry k-point paths employed for electronic band structure calculations were generated and illustrated using Xcrysden.

## 3. Results and discussion

### 3.1. Structure and stability of pristine $MgAlB_2$ monolayer

#### 3.1.1. Thermodynamic stability: formation energy

The bottom layer of divalent Mg atoms in the $Mg_2B_2$ monolayer is replaced by a layer of trivalent Al atoms to form the $MgAlB_2$ Janus MBene. We first optimized the unit cell of hexagonal Janus $MgAlB_2$. The optimized geometry of the pristine Janus $MgAlB_2$ monolayer belongs to the *P6mm* space group. In Janus $MgAlB_2$, a honeycomb boron layer is sandwiched between two hexagonally arranged atomic layers of Mg and Al on opposite sides of boron layer. After performing geometry optimization, the optimized lattice parameters were found to be 'a' = 'b' = 3.02 Å, having an interlayer distance, 'd' = 3.4 Å, as shown in **Fig. 1** below. The honeycomb boron layer is composed of B atoms arranged in a hexagonal network, with a B-B bond length of 1.72 Å. In **Fig. 1**, we have shown the optimized geometries of Janus $MgAlB_2$, pristine $Mg_2B_2$, and pristine $Al_2B_2$. The lattice parameters for $Mg_2B_2$were found to be 'a' = 'b' = 3.11 Å, having an interlayer distance, 'd' = 3.32 Å. The unit cell of $Mg_2B_2$ belongs to the *P6/mmm* space group. These results of $Mg_2B_2$ are the same as those of reported by Sun et al. [57] The in-plane lattice parameters of $Al_2B_2$ were found to be 'a' = 'b' = 2.91 Å, having an interlayer distance, 'd' = 3.44 Å. The unit cell of $Al_2B_2$ belongs to the *P6/mmm* space group. These results of $Al_2B_2$ are similar to those reported by Zou et al. [59] The formation energy of the $MgAlB_2$ monolayer was calculated to assess its thermodynamic stability and the possibility of experimental synthesis. The formation energy per atom was obtained from the following equation:[71]

$$E_{form} = (E_{MgAlB_2} - E_{Mg} - E_{Al} - 2E_B)/4. \quad (1)$$

In this expression, $E_{MgAlB_2}$ is the total energy of the $MgAlB_2$ unit cell. The quantities $E_{Mg}$, and $E_{Al}$, and $E_B$, refer to the chemical potentials of magnesium, aluminum, and boron atoms, respectively, obtained from their most stable bulk reference states. The factor of 4 is the total number of atoms present in the unit cell.

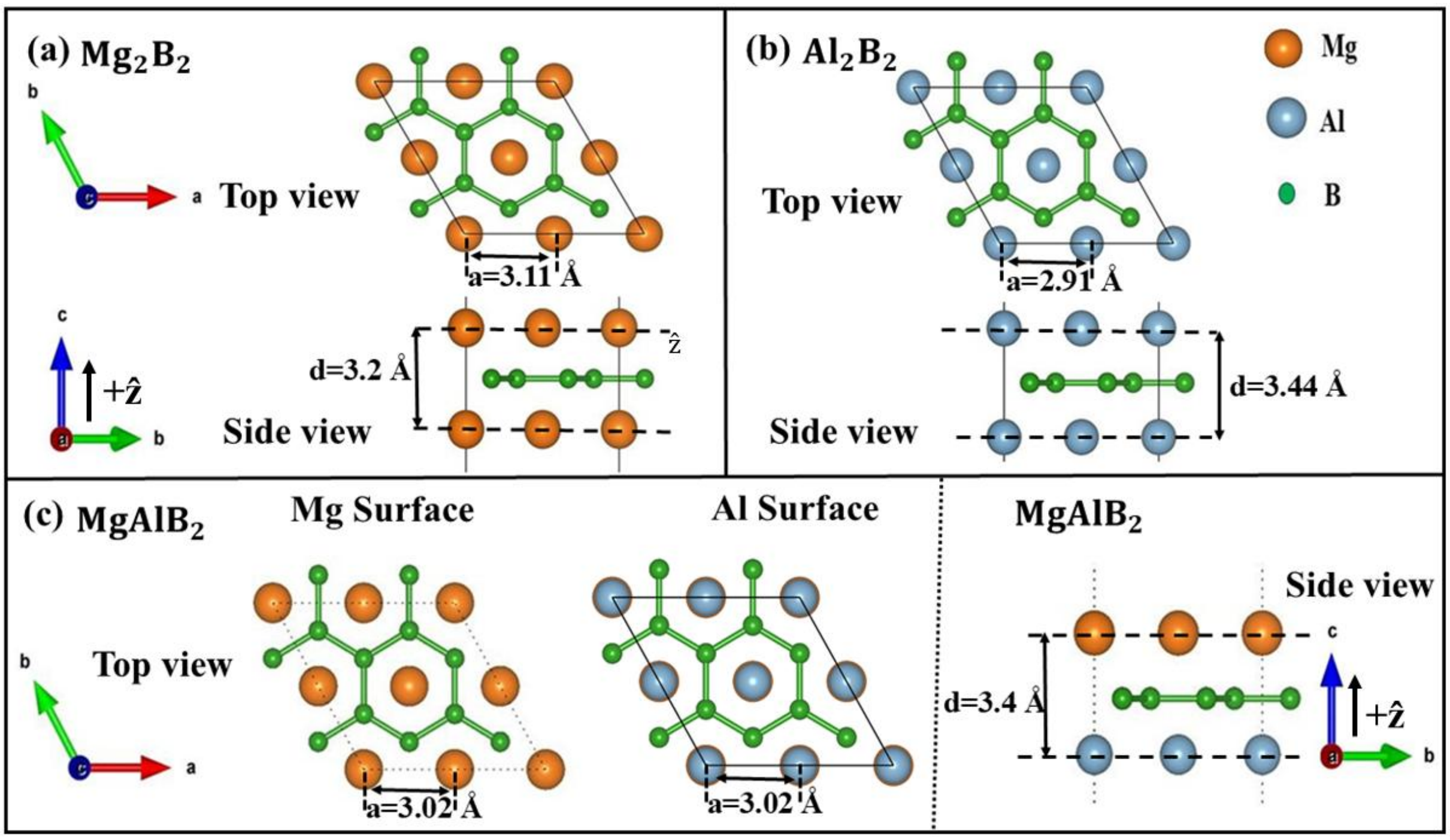

**Fig. 1** Top and side views of (a) $Mg_2B_2$, (b) $Al_2B_2$, and (c) $MgAlB_2$ with upper (Mg-surface) and lower surface (Al-surface).

For the calculation of formation energy, we have used a hexagonal closed pack (hcp) phase of magnesium [72], a face-centered cubic (fcc) phase of aluminum [73], and a rhombohedral phase of boron [74], in finding the chemical potential of the respective atoms. In **Table S1**, we have shown the different stable phases of the constituent elements of $MgAlB_2$, and the Li atom, along with energy per atom. We obtained the formation energy of Janus $MgAlB_2$ as -0.067 eV. We have also calculated the cohesive energy of the Janus $MgAlB_2$ taking the energy of isolated constituent atoms as reference. The calculated cohesive energy is - 4.45 eV/atom. The negative formation energy and cohesive energy indicate that the formation of the $MgAlB_2$ monolayer is energetically favorable relative to its constituent elements, thereby indicating its potential experimental accessibility under appropriate synthesis conditions. This negative formation energy confirms that the elements are strongly bound internally compared to their bulk phase. This result reinforces the notion that $MgAlB_2$ is a thermodynamically stable material. The introduction of elemental heterogeneity through the addition of trivalent Al atoms breaks the out-of-plane symmetry of the pristine $Mg_2B_2$, giving rise to a dipole moment and a finite electric field. We have calculated the dipole moment and the planar average potential along the z-axis of the Janus $MgAlB_2$, pristine $Mg_2B_2$, and pristine $Al_2B_2$, respectively. Notably, we define the positive z-axis (+$\hat{z}$ direction) as pointing from the Al surface to the Mg surface of the Janus $MgAlB_2$ monolayer (see **Fig. 1)**. The Janus $MgAlB_2$ possesses an out-of-plane dipole moment of +0.39 $\hat{z}$ Debye, whereas the pristine $Mg_2B_2$ and $Al_2B_2$ possesses zero dipole moment as shown in **Table 1**. For $MgAlB_2$, the presence of a finite out-of-plane dipole moment and a nonzero electric field is evident from the planar-averaged potential in **Fig. 2(a)**. The planar averaged electrostatic potential of $MgAlB_2$ exhibits a pronounced asymmetry along the out-of-plane direction, indicating the formation of a dipole moment across the monolayer. From the **Fig. 2(a)**, it is shown that the electrostatic potential profile is not symmetric around the slab region, resulting in a finite vacuum-level difference $\Delta\Phi = 0.47$eV between the two surfaces. The built-in electric field can be calculated from the slope of the two minima of the electrostatic potential of the Mg and Al surfaces. The calculated electric field is found to be 2.55 $eVÅ^{-1}$ (-$\hat{z}$) directed from Mg to Al surface originated from the electronegativity difference between Mg (electronegativity according to Pauling's scale, $\chi$ = 1.31) and Al (electronegativity according to Pauling's scale, $\chi$ = 1.61). In the **Fig. S3**, we have shown the electrostatic profile of both pristine $Mg_2B_2$ and $Al_2B_2$. The electrostatic potential profiles of both monolayers show a symmetric charge distribution, indicating a zero built-in electric field and a zero dipole moment.

**Table 1** Dipole moment of different monolayers.

| Monolayer | Dipole moment (Debye) |
|---|---|
| $MgAlB_2$ | +0.39 $\hat{z}$ |
| $Mg_2B_2$ | 0 |
| $Al_2B_2$ | 0 |

We also calculated the electron localization function (ELF) maps to obtain an in-details picture of the electronic phenomena and bond characteristics. We have shown the ELF maps of the (001) and the (100) planes of Janus $MgAlB_2$ in **Fig. 2(b)**. The likelihood of localisation of an electron can be demonstrated through the range of ELF. ELF values range from 0 to 1.0. ELF = 0 in the plot represents very low electron density, ELF = 0.5 represents fully delocalized electrons, ELF = 1.0 represents completely localized electrons. Additionally, we have shown the ELF maps of the (001) and the (100) planes of $Mg_2B_2$ in **Fig. S3(c)** and $Al_2B_2$ in **Fig. S3(d),** respectively. The ELF maps of $MgAlB_2$, $Mg_2B_2$, and $Al_2B_2$ reveal pronounced electron accumulation around the B (boron) atoms and along the B-B bonds, signifying the stronger covalent character of the boron network. The ELF maps of $Al_2B_2$ exhibits relatively stronger electron localization near the Al-B region due to the higher electronegativity of Al, indicating stronger Al-B hybridization compared to the Mg-B region in $Mg_2B_2$. Furthermore, ELF maps $Mg_2B_2$ show greater electron depletion around Mg atoms, whereas, ELF maps of $Al_2B_2$ indicate a significant accumulation of the charge on the Al-surface. The substitution of Al in Janus $MgAlB_2$ modifies the electronic environment by inducing asymmetric charge redistribution across the monolayer. Because Al is more electronegative than Mg, enhanced electron accumulation appears near the Al-terminated face, whereas comparatively reduced electron localization is observed on the Mg-terminated face. This asymmetric charge distribution can create a surface-confined electron layer and is expected to influence the surface electrostatic potential, Li adsorption behavior, and Li-ion diffusion.

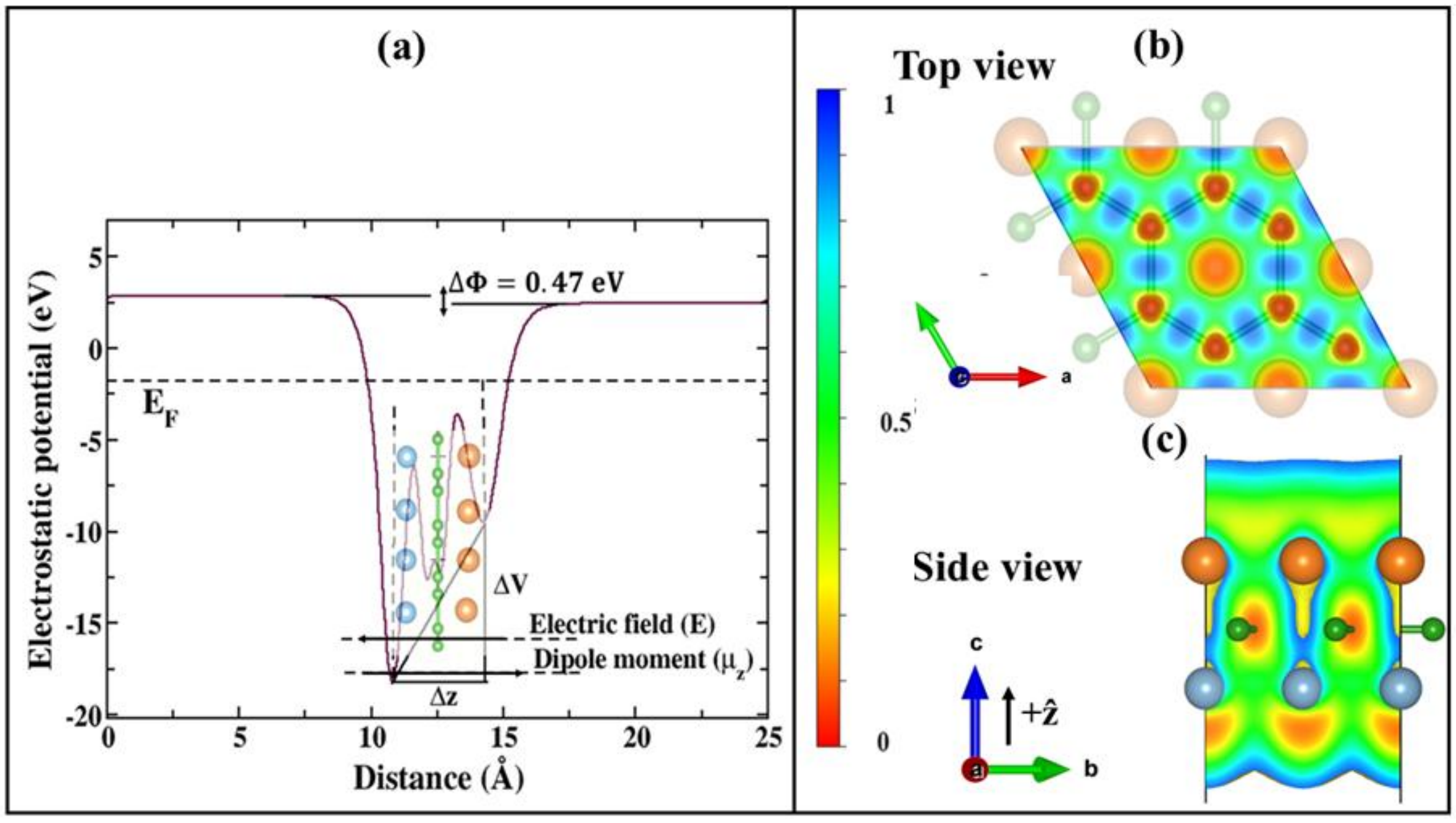

**Fig. 2** (a) Average electrostatic potential of Janus $MgAlB_2$ along the z-axis, (b) top view and (c) side view of ELF map of Janus $MgAlB_2$.

### 3.1.2. Dynamical stability: phonon dispersion

Phonon dispersion calculations were performed using the finite-displacement (FD) method as implemented in VASP in combination with the Phonopy package on a $4 \times 4 \times 1$ supercell. The phonon band structure, total phonon density of states (DOS), and atom-projected phonon density of states (PDOS) were subsequently obtained. Our results show that the calculated phonon frequencies of pristine $MgAlB_2$ are positive in the whole Brillouin zone. Phonon dispersions (see **Fig. 3(a)**) show no imaginary modes throughout the Brillouin zone, confirming the dynamical stability of the Janus $MgAlB_2$. Such dynamical stability plays a vital role in maintaining the structural integrity during repeated lithiation/delithiation cycle process and strain effect. [75] The phonon spectra exhibit the high-frequency optical branch extending up-to 28 THz (933.9 cm$^{-1}$). The high-frequency optical modes in the range of 20-28 THz predominantly originates from the B atoms, as confirmed by the phonon projected density of states (PDOS) in **Fig. 3(b)**. The appearance of these high-energy optical branches is attributed to the strong covalent B-B bonding within the monolayer, indicating in-plane bonding interactions. The appearance of such high frequency phonon spectrum of $MgAlB_2$ exceeds those of the reported maximum phonon frequencies of $FeB_2$ (854 cm$^{-1}$)[76], TiC (810 cm$^{-1}$)[77], silicene (510 $cm^{-1}$)[78], and $MoS_2$ (473 $cm^{-1}$)[79]. Moreover, the steep dispersion of the acoustic branches near the gamma (Γ) point suggests relatively high phonon group velocities, which may contribute to efficient in-plane thermal transport and structural stability during lithiation/delithiation cycles.

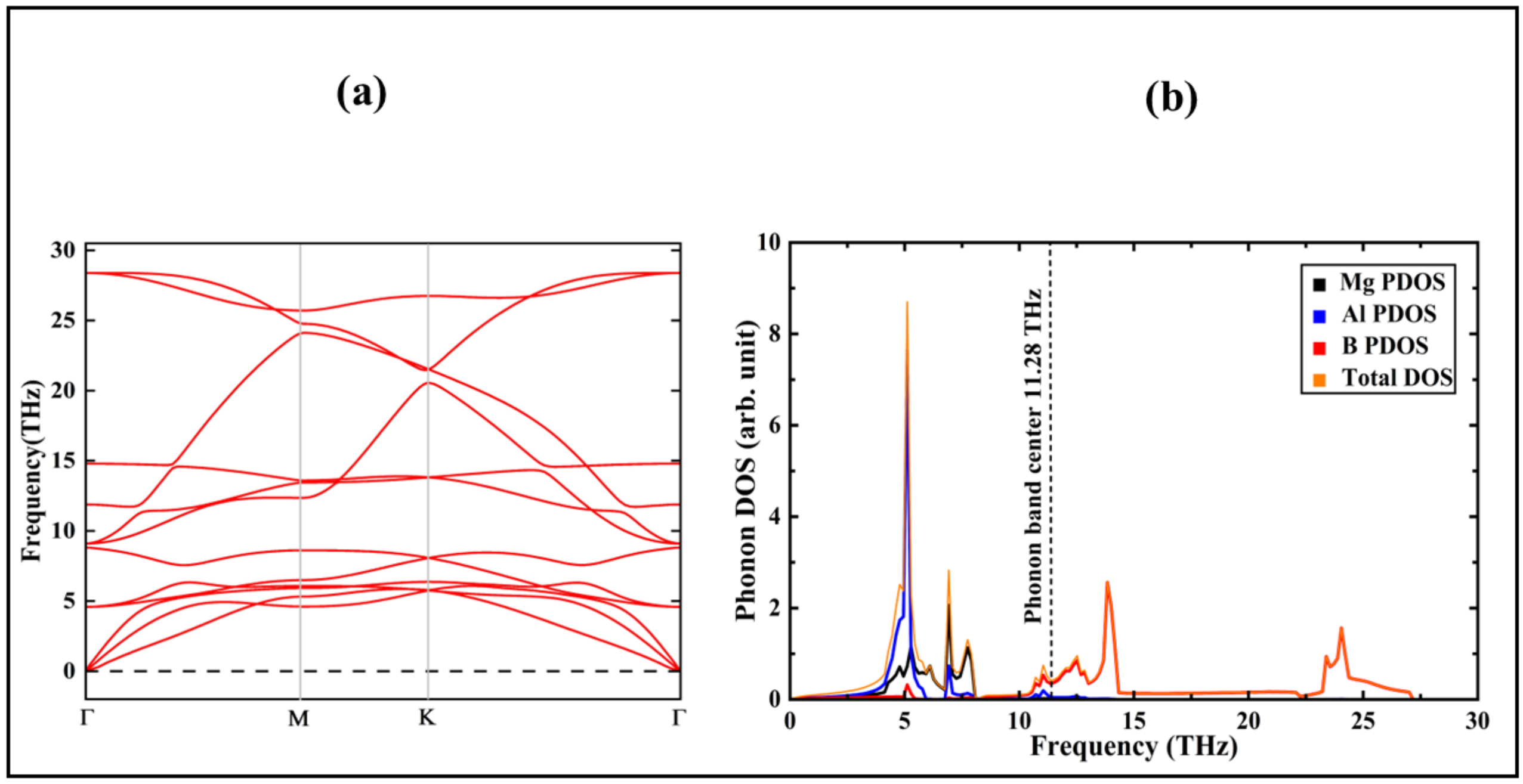

**Fig. 3** (a) Phonon dispersion spectra of the pristine Janus $MgAlB_2$ along the high symmetry directions, and (b) phonon density of states with average phonon band center weighted over total phonon density of states (DOS).

We have also calculated the phonon band structure, atom-projected density of states of a single lithium adsorbed $MgAlB_2$ monolayer at both Mg and Al surfaces on a supercell of 2x2x1. We define the 'phonon band center', i.e., the average phonon frequency weighted by the DOS, to quantify the average vibrational frequency of the system. Mathematically, it is written as:[80]

$$\omega_{ave} = \frac{\int \omega \times g(\omega) d\omega}{\int g(\omega) d\omega}, \quad (2)$$

where, here $g(\omega)$ represents the phonon density of states. Also, we have calculated the lithium phonon band center weighted over the lithium projected density of states. The 'lithium projected phonon band center', which can be considered as the centroid of the Li-projected phonon DOS, governs the dynamical stiffness of Li within the host lattice and plays a crucial role in determining lithium-ion mobility. A clear correlation is observed between lattice vibrational softening and the diffusion barrier governing Li ion diffusion. The total phonon band center of pristine $MgAlB_2$ is calculated to be 11.28 THz (see **Fig. 3(b)**). Upon Li adsorption, the total phonon band centers decrease to 10.01 and 10.11 THz for the Mg- and Al-terminated surfaces, respectively, as shown in

**Fig. 4**. Furthermore, the Li-projected phonon band centers are found to be 4.32 THz and 7.36 THz on the Mg and Al surfaces, respectively. The reduction in both the total and Li-projected phonon band centers relative to the pristine monolayer indicates a softening of the vibrational spectrum and a decrease in the average vibrational frequency. Such phonon softening is generally associated with weaker restoring forces acting on diffusing ions, thereby facilitating ion migration through a more softening lattice environment. Consistent with this picture, the Mg-terminated surface exhibits the lowest Li-projected phonon band center (4.32 THz) and an ultralow Li-ion diffusion barrier of 17.1 meV (see **Section 5**). This result suggests that low-frequency vibrational modes effectively assist Li-ion hopping between neighboring adsorption sites, leading to nearly barrier-free diffusion and excellent rate performance. In contrast, the higher Li-projected phonon band center on the Al surface (7.36 THz) reflects a comparatively stiffer local vibrational environment arising from stronger Li-host interactions. The increased lattice rigidity is accompanied by a larger diffusion barrier of 43.6 meV (see **Section 5**), indicating relatively less favorable Li-ion mobility compared with the Mg surface. Nevertheless, the barrier remains sufficiently low to support rapid ion transport. Overall, the strong correspondence between the phonon band centers and the calculated migration barriers highlights the important role of lattice dynamics in governing Li diffusion. The coexistence of stable Li adsorption and exceptionally fast ion transport favors Janus $MgAlB_2$ as a better anode material for LIBs.

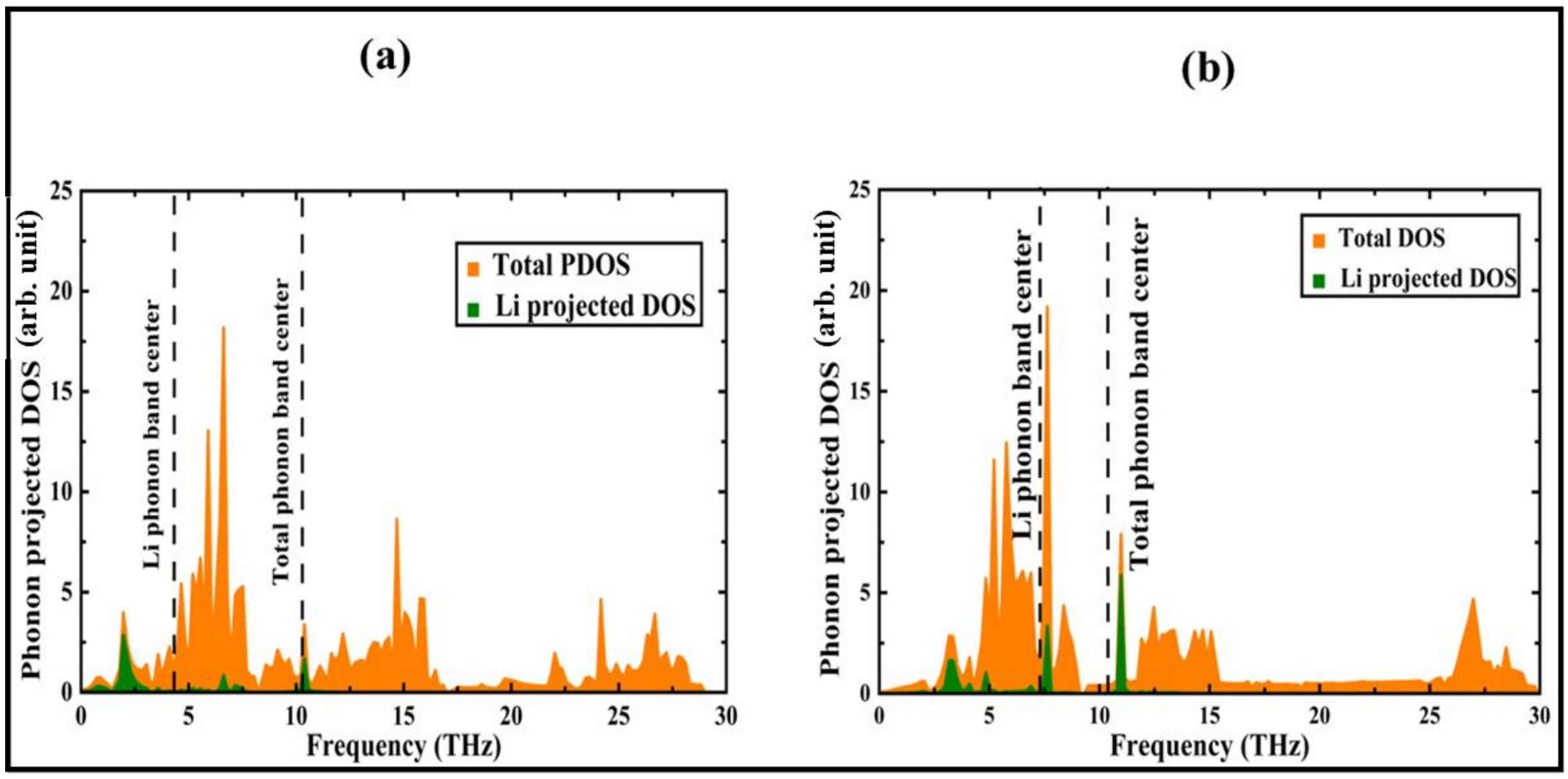

**Fig. 4:** Total and lithium-projected phonon DOS of single adsorbed Janus $MgAlB_2$ (a) at S1-site on the Mg surface, and (b) at S4-site on the Al surface, respectively.

### 3.1.3 Mechanical stability: elastic constants

The mechanical stability of the pristine $MgAlB_2$ monolayer was assessed through the calculation of its in-plane elastic constants. We examined the mechanical stability of the pristine $MgAlB_2$ monolayer by calculating linear elastic constants. The material meets Born-Huang's mechanical stability criterion for a 2D hexagonal system, $C_{11} > 0$, $C_{11}C_{12} - C_{12}^2 > 0$, and $C_{66} = \frac{C_{11}-C_{12}}{2} > 0$. [81] The obtained 2D linear elastic constants for the $MgAlB_2$ monolayer are $C_{11}$= 179.450 $Nm^{-1}$, $C_{22}$ = 179.450 $Nm^{-1}$, $C_{12}$ = 51.649 $Nm^{-1}$, $C_{44}$ = 63.90 $Nm^{-1}$. This robust mechanical stability confirms that the structure can withstand distortions during recursive lithiation and delithiation, removing any permanent structure-breaking or cracking. The mechanical stability of the Janus $MgAlB_2$ monolayer during lithiation and delithiation is expected to facilitate enhanced Li storage. To assess the extent of possible structural deformation during charge/discharge cycling, the angle-dependent Young's modulus, Y(θ), and Poisson's ratio, ν(θ), were derived from the elastic constants using the following relations:[13]

$$Y(\theta) = \frac{C_{11}C_{22}-C_{12}^2}{C_{11}\cos 2\theta + C_{22}\sin 2\theta + \left(\frac{C_{11}C_{22}-C_{12}^2}{C_{44}} - 2C_{12}\right)\cos^4\theta \sin^4\theta}, \quad (3)$$

$$\vartheta(\theta) = \frac{(C_{11}+C_{22}-\frac{(C_{11}C_{22}-C_{12}^2)}{C_{44}})\sin^2\theta\cos^2\theta - C_{12}(\cos^4\theta+\sin^4\theta)}{C_{11}\cos^4\theta + C_{22}\sin^4\theta + (\frac{C_{11}C_{22}-C_{12}^2}{C_{44}} - 2C_{12})\sin^2\theta\cos^2\theta} \quad . \quad (4)$$

The angle-dependent functions are plotted in **Fig. 5**. The polar plot in **Fig. 5(a)**, demonstrating the variation of the Young's modulus ($Nm^{-1}$), as a function of orientation (θ), shows a circular variation, indicating the isotropic nature of the material. Our calculated Young's modulus, with a value of 163 $Nm^{-1}$, deforms uniformly upon application of strain, which originates due to its intrinsic hexagonal symmetry. Our calculated Young's modulus, can be compared to that of other transition metal chalcogenides, such as $MoS_2$ (170 $Nm^{-1}$) [82], and lower than that of graphene (approximately 340 $Nm^{-1}$). This in-plane uniform Young's modulus indicates that the material possesses a moderate stiffness with good stability, making the structure beneficial for smooth lithium-ion diffusion. Moreover, this keeps the geometry intact from structural cracking during lithium charge/discharge cycles. The polar plot of Poisson's ratio in **Fig. 5(b)** reflects an isotropic nature, with a small value of 0.2, indicating a little lateral expansion/contraction upon application of lateral strain. On the whole, $MgAlB_2$ can escape from significant deformation during charging/discharging cycles, making it more beneficial for use as an electrode in LIBs.

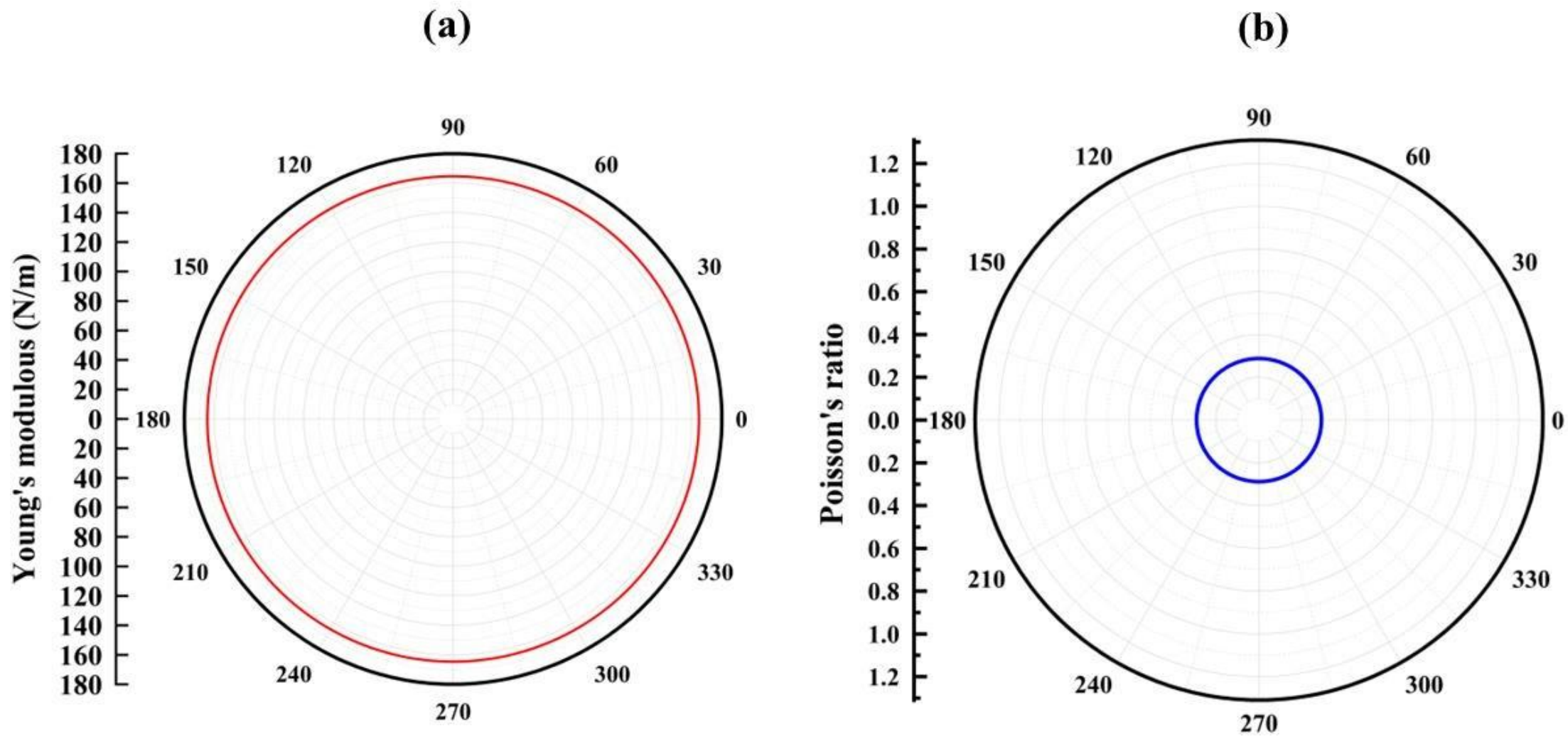

**Fig. 5** Direction dependent variations of (a) Young's modulus, Y($\theta$), and (b) Poisson's ratio, ν($\theta$), for the Janus $MgAlB_2$ monolayer.

### 3.1.4 Thermal stability: finite temperature AIMD

Ab initio molecular dynamics (AIMD) simulations were conducted on the pristine Janus $MgAlB_2$ structure at an elevated temperature of 500K to find its thermal stability. To ensure computational accuracy, we had performed AIMD simulation on a 2×2×1 supercell with a time step of 1 fs for a total simulation time of 10 ps. Temperature regulation throughout the simulation was achieved using the Nosé-Hoover thermostat algorithm, which effectively maintains constant temperature conditions through the introduction of an artificial heat bath coupled to the system. Upon completion of the AIMD simulation, our analysis revealed that the final structural configuration maintained its integrity and stability, as clearly illustrated in the **Fig. S16** (see in **SI**). The absence of significant energy fluctuations or any structural distortion during the simulation indicates the thermal stability of Janus $MgAlB_2$. Initially, the atomic position fluctuations demonstrated considerable magnitude, indicating high thermal motion. However, as simulation time advanced, these vibrational amplitudes gradually diminished and stabilized as depicted in the right panel of **Fig. S16** (see in **SI**). This progressive reduction in fluctuation magnitude serves as strong evidence that the system successfully reached thermal equilibrium, with the atomic configuration naturally evolving toward the minimum energy state. The final configuration after a 10 ps AIMD simulation (see **Fig. S16** in **SI**) shows no noticeable structural deformation or bond dissociation, confirming the thermal stability of pristine $MgAlB_2$. The thermal stability of the material demonstrates that the structure may remain intact during repeated charging/discharging cycles, which further support multilayer storage of lithium. The retention of structural integrity at elevated temperatures demonstrates the robustness of the Janus $MgAlB_2$ monolayer and supports its potential application as an anode material.

### 3.2 Electronic properties

The electronic properties of the Janus $MgAlB_2$ monolayer were examined through its electronic band structure and projected density of states (PDOS), as illustrated in **Fig. 6**. The calculated electronic band structure reveals that Janus $MgAlB_2$ possesses an intrinsic metallic nature, as multiple energy bands cross the Fermi line. The absence of an electronic band gap indicates the availability of continuous conducting channels, which facilitate efficient electron transport throughout the material. Furthermore, the PDOS analysis demonstrates significant orbital hybridization among the constituent atoms near the Fermi line, suggesting strong electronic interactions and enhanced charge delocalization. Such metallic behavior is highly advantageous for electrochemical energy-storage applications, as it promotes rapid charge transfer kinetics and reduces electronic transport resistance. The PDOS analysis in **Fig. 6(b)** further clarifies the orbital origin of the electronic states. The electronic states near the Fermi level originate from the B-2p, Al-3p, Mg-3s, and Mg-3p orbitals. Consistent with the band structure, the finite density of states near the Fermi level confirms the metallic character of $MgAlB_2$ and the availability of electronic conduction channels. To elucidate the electronic modification induced by the Janus structure, we further analyzed the electronic structures of the parent counterparts, $Mg_2B_2$ and $Al_2B_2$ (see **Fig. S4 and S5**).

In $Mg_2B_2$, the electronic states in the vicinity of the Fermi level are predominantly derived from the B-2p orbitals, whereas the Mg-3s states contribute negligibly small (see **Fig. S4)**. This indicates that the boron framework largely determines the low-energy electronic structure of the system. However, despite the significant B-2p contribution near the Fermi level, the boron atoms are not directly exposed at the surface. Since Li adsorption occurs on the Mg-terminated surfaces, the weak participation of Mg states near the Fermi level may limit Li-substrate interactions. Furthermore, the electronic states near the Fermi level remain largely localized within the boron network, which may restrict electronic delocalization and consequently hinder efficient charge transport during Li-ion migration. To overcome these limitations, we explored a Janus design strategy by replacing one of the Mg layers with an Al layer. The motivation behind this approach is to introduce additional metal-derived electronic states near the Fermi level and thereby enhance orbital hybridization and electronic delocalization. This expectation is supported by the electronic structure of $Al_2B_2$, where the states around the Fermi level originate from a combination of Al-3s, Al-3p, and B-2p orbitals (see **Fig. S5**). In the Janus $MgAlB_2$ monolayer, the electronic states near the Fermi level arise from the hybridized contributions of B-2p, Mg-3s, Mg-3p, and Al-3p orbitals. The involvement of multiple atomic species in the low-energy electronic structure results in stronger orbital mixing and a more delocalized distribution of electronic states. Compared with $Mg_2B_2$, where the Fermi-level states are primarily confined to the boron framework, $MgAlB_2$ exhibits a more interconnected electronic network that provides continuous conduction pathways across the monolayer. Such enhanced electronic delocalization is expected to facilitate charge transfer, strengthen Li-substrate interactions, and promote efficient electrochemical performance during lithium storage and transport. In case of $Al_2B_2$ there are multiple energy bands across the Fermi line including Dirac-like linear dispersions (between K-Γ high-symmetry points in the Brillouin zone) and several nodal crossing points near the Fermi level (between M-K) (see **Fig. S5(a)).** In $Mg_2B_2$ both these Dirac points were unoccupied. Upon formation of Janus $MgAlB_2$ by the substitution of Al in $Mg_2B_2$ makes Fermi level upshifted. Thus, one of the Dirac points near the K-point gets occupied in $MgAlB_2$. It is to be noted that, the nodal/Dirac points presented in parent $Mg_2B_2$ and $Al_2B_2$ remain preserved in the Janus $MgAlB_2$.

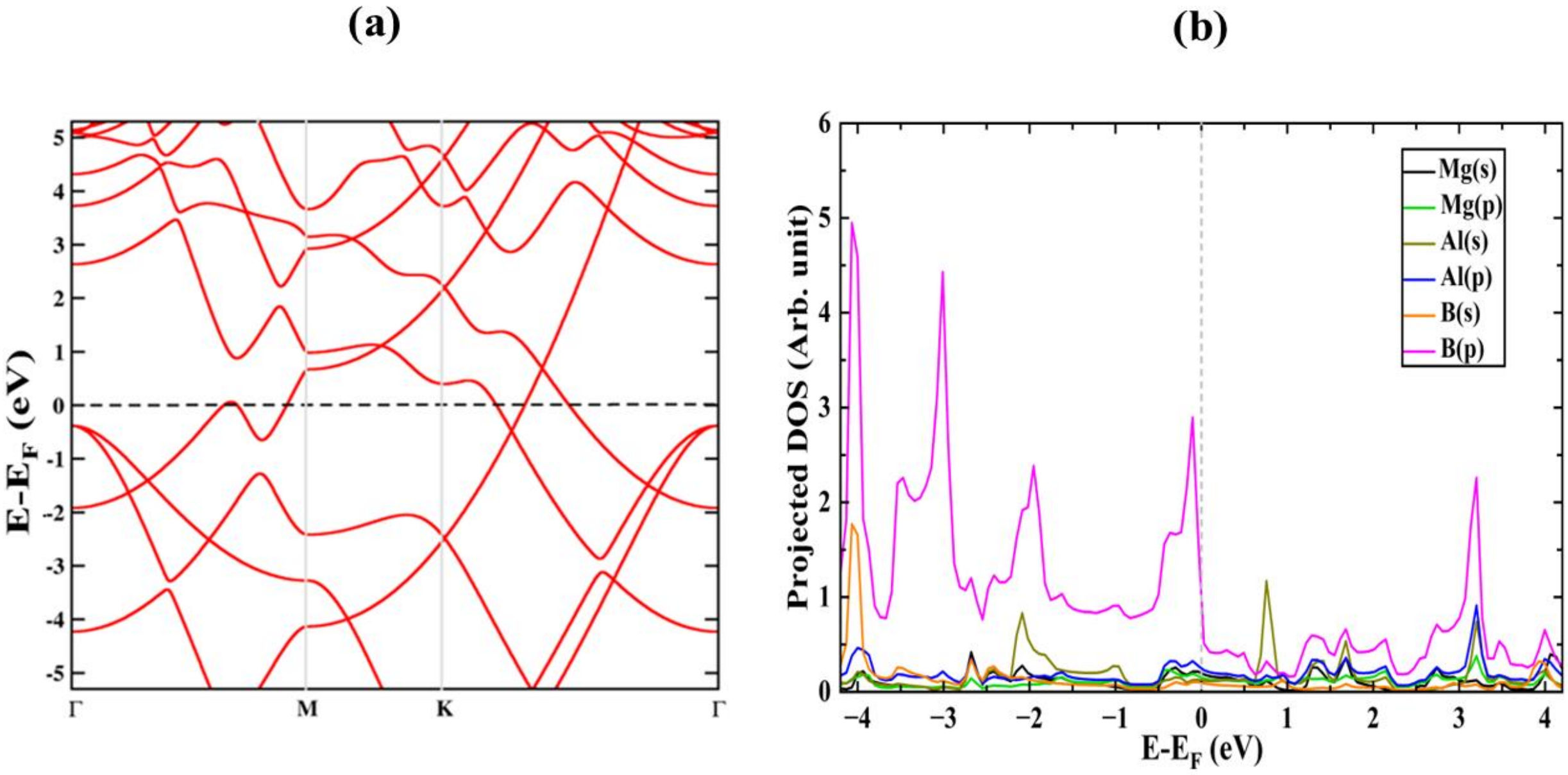

**Fig. 6** (a) Band structure and (b) PDOS of Janus $MgAlB_2$ unit cell.

### 3.3 Adsorption of the single Li atom on the $MgAlB_2$ monolayer

The above-mentioned stabilities and properties favored the potentiality of the Janus $MgAlB_2$ as an anode material. To study the electrochemical properties, we have investigated the adsorption behavior of a single lithium atom on the Janus $MgAlB_2$. We studied the adsorption property of the lithium atom on the Janus $MgAlB_2$ to determine the key parameters, namely, theoretical storage capacity, open circuit voltage (OCV), and diffusion barrier. The substitution of one Mg layer by Al breaks the structural symmetry of the parent MBene, leading to inequivalent chemical environments on the two surfaces of the Janus $MgAlB_2$ monolayer. This structural asymmetry effect results in distinct Li adsorption characteristics compared to pristine $Mg_2B_2$ and $Al_2B_2$ monolayers. The substitution of the more electronegative trivalent Al atom induces asymmetry, leading to the formation of two distinct adsorption basins and an adsorption-suitable chemical environment for lithium adsorption, in contrast to the single adsorption basin observed in both the pristine $Mg_2B_2$ and $Al_2B_2$ monolayers. Due to the origin of an asymmetric charge distribution on the incorporation of higher electronegative trivalent Al, a dipole moment is generated, which gives rise to site-dependent adsorption. We took a 2×2×1 supercell of Janus $MgAlB_2$ for single Li adsorption at different sites. We identified six adsorption sites viz: Site-1 (S1; top of boron site), Site-2 (S2; top of Mg site), and Site-3 (S3; top of bridge center of two boron atoms on the hexagon) along the Mg surfaces, and three similar sites along the Al surfaces as shown in **Fig. 7(a)**, and Site-4(S4; top of boron site), Site-5(S5; top of Al site) and Site-6(S6; top of bridge center of two boron atoms lies on the hexagon) as shown in **Fig. 7(b)**. We have calculated the adsorption energies $E_{ave}$ of a single lithium atom at different sites on both surfaces. The average adsorption energy ($E_{ave}$) is calculated by using the following equation:[58]

$$E_{ave} = \frac{\left(E_{MBene@Li_p} - E_{MBene@Li_{p-1}} - rE_{Li}\right)}{r}, \quad (5)$$

where $E_{MBene@Li_p}$ is the energy of the p layers of Li adsorbed MBenes ($MgAlB_2/Mg_2B_2/Al_2B_2$) compound, and $E_{MBene@Li_{p-1}}$ is the energy of the (p-1) layers of Li adsorbed MBenes and $E_{Li}$ is the energy of a single lithium atom in its stable bcc phase [49]. Here r is the number of lithium atoms adsorbed on the monolayer in the p layer. For single lithium adsorption, p = 1 and r = 1. Using this **eq. (5)**, we have calculated the adsorption energy of single Li atom at different adsorption sites and have shown in **Fig. 7**. The calculated adsorption energies at all the symmetric sites are negative. This negative adsorption energy indicates that the adsorption process is an exothermic process and energetically favored. Among these sites, the strongest adsorption occurs at the S1-site among the three sites on the Mg surface, with the lowest adsorption energy of -0.32 eV, followed by the S3-site. S4-site shows the strongest binding among all six sites, including three sites on the Al surface with the lowest adsorption energy of -0.61 eV. The adsorption energies, adsorption heights, bond distances, the amount of transferred charge, and the dipole moment upon adsorption of a single lithium at different sites are shown in **Table 2.** Upon incorporation of a trivalent Al atom in $Mg_2B_2$, a strong intrinsic dipole moment along the out-of-plane direction is generated, having a value of +0.39 $\hat{z}$ Debye directed from the Al surface toward the Mg surface. This dipole moment is generated due to the difference in electronegativity between the trivalent Al (electronegativity, $\chi$=1.61in Pauling's scale) and the divalent Mg (electronegativity, $\chi$ =1.31 in Pauling's scale), giving rise to an electron-rich Al surface and the electron-scarce Mg surface. The 3p orbital of Al hybridizes strongly with the B-2p orbital (see **Fig. 6(b)**), making an electron-rich Al-B region that creates energetically favored adsorption sites for the lithium atoms. The relatively stronger binding of the lithium atom on the Al surface can be analyzed with the help of the ELF maps of Janus $MgAlB_2$. The ELF map (see **Fig. 2(c)**) demonstrates relatively larger electron localization near the Al-B region compared to the Mg surface. This larger likelihood of electron density near the Al-B region occurs due to the stronger hybridization between the Al-3p and B-2p orbitals (see **Fig. 6(b)**).

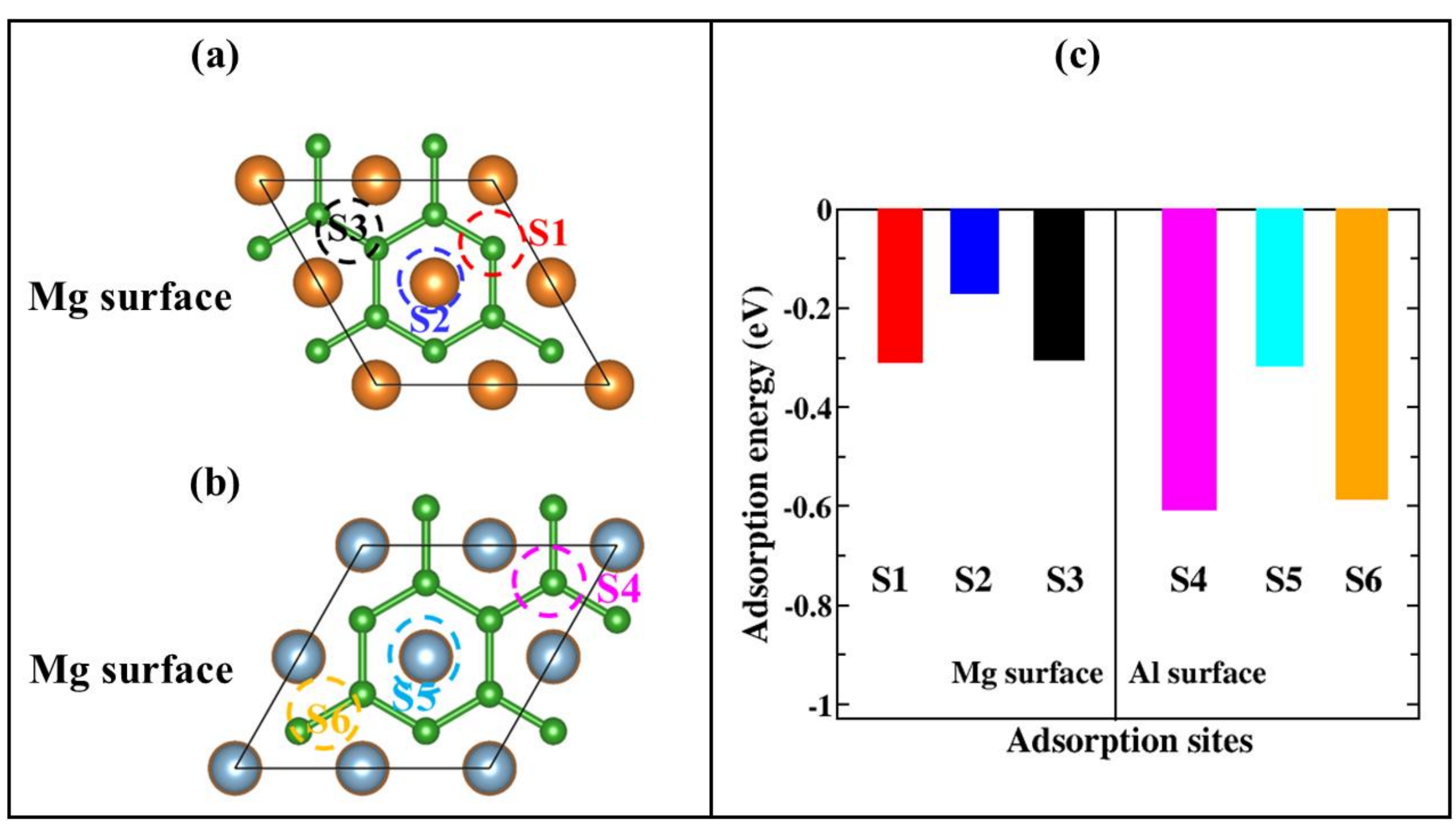

**Fig. 7** (a) Top view of adsorption sites on the upper Al surface, (b) top view of adsorption sites on the upper Mg surface, (c) histogram of adsorption energies at different adsorption sites on the Janus $MgAlB_2$ monolayer.

Whereas, relatively weaker electron localization near the Mg surface indicates an ionic character, which results in comparatively weaker lithium adsorption. This weaker adsorption on the Mg surface is further confirmed by the EDD plot (see **Fig. 8**), which demonstrates the occurrence of (relatively weaker than at the S1-site on the Mg surface) a charge transfer of 0.827 $e^-$ from the lithium atom. Moreover, the addition of Al thus opens additional three new adsorption sites that further open up the possibility of multilayer adsorption of lithium atoms compared to its parent symmetric $Mg_2B_2$ monolayer. Additionally, the dipole moment stabilizes further adsorption of lithium atoms on both surfaces by creating an internal electric field that causes more electrostatic interaction through enhanced charge transfer. Also, we have calculated the dipole moment to gain an in-details picture of the electrostatic interactions at different sites after the adsorption of a single lithium. The obtained dipole moments, as detailed in **Table 2,** demonstrate that the adsorption of a single lithium on the Al surface yields a dipole moment of -3.21 $\hat{z}$ Debye at the S4-site, whereas the obtained dipole moment is +2.6 $\hat{z}$ Debye at the S1-site on the Mg surface. Thus, surface polarization effect results in a substantial charge transfer, which stabilizes the lithium atoms on the surface. This built-in dipole moment has a significant role in storing and stabilizing multiple lithium atoms on the Janus $MgAlB_2$ surface through electrostatic interaction.

To get deep insight into the interaction between the lithium atom and the bare monolayer, we determined the amount of transferred charge from the lithium atom to the monolayer and the orbital hybridization of the constituent atoms of the lithium adsorbed monolayer. To elucidate the interaction, Bader charge analysis was employed to quantify the electronic charge transferred between the adsorbed Li atom and the pristine monolayer. [83]

**Table 2** The adsorption energies, adsorption heights, bond distances, amount of charge transfer, and dipole moments at different sites on the Janus $MgAlB_2$ .

| Adsorption sites | Adsorption energy (eV) | Adsorption height (From hexagon center) (Å) | Charge transfer ($e^-$) | Dipole moment (Debye) |
|---|---|---|---|---|
| S1 (Mg side) | -0.32 | 4.129 | 0.827 | +2.60 ẑ |
| S2 (Mg side) | -0.169 | 4.05 | 0.824 | +2.24 ẑ |
| S3 (Mg side) | -0.31 | 4.13 | 0.826 | +2.60 ẑ |
| S4 (Alside) | -0.64 | 3.65 | 0.853 | -3.21ẑ |
| S5 (Al side) | -0.26 | 3.94 | 0.839 | -2.32 ẑ |
| S6 (Al side) | -0.57 | 3.78 | 0.843 | -3.20 ẑ |

To further elucidate the nature of the Li-substrate interaction, the electron density difference (EDD) was computed to visualize the redistribution of charge upon adsorption. The EDD was obtained by subtracting the charge densities of the isolated constituents from that of the fully relaxed adsorption system according to the following equation:[52]

$$\rho_{net} = \rho_{MBene@Li} - \rho_{MBene} - \rho_{Li}. \quad (6)$$

In the above expression, $\rho_{MBene@Li}$, $\rho_{MBene}$, and $\rho_{Li}$ represent the electron densities of the Li-adsorbed MBene systems, the pristine MBene monolayer, and the isolated Li atom, respectively. We have plotted the EDD plots of Li adsorbed monolayers at different adsorption sites to understand the underlying interaction of the adsorption process on the Mg and Al surfaces (see **Fig. 8** and **Fig. 9)**, respectively. In the EDD plots, we have shown the spatial distribution of charge accumulation (yellow colour) and charge depletion (cyan colour) on both surfaces. We have also shown the EDD plots at different sites of both the pristine $Mg_2B_2$ and $Al_2B_2$ in **Fig. S6** and **Fig. S7** (see **SI**), respectively. The EDD maps reveal that the charge transferred from the adsorbed Li atom is redistributed over the surface of the Janus monolayer rather than remaining localized at the adsorption site. Charge accumulation is predominantly observed in the interfacial region between Li and the substrate, while charge depletion occurs around the Li atom. These features indicate net electron donation from Li to the monolayer and confirm the charge-transfer character of the adsorption process.

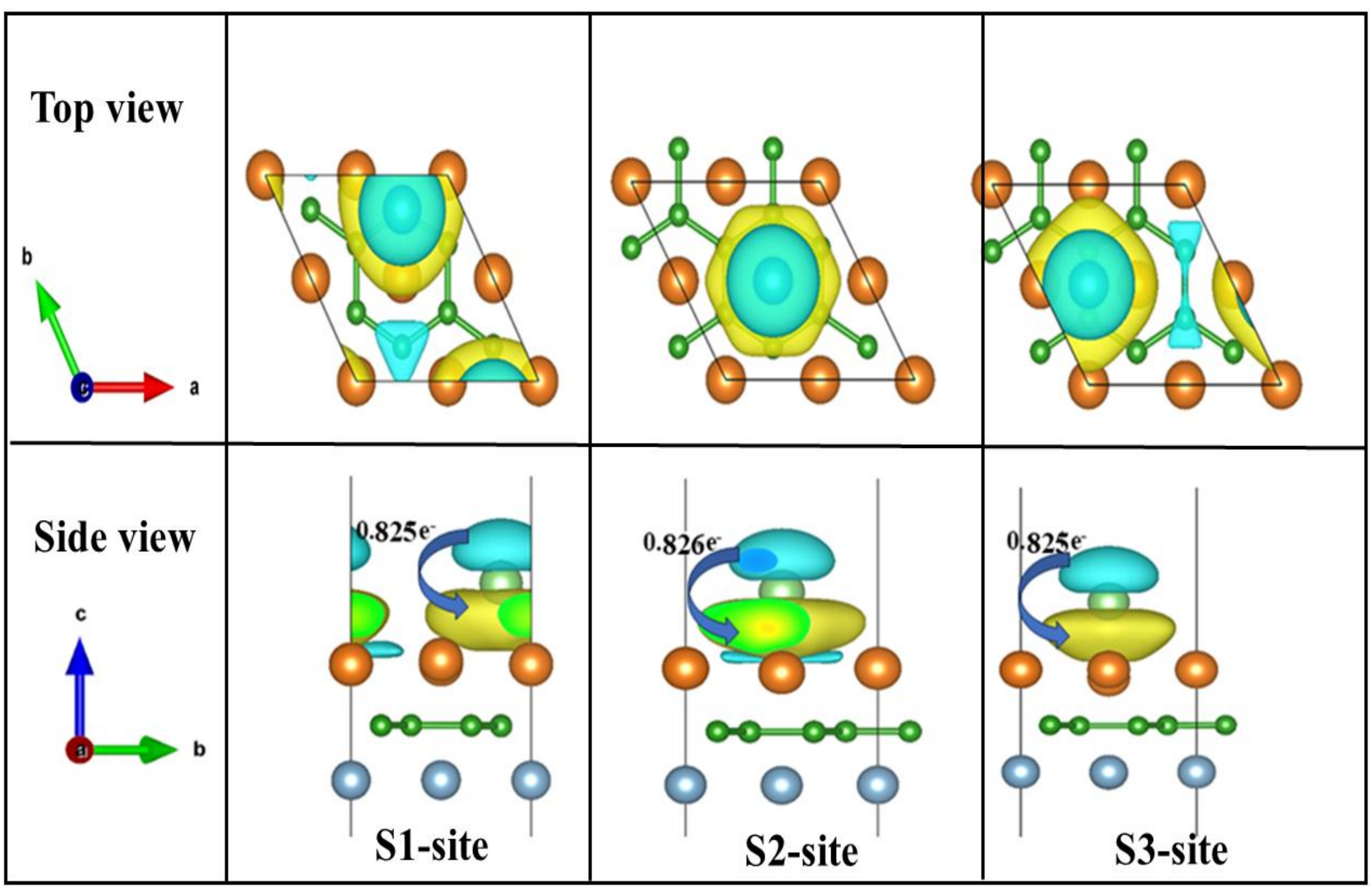

**Fig. 8** Top view & side view of the EDD plots for a Li atom adsorbed at the different sites on the Mg surface of the Janus $MgAlB_2$. In the EDD plots, the cyan and yellow colours represent regions of electron depletion and accumulation, respectively. The iso-surface value is set to 0.01 $e\text{Å}^{-3}$.

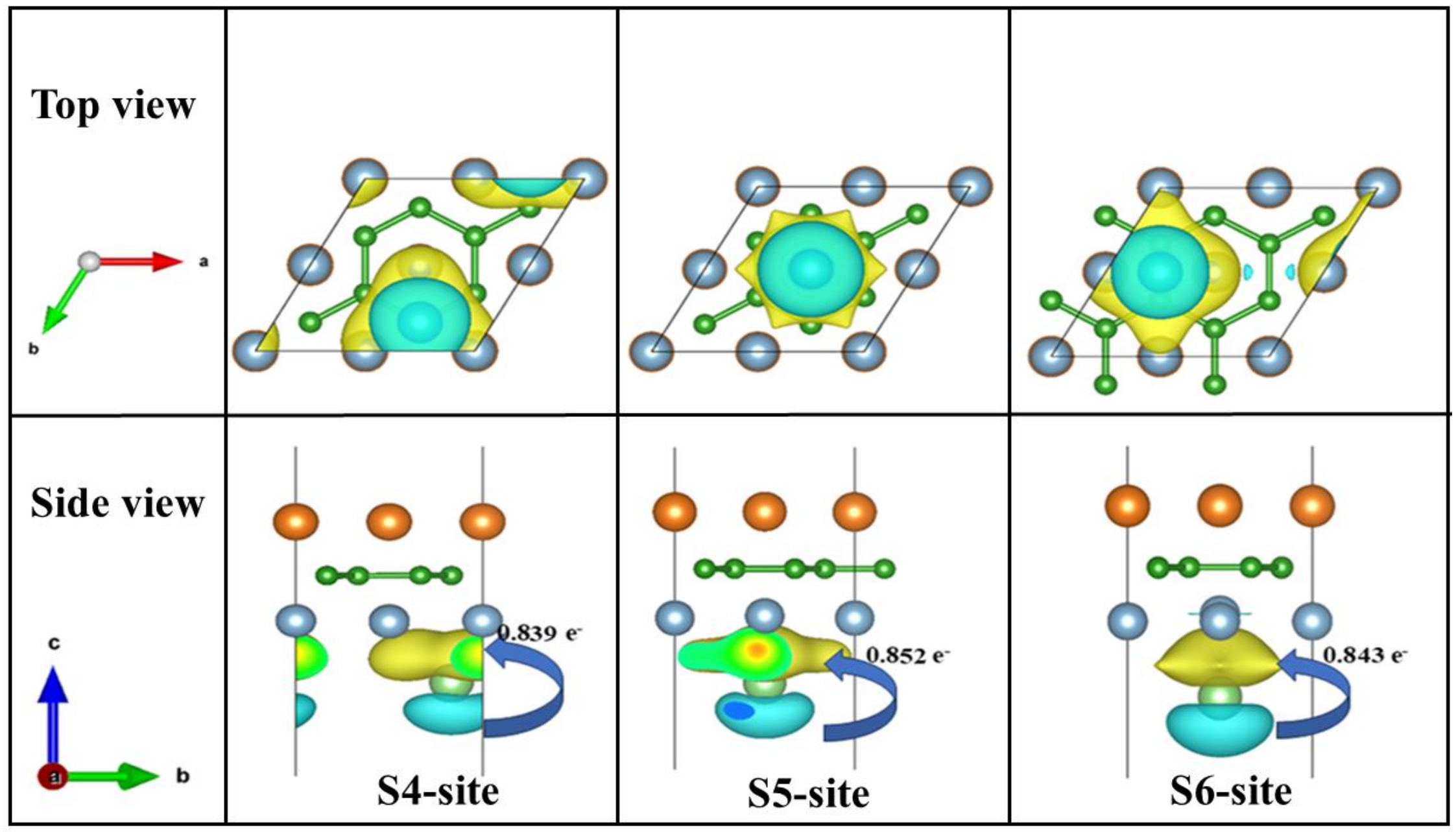

**Fig. 9** Top view & side view of the EDD plots for a Li atom adsorbed at different sites on the Al surface of the Janus $MgAlB_2$. In the EDD plots, the cyan and yellow colours represent regions of electron depletion and accumulation, respectively. The iso-surface value is set to 0.01 $e\text{Å}^{-3}$.

To quantify the amount of transferred charge, Bader charge analysis was carried out for all adsorption configurations. From the Bader charge analysis, it is observed that the amount of transferred charge at different sites for both surfaces is all more than 0.8 $\mathbf{e^-}$. Our study demonstrates that Li is an electron donor, whereas the monolayer is an electron acceptor. Notably, it is observed that there is an occurrence of pronounced charge transfer of amount 0.85 $\mathbf{e^-}$ to the monolayer at the S4-site, which is the most energetically stable site on the Al surface of the Janus $MgAlB_2$. This transferred charge is larger than the charge transfer of an amount of 0.81 $\mathbf{e^-}$ at the S1-site on the Mg surface in the parent $Mg_2B_2$ monolayer, as shown in **Fig. S6** (see **SI**), indicating the influence of asymmetric charge redistribution on the addition of Al. Similarly, the amount of transferred charge is 0.85 $\mathbf{e^-}$ at the S4-site on the Al. This transferred charge is larger than the transferred charge of an amount of 0.84 $\mathbf{e^-}$ at the S4-site in the parent $Al_2B_2$ monolayer, as shown in **Fig. S7** (see **SI**). In the pristine $Mg_2B_2$, the charge distribution is symmetric on both surfaces, indicating identical adsorption at different sites. But the addition of an Al atom to Janus $MgAlB_2$ facilitates an asymmetric charge redistribution that makes the electron-rich Al surface and the electron-scarce Mg surface. The addition of Al increases strong hybridization between Al-2p and B-2p orbitals that enhances more electronic states near the Fermi level. This enhanced PDOS increases electronic conductivity and triggers delocalization of electrons, which opens a smooth channel for lithium diffusion, giving rise to a lower diffusion barrier. The EDD distribution indicates that the transferred charge is distributed over the surface of the Janus $MgAlB_2$ monolayer rather than being localized at a specific adsorption site. Together with the Bader charge analysis, this finding demonstrates significant electron transfer from Li to the substrate and a robust Coulombic interaction between the lithium atom and the substrate. These characteristics contribute to the stability of Li adsorption and support the suitability of $MgAlB_2$ as a potential anode material for lithium-ion batteries.

### 3.4 Electronic properties of a single Li adsorbed on the Janus $MgAlB_2$ monolayer

Electronic conductivity determines the rate capability of the anode materials. To assess the impact of lithiation on the electronic transport properties, the band structure and projected density of states (PDOS) of the single lithium adsorbed $MgAlB_2$ monolayer were calculated. This analysis provides insight into the electronic conductivity after lithiation that helps elucidate the ability of the monolayer to accommodate additional Li atoms at higher concentrations. Evaluation of the density of states gave an in-detailed understanding of the hybridization between orbitals of different atoms. We have plotted the band structure and PDOS of a single lithium adsorbed Janus $MgAlB_2$ at the S1-site and S4-site in **Fig. 10** and **Fig. 11**, respectively. While the band structure and the PDOS plots of a single Li adsorbed on the Mg and Al surfaces of the Janus $MgAlB_2$ at the other stable adsorption sites are shown in **Fig. S8** and **Fig. S9** (see **SI),** respectively. The plot in **Fig. 10** and **Fig. 11** demonstrates that the bands cross the Fermi line, indicating the metallicity is retained after lithium addition. The PDOS plots in **Fig. 10** and **Fig. 11** demonstrate that a strong orbital contribution comes from Li-2s, Mg-3s, Al-3p, and B-2p orbitals near the Fermi level, indicating significant hybridization and moderate electronic interaction between the Li atom and the matrix monolayer. The PDOS plot and EDD plot for a single lithium adsorbed monolayer at the most stable adsorption site demonstrate that the transferred charge is delocalized. This delocalized charge provides a smooth diffusion channel and may enhance multilayer adsorption. Intrinsic metallic nature and significant density of states near the Fermi level show that the system explicitly keeps its metallic nature after lithium adsorption.

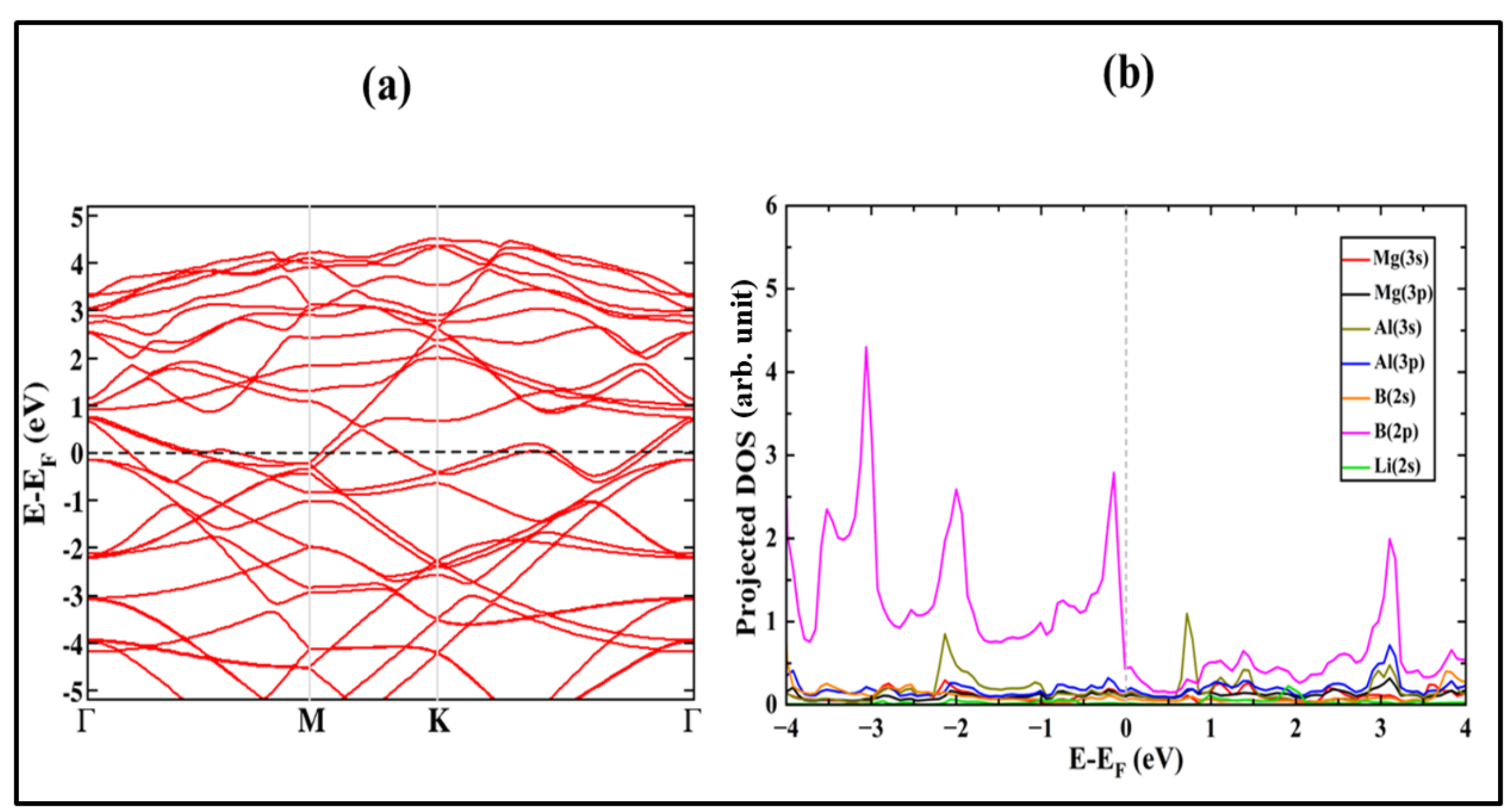

**Figure 10**. (a) Band structure and (b) PDOS of a single Li adsorbed Janus $MgAlB_2$ at S1-site along the Mg surface.

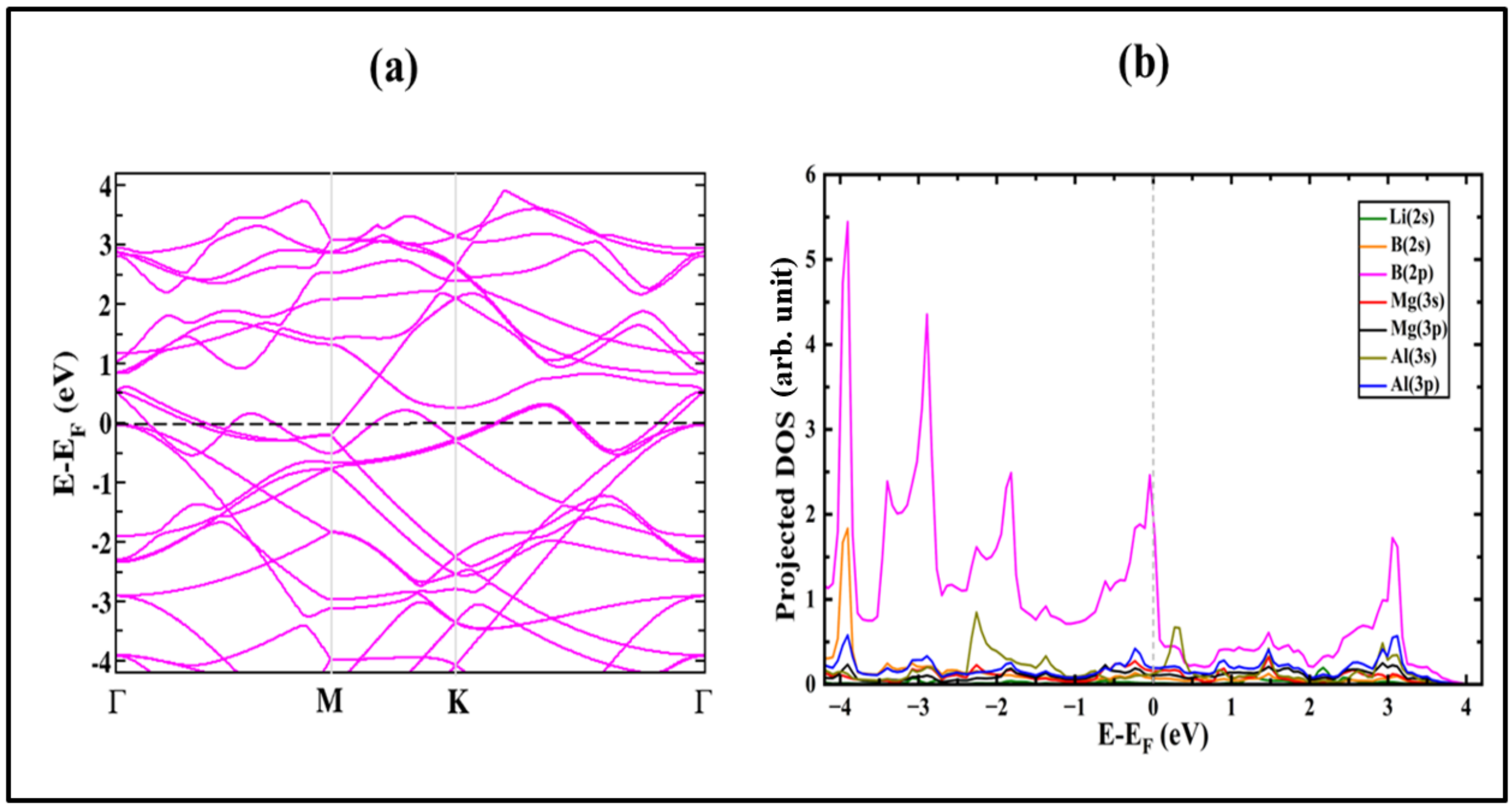

**Figure 11. (a)** Band structure and (b) PDOS of a single Li adsorbed on the Janus $MgAlB_2$ at the S4 site along the Al surface.

## 4. Performance of $Mg_2B_2$ and $Al_2B_2$ as an anode material for LIBs

Before investigating the electrochemical properties of the Janus $MgAlB_2$ monolayer, we first examine the lithium-storage characteristics of its parent compounds, $Mg_2B_2$ and $Al_2B_2$. To assess their electrochemical behavior, a single Li atom was adsorbed on the three high-symmetry adsorption sites of a $2 \times 2 \times 1$ supercell of $Mg_2B_2$: the S1-site (top of the B atom), the S2-site (top of the Mg atom), and the S3-site (top of the bridge position between two neighboring B atoms in the hexagonal network). We have used **eqn. (5)** to calculate the adsorption energies for a single lithium atom. We found that the adsorption energy is the lowest at the S1-site. In the **Fig. S10** (see **SI**), we have shown the different adsorption sites and histogram of corresponding adsorption energies at these sites for both monolayers. The histogram in **Fig. S10** demonstrates that the adsorption energy trend is $E_{S1} < E_{S3} < E_{S2}$. These values of adsorption energies indicate that the substrate stabilizes the lithium atom after adsorption. Subsequently, we performed multilayer adsorption for all possible configurations of Li atoms on both surfaces of the pristine $Mg_2B_2$ monolayer. Similarly, we had performed adsorption of a single lithium at the three adsorption sites on $Al_2B_2$, which are similar to the adsorption sites of the pristine $Mg_2B_2$. The adsorption energy at these three different adsorption sites, viz. S4-site (top of the B atom), S5-site (top of the Al atom), S6-site (top of the bridge position between two neighboring B atoms in the hexagonal network) follow the trend $E_{S4} < E_{S6} < E_{S5}$, with S4-site coming out to be the most stable adsorption site. Bader charge transfer analysis and EDD plot in **Fig. S7** (see **SI)** show that lithium atoms transfer an amount of charge of more than 0.84$e^-$ to the monolayer at all three sites on the Al surface. EDD plots and the amount of Bader charge transfer are shown in **Fig. S7** (see **SI)** for Mg surface. The presence of excess electrons of Al (electronegativity 1.61 in Pauling's scale) atoms in $Al_2B_2$ creates an electron layer on the Al surface which facilitates relatively stronger binding of lithium and also stabilizes the lithium atom. The relatively stronger binding of the lithium atom on the $Al_2B_2$ leads to a relatively deeper potential energy well (trap), which results in a relatively higher diffusion barrier in comparison to pristine $Mg_2B_2$. $Mg_2B_2$ shows a relatively smaller diffusion barrier of 24.9 meV, whereas, $Al_2B_2$ shows a relatively larger diffusion barrier of 32.9 meV. We have shown the lowest diffusion barrier path among two possible diffusion paths on both the monolayers in **Fig. S11** (see in **SI**).

Using **eq. (5)** we have calculated the average adsorption energy as a function of number of Li atoms and plotted its variation with the number of adsorbed Li atoms for both pristine $Mg_2B_2$ and $Al_2B_2$ monolayers in **Fig. S12** (see SI). As shown in **Fig. S12**, the average adsorption energy remains negative for both systems up to the adsorption of 8 Li atoms, corresponding to the completion of a single adsorption layer. This indicates that single-layer lithiation is energetically favourable in both monolayers. For $Mg_2B_2$, the average adsorption energy reaches -0.40 eV when the first adsorption layer is fully occupied by 8 Li atoms. To investigate the possibility of further lithiation, additional Li atoms were introduced to initiate second-layer adsorption. However, the average adsorption energy becomes positive once the number of Li atoms exceeds eight. In particular, a sharp increase to +0.12 eV is observed upon the adsorption of the ninth Li atom, as shown in **Fig. S12(a),** indicating that further Li adsorption is energetically unfavourable. A similar trend is observed for the $Al_2B_2$ monolayer. As the Li concentration increases, the average adsorption energy initially varies and eventually reaches a minimum value of -0.52 eV at eight adsorbed Li atoms, corresponding to complete occupation of the first adsorption layer. Upon the addition of a ninth Li atom, a clear discontinuity appears in the adsorption-energy curve (see **Fig. S12(b)**), and the average adsorption energy increases to +0.03 eV. Furthermore, several optimized configurations associated with second-layer adsorption exhibit noticeable structural distortions, indicating reduced thermodynamic stability. In **Fig. S13 and Fig. S14** we have shown the optimised geometries of maximum Li adsorbed 1$^{st}$ layer (8 atoms) and 2$^{nd}$ layer (16 atoms) of $Mg_2B_2$ and $Al_2B_2$. The abrupt change in adsorption energy after first-layer saturation signifies the completion of the energetically favourable adsorption regime. Once all available adsorption sites are occupied, additional Li atoms cannot establish sufficiently strong interactions with the substrate and instead experience enhanced Li-Li repulsion. As a result, second-layer adsorption becomes energetically unfavourable. The adsorption energies at first-layer saturation, namely -0.40 eV for $Mg_2B_2$ and -0.52 eV for $Al_2B_2$, therefore define the maximum stable lithiation limits of the two monolayers. Beyond these concentrations, excess Li atoms are more likely to aggregate into Li-rich clusters rather than remain stably adsorbed on the surface. These results demonstrate that both $Mg_2B_2$ and $AL_2B_2$ can accommodate only a single stable Li adsorption layer, with increasing electrostatic repulsion at higher Li concentrations preventing further favorable adsorption. Thus, we have achieved a storage capacity of 763.34 $mAhg^{-1}$for $Mg_2B_2$ and 709.17 $mAhg^{-1}$ for $Al_2B_2$.

## 5. Li diffusion on the Janus $MgAlB_2$ monolayer

The $Li^+$ ion mobility on the anode surface determines the charging/discharging rate, which is the critical factor for the evaluation of the efficiency of the monolayer as a better anode material. The ionic diffusivity of Li-ion on the monolayer was determined by evaluating the diffusion barrier using the CI-NEB method. We have taken 7 images for NEB calculation with a cut off energy of 750 eV, an energy convergence criterion of $1 \times 10^{-10}$ eV, a force convergence criterion of $1 \times 10^{-2}$ eV/Å. For the NEB calculation, we used the DFT-D3 (Becke-Johnson damping) dispersion correction to include weak van der Waals (vdW) interactions. After calculating the adsorption energy, it is found that the most stable site is the S1-site, followed by the S3-site, on the Mg surface. Similarly, the S4-site is the most stable site, followed by the S6-site, on the Al surface. Since the S1-site is the energetically most stable site, we have determined the lithium diffusion barrier by considering a diffusion pathway between two adjacent S1-sites on the Mg surface. There are two possible migration pathways, viz. (i) path1, i.e., along the S1-S3-S1 pathway, and (ii) path2, i.e., along the S1-S2-S1 pathway. Similarly, S4-site being the most stable site on the Al surface, there are two possible diffusion pathways, viz. (i) path1, i.e., along the S4-S6-S4 pathway, and (ii) path2, i.e., along the S4-S5-S4 pathway. In **Fig. 12**, we have shown the diffusion pathways and the corresponding diffusion barriers of Janus $MgAlB_2$ on both the Mg and Al surfaces. Among these, the lithium atom faces a very low diffusion barrier of 17.1 meV along the path1 (S1-S3-S1 pathway), indicating a faster lithium-ion mobility. On the Al surface, CI-NEB results show the lowest diffusion barrier of 43.7 meV along path1(S4-S6-S4 pathway) among two diffusion pathways. This diffusion barrier is also very low, indicating that the Li-ion transport on the Al surface of the Janus structure also has a very high mobility. The low diffusion barrier indicates high Li-ion mobility on the monolayer surface, a faster ion transport during electrochemical cycling. This fast diffusion kinetics is expected to lead an excellent rate capability and an improved cycling performance in LIBs. Specifically, the diffusion barrier of 17.1 meV along the S1-S3-S1 pathway on the Mg surface is very low, showcasing a level of lithium-ion mobility that is significantly better than many other materials used for anode applications. Such a low diffusion barrier makes $MgAlB_2$ a potential anode material for high-rate LIBs applications. Furthermore, when compared to its parent structures, $Mg_2B_2$ and $Al_2B_2$, the diffusion barrier in $MgAlB_2$ has reduced, further emphasizing its potential as an alternative anode material. The low diffusion barrier on both Mg and Al surface can be explained through Li-substrate interaction/chemical bonding analysis. We had performed chemical bonding analysis through the determination of integrated crystal orbital Hamilton population (ICOHP) values using the LOBSTER package. [68] In **Table 3**, we have provided the ICOHP values for interactions of Li with metal atoms at different adsorption sites of Janus $MgAlB_2$.The more negative ICOHP value (-0.66 eV) at the S4 site on the Al surface reflects stronger orbital interactions between the Li and the host monolayer. This stronger bonding is consistent with the favorable adsorption energetics at the S4-site and may contribute to the increased stability of Li adsorption at this site. [84] The incorporation of a trivalent Al atom into the pristine $Mg_2B_2$ monolayer modifies the local chemical environment and gives rise to distinct B-Al and B-Mg bonding characteristics. The stronger B-Al interaction helps in maintaining the structural integrity of the Janus framework, whereas the comparatively weaker B-Mg interaction leads to a delocalized electronic environment, as evidenced by the ELF distributions in **Fig. 2(b)** and **Fig. 2(c)**. The ELF reveals pronounced electronic delocalization along the B-B covalent network, generating a relatively uniform electron-rich environment that reduces the likelihood of localized trapping sites for diffusing Li ions on the Mg-terminated surface. In contrast, the more localized electronic distributions in pristine $Mg_2B_2$and $Al_2B_2$ lead to comparatively higher Li-ion diffusion barriers of 24.9 meV and 32.9 meV, respectively, whereas the Janus $MgAlB_2$ monolayer exhibits an ultralow diffusion barrier of only 17.1 meV. The diffusion-energy profiles of pristine $Mg_2B_2$and $Al_2B_2$are presented in **Fig. S11** (see SI). This very low diffusion barrier observed in $MgAlB_2$ is among the lowest diffusion barriers when comparing with other 2D materials, including other MBenes. To check the robustness of this value of diffusion barrier along path1(S1-S3-S1) on the Mg surface, we have calculated diffusion barriers for several possible diffusion routes on the Mg surface of the Janus $MgAlB_2$ using the CI-NEB method. We have shown different diffusion pathways and corresponding diffusion barriers along different possible S1-S3-S1 pathway on the Mg surface in **Fig. S15 (**see **SI)**. The calculated diffusion barriers for different migration paths were found to be very low, ranging from 16.7 to 17.4 meV, indicating nearly path independent and ultrafast lithium-ion diffusion across the surface. The variation in the diffusion barrier along different paths are less than 0.4 meV. In **Table 4**, we have shown the diffusion barrier, OCV, and specific capacity of various two-dimensional (2D) materials. For comparison, other MBene materials often show diffusion barriers in the range of 50-100 meV, indicating that $MgAlB_2$ offers superior lithium-ion mobility. For example, $Mo_2B_2$ has a diffusion barrier of 270 meV [37], indicating slower lithium-ion mobility compared to $MgAlB_2$. Similarly, $Ti_2B_2$ has a diffusion barrier of 17 meV [39], which is almost same as that of $MgAlB_2$. When we expand the comparison to include other anode materials commonly used in lithium-ion batteries, $MgAlB_2$'s diffusion barrier remains competitive. For instance, graphite, a widely used anode material, has a diffusion barrier of around 450-1200 meV [86], which is significantly higher than that of $MgAlB_2$. Our calculated diffusion barrier shows superior rate performance over $W_2C$ (45 meV) [84], silicene (0.14-0.26 eV) [55], graphene (0.32 eV) [15], MoSSe (0.24-0.29 eV) [53], $V_2COS$ (0.30-0.23 eV) [52], BeN (0.30 eV) [85] . This highlights the advantages of $MgAlB_2$ in terms of lithium-ion mobility, offering faster ion diffusion than both graphite and silicon. This comparison underscores a favorable potential of $MgAlB_2$ for potential anode used for LIBs, where fast ion diffusion is critical for improving battery charging times and overall efficiency.

**Table 3** ICOHP values for Li - metal atoms interactions at different adsorption sites of Janus $MgAlB_2$.

| | $MgAlB_2$ (Along Mg surface) | | | $MgAlB_2$ (Along Al surface) | | |
|---|---|---|---|---|---|---|
| ICOHP (in eV) | S1-site | S2-site | S3-site | S4-site | S5-site | S6-site |
| | -0.299 | -0.264 | -0.245 | -0.66 | -0.45 | -0.403 |

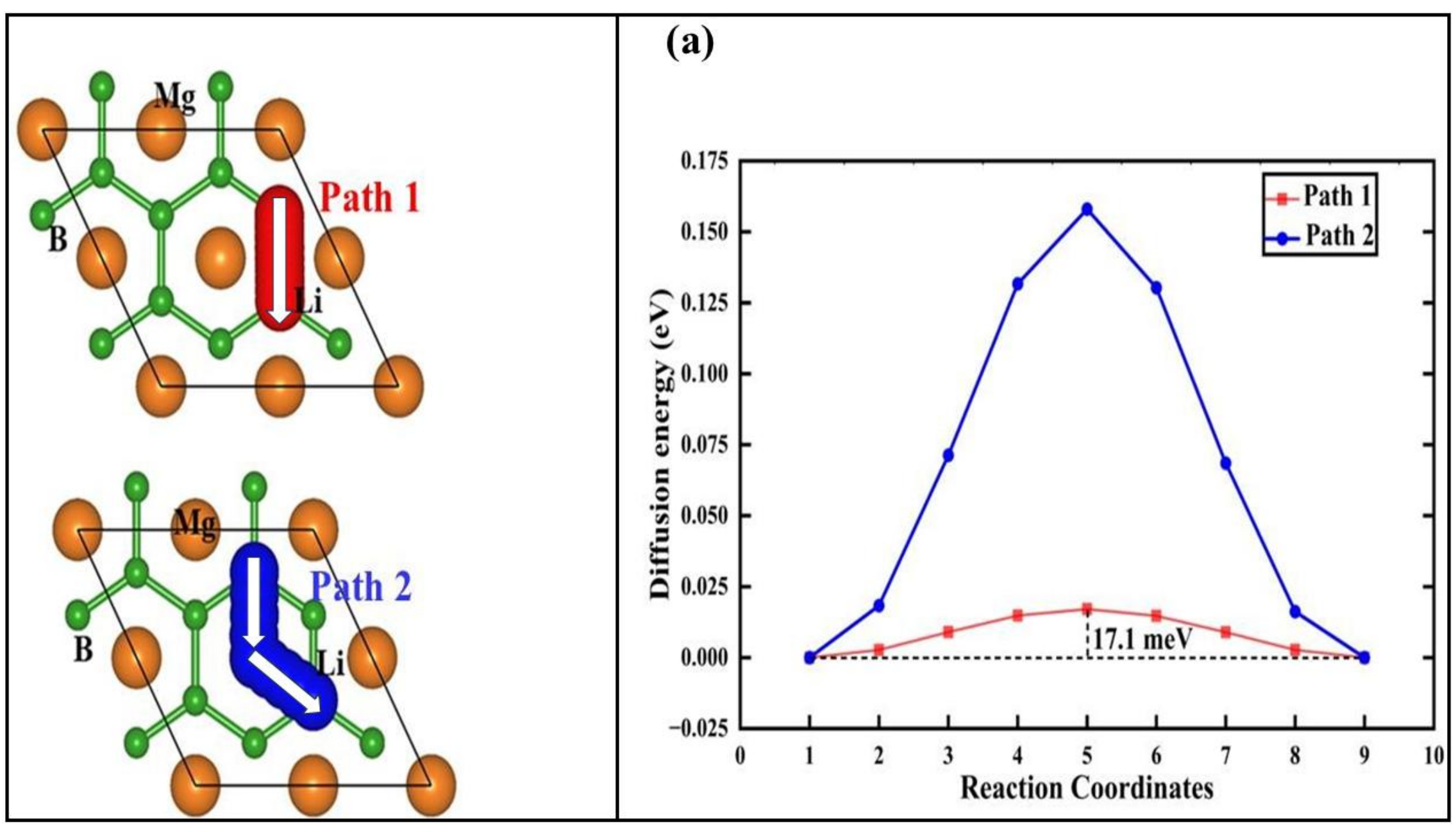

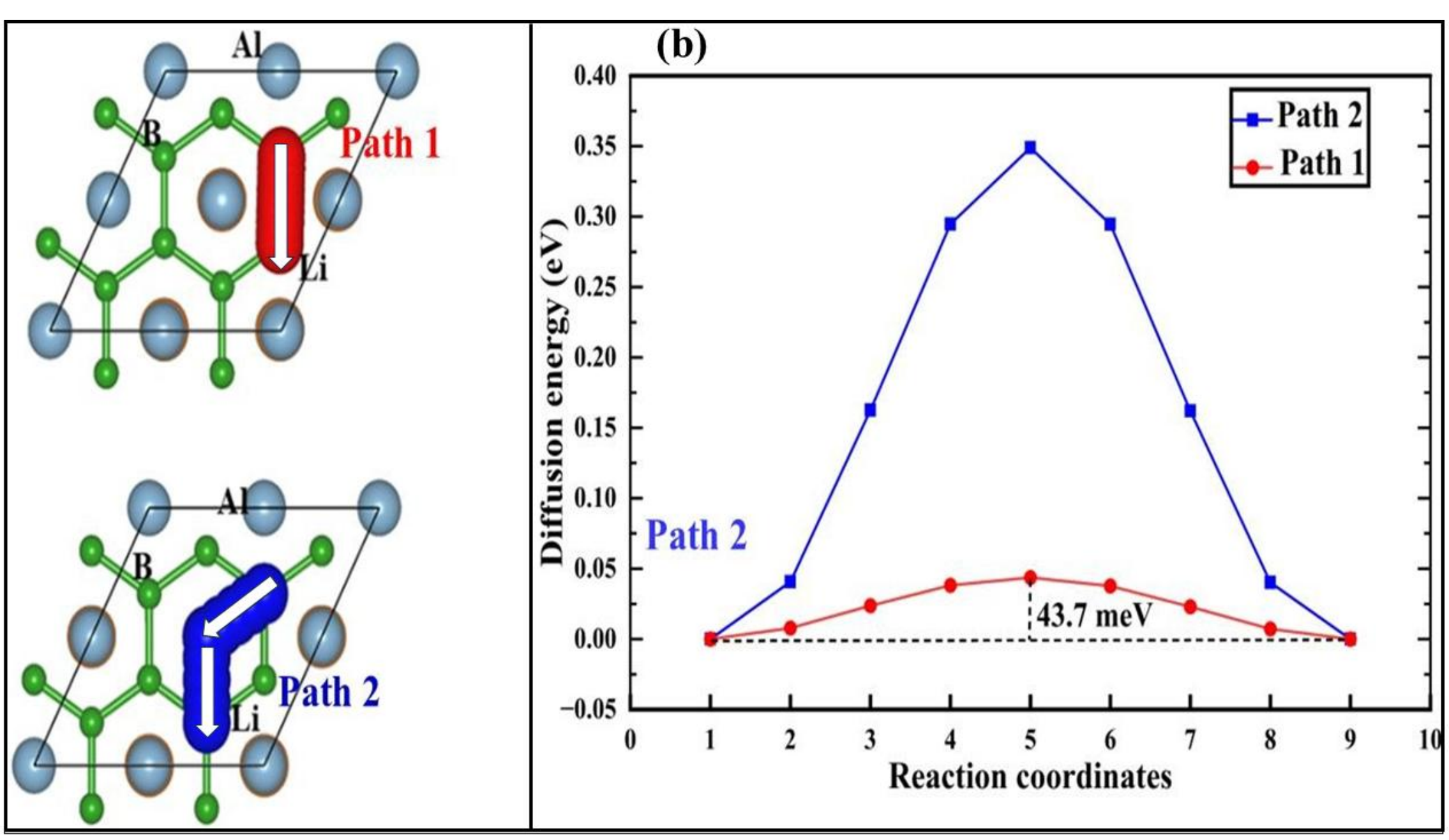

**Fig. 12** Li ion migration through two different paths and their corresponding diffusion barrier on (a) Mg surface and (b) Al surface of the Janus $MgAlB_2$ monolayer.

Additionally, we determined the diffusivity at room temperature. The diffusion rate of $Li^+$ ions is the key factor in determining the overall efficiency and the performance of LIBs. A higher ionic diffusion rate enhances ion transport kinetics, which is crucial for improving the rate capability and the overall battery performance. In general, a higher diffusion coefficient is associated with improved rate capability and enhanced battery performance. The diffusion coefficient (D) has been derived using the Arrhenius equation provided below:

$$D = r^2 \vartheta_0 e^{-\frac{E_{db}}{k_B T}} . \quad (7)$$

In the above expression, r denotes the migration distance between two equivalent adsorption sites along the diffusion pathway, which is obtained from the optimized migration geometry. The parameter $\vartheta_0$ represents the attempt frequency (phonon frequency) associated with atomic vibrations and is typically of the order of 10-100 THz. In the present work, $\vartheta_0$ is taken to be $10^{13}$Hz. Here, $E_{db}$, $K_B$, and T are the diffusion energy barrier, the Boltzmann constant and absolute temperature, respectively. We considered the S1-S3-S1 pathway as the $Li^+$ ion migration pathway for estimating the diffusion coefficient. At room temperature (T = 300) K), the thermal energy $K_B T$ is approximately 0.025 eV. Using the nearest migration distance as 4.85 Å, the calculated the diffusion coefficient is calculated to be 3.43×$10^{-10}$ $cm^2 s^{-1}$. Our calculated diffusion coefficient puts Janus $MgAlB_2$ in the same row among the benchmark 2D monolayer such as a-silicon ($D_a$ of a-Si $10^{-12}$ $cm^2 s^{-1}$) [87], c-silicon ($D_c$ of a-Si $3.6 \times 10^{-12}$ $cm^2 s^{-1}$). This value of diffusion coefficient is comparable to lithium-ion diffusion coefficient in graphite which lies in the range $10^{-10}$ $cm^2 s^{-1}$ to $10^{-11}$ $cm^2 s^{-1}$. [86, 88] The relatively large diffusion coefficient is consistent with the ultra-low migration barrier obtained from the NEB calculations, indicating highly favorable Li-ion transport kinetics on the monolayer surface. Such rapid diffusion behavior is advantageous for fast charge/discharge processes and high-rate electrochemical performance.

## 6. Multilayer adsorption, theoretical capacity, and voltage profile of the Janus $MgAlB_2$ monolayer

The theoretical storage capacity of an electrode material is governed by its ability to store a large number of metal ions while maintaining structural stability. To understand the Li storage mechanism, thermodynamic stability of Li adsorption, and evaluate its maximum lithium-ion storage capacity, the average adsorption energy was calculated by sequentially adding the Li atoms on both surfaces of the Janus $MgAlB_2$ monolayer. We have performed multilayer adsorption of lithium atoms and determined the energetic stability of multilayer adsorption by calculating average adsorption energy using **eq. (5).** Multilayer adsorption of Li atoms is performed on a 2×2× 1 supercell. Here all possible configurations for multilayer Li adsorption are considered, and we had added lithium atoms on both the surfaces of the monolayer in a stoichiometric ratio $Mg_4Al_4B_8Li_n$. For layer 1, n = 8, all the lithium atoms were stored at the S1 and S4 sites on both the Mg and Al surfaces which are the most energetically favored sites on these surfaces correspondingly. We stored lithium atoms in the stoichiometric formula $Mg_4Al_4B_8Li_n$, i.e., on a 2×2×1 supercell of $MgAlB_2$ for different number of lithium atoms, n ranging from =1, 2, 3, …15, 16 ....,24.

To determine the maximum Li-storage capacity of the Janus $MgAlB_2$ monolayer, we calculated the average adsorption energy ($E_{ave}$) for different Li concentrations and plotted its variation as a function of the number of adsorbed Li atoms (n) in **Fig. 13(a)**. The optimized structures corresponding to the first and second lithiation layers are shown in **Fig. 14**. For all Li concentrations up-to n = 16, the average adsorption energy remains negative, indicating that Li adsorption is energetically favourable up to the formation of two complete adsorption layers. The strongest Li binding occurs for the first adsorbed atom with $E_{ave}$= -0.64 eV. As the Li coverage increases, the adsorption energy gradually becomes less negative and reaches -0.46 eV upon completion of the first adsorption layer (n = 8). Subsequent Li atoms occupy the next most favorable adsorption sites, leading to the formation of a second Li layer on both surfaces of the monolayer. Although $E_{ave}$ continue to increase with increasing Li concentration due to enhanced Li-Li electrostatic repulsion, it remains negative throughout the second-layer adsorption regime and attains a value of -0.12 eV at n = 16. These results demonstrate that the Li-substrate interaction remains sufficiently strong to stabilize double-layer lithiation. The favorable adsorption behavior of $MgAlB_2$ can be attributed to its intrinsic Janus asymmetry. The pristine monolayer possesses a built-in dipole moment of +0.39 $\hat{z}$ Debye arising from the chemical inequivalence of the Mg and Al surfaces. This internal polarization enhances Li binding, resulting in a significantly stronger adsorption energy for a single Li atom than those obtained for pristine $Mg_2B_2$ (-0.25 eV) and $Al_2B_2$ (-0.42 eV), both of which possess zero intrinsic dipole moment. Upon Li adsorption, charge transfer from Li to the monolayer modifies the dipole moment, while the residual polarization remains sufficiently strong to support further Li uptake. Even after completion of the first adsorption layer, $MgAlB_2Li_8$ retains a finite dipole moment of +0.15 $\hat{z}$ Debye, whereas the corresponding fully lithiated $Mg_2B_2$and $Al_2B_2$ monolayers remain nonpolar. This residual polarization facilitates the adsorption of additional Li atoms and enables stable double-layer lithiation. Beyond n = 16, the adsorption process becomes energetically unfavorable. The addition of further Li atoms results in a substantial reduction of the internal polarization (+0.08 $\hat{z}$ Debye) due to electrostatic screening, while Li-Li Coulomb repulsion becomes increasingly dominant. Consequently, the average adsorption energy approaches positive values and reaches approximately +0.06 eV upon attempted third-layer adsorption (n = 24). The formation-energy curve shown in **Fig. 13** further supports this conclusion, remaining negative only within the range $1 \leq n \leq 16$. Therefore, the Janus $MgAlB_2$ monolayer can stably accommodate two complete Li layers, corresponding to sixteen Li atoms in the $2 \times 2 \times 1$ supercell. Additional Li atoms are expected to form Li-rich clusters or metallic Li deposits rather than remain stably adsorbed on the monolayer surface. Overall, the lithiation behavior is governed by the competition between dipole-enhanced Li adsorption and Li-Li electrostatic repulsion, with the former dominating at low and intermediate Li concentrations and the latter limiting further Li uptake at high coverage.

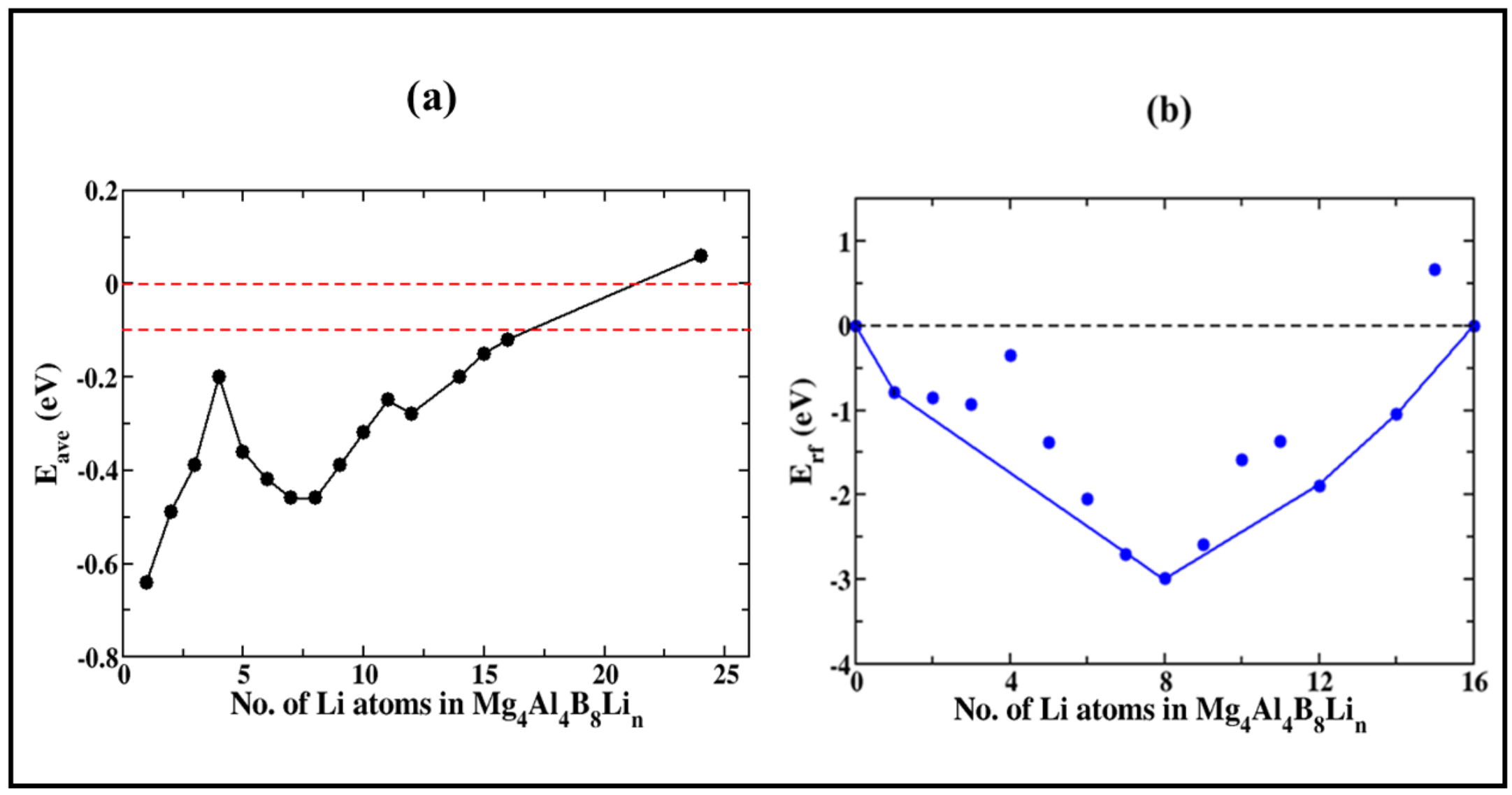

**Figure 13.** (a) Average adsorption energies (b) relative formation energies as a function of number of Li atoms adsorbed on the $2 \times 2 \times 1$ supercell of Janus $MgAlB_2$ monolayer. Here 'n' represents the number of Li atoms adsorbed on the monolayer. Notably, for the formation energy calculation we have considered $n_{max}$= 16 corresponds to m = 1.0.

We also calculated the maximum storage capacity for the optimum Li adsorption by using the formula written below:[57]

$$C = \frac{z x_{max} F}{W_{MgAlB_2}} , \tag{8}$$

where z is the number of valence electrons (z =1 for Li), $x_{max}$ is the highest concentration of lithium atom in a unit cell of the monolayer, F is Faraday constant (26801 $mAhmol^{-1}$) and $W_{MgAlB_2}$ refers to the molecular weight of the monolayer. A maximum of four Li atoms can be stably stored per unit cell of the Janus $MgAlB_2$ monolayer. This corresponds to a theoretical specific capacity of 1470.24 $mAhg^{-1}$ upon complete double-layer lithiation, which is nearly twice the capacity of the parent $Mg_2B_2$ [57] and $Al_2B_2$ monolayers. Furthermore, this value exceeds those reported for many representative two-dimensional anode materials (see **Table 4**), underscoring the effectiveness of Janus engineering in enhancing lithium-storage performance.

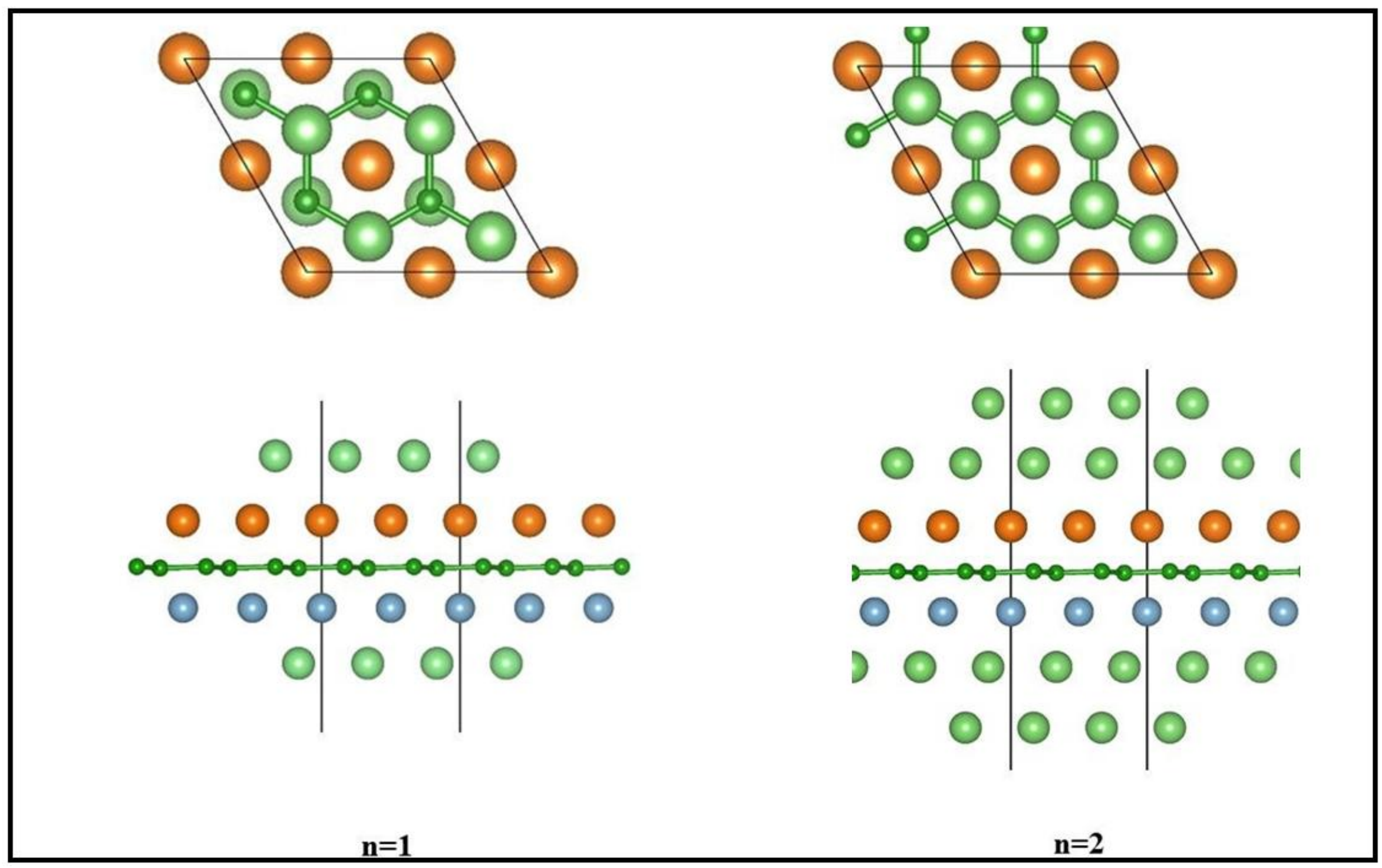

**Figure 14.** The top and side views of optimized $MgAlB_2$ with maximum Li adsorption. The left and right panels represent the first and second layers, respectively.

The storage capacity we obtained more than other materials mentioned in the below **Table 4**. Volume expansion and structural deformation upon lithiation of anode material are crucial factors for determining the long term cyclability of anode materials, as excessive volume changes can induce mechanical degradation and lead to poor cycling performance. After the maximum number of lithium adsorption on the Janus $MgAlB_2$ monolayer, we evaluated the variation of the lattice constants and the thickness of the Janus monolayer. The lattice constant of the 2×2×1 supercell of the pristine Janus $MgAlB_2$ was 6.02 Å. The in-plane area (A) was found to be, A= $(\sqrt{3}/2)$×6.02×6.02 Å$^2$= 31.28 Å$^2$. The thickness (d) was 3.32 Å. After the calculated lattice constant of the pristine Janus monolayer after double layer adsorption of lithium was found to be 6.04 Å, while the calculated thickness of the Janus monolayer was found to be 3.40 Å. The calculated in-plane area after maximum adsorption of lithium was found to be, A'= $(\sqrt{3}/2)$×6.04×6.04 Å$^2$= 31.59 Å$^2$. After the maximum number of lithium adsorption, the Janus monolayer showed an in-plane area expansion of 0.66%, and a volume expansion of 3.7%. Such a small volume change indicates that the Janus $MgAlB_2$ monolayer can effectively store a significant amount of Li with minimum structural distortion. The low volume expansion suggests excellent mechanical robustness during repeated lithiation/delithiation cycles and is expected to contribute to enhanced cycling stability and long-term electrochemical durability.

**Table 4** Diffusion barrier, OCV and specific storage capacity of various 2D materials & graphite.

| Various 2D materials and graphite | Diffusion barrier (meV) | OCV(V) | Specific capacity ($mAhg^{-1}$) |
|---|---|---|---|
| $MgAlB_2$ (This work) | 17.1 | 0.27 | 1470.24 |
| $Mg_2B_2$ [57] | 27 | 0.275 | 764 |
| $Ca_2B_2$ [57] | 16 | 0.395 | 859 |
| $Sc_2B_2$ [99] | 3 | 0.41 | 480 |
| $Ti_2B_2$ [39] | 17 | 0.526 | 456 |
| $ScTiB_2$ [54] | 14 | 0.15-0.48 | 937 |
| $Zr_2B_2$ [90] | 17 | 0.236 | 526 |
| $Ti_2B$ [91] | 23.1 | 0.381 | 503.1 |
| $Sc_2C$ [92] | 18 | 0.27 | 462 |
| $Mn_2C$ [93] | 24 | 0.15 | 887 |
| $Ti_2BN$ [94] | 24 | 0.24 | 889 |
| $Nb_2C$ [95] | 32 | 0.25 | 542 |
| $BC_3$ [96] | 40 | 0.48 | 883 |
| $Ca_2C$ [97] | 27 | 0.1 | 582 |
| $Mo_2C$ [98] | 35 | 0.14 | 526 |
| $Ti_2N$ [89] | 17 | 0.53 | 487 |
| Graphite [88] | 450-1200 | 0.32 | 372 |
| $Mg_3B_4$ [47] | 22 | 0.36 | 462 |
| $Ca_3B_4$ [57] | 7 | 0.343 | 328 |
| $Mg_4B_6$ [57] | 23 | 0.423 | 331 |
| $Ca_4B_6$ [57] | 29 | 0.341 | 331 |
| MoSSe [53] | 240 | 0.62 | 776.5 |

## 7. Theoretical open-circuit voltage of $MgAlB_2$ monolayer

The electrochemical performance of an anode material is primarily governed by its lithium-storage capacity and open-circuit voltage (OCV), which together determine the achievable energy density and operating characteristics of a lithium-ion battery. Therefore, a detailed evaluation of these quantities is essential for assessing the suitability of Janus $MgAlB_2$ as a potential anode material. The lithiation process can be described by the half-cell reaction to illustrate the charging/discharging process of the $MgAlB_2$ monolayer:

$$MgAlB_2 + xLi^{1+} + xe^- \leftrightarrow Li_xMgAlB_2. \quad (9)$$

The average OCV is related to the Gibbs free energy change (ΔG) and can be determined by the following equation:

$$\Delta G = \Delta E + P\Delta V - T\Delta S. \quad (10)$$

For solid-state electrode systems, the (PΔV) contribution is negligibly small because lithiation induces only a minor volume change. Likewise, the entropy term (TΔS) at room temperature is typically on the order of a few tens of meV, which is much smaller than the characteristic adsorption energies of Li atoms. Consequently, the Gibbs free-energy variation change (ΔG can be approximated by the total internal energy change (ΔE) energy difference obtained from first-principles calculations, i.e., $\Delta G \approx \Delta E$. Under this approximation, the OCV can be directly derived from the energetics of Li adsorption. Consequently, the average OCV is computed from the average adsorption energy $E_{ave}$. To construct the voltage profile, the convex-hull formalism was employed to identify the thermodynamically stable intermediate lithiation states. The average OCV between two adjacent stable compositions containing $n_1$ and $n_2$ adsorbed Li atoms was calculated. To calculate the OCV for different Li atom concentrations, we use the following equation: [101]

$$OCV = [E_{Li_{n_1}@MgAlB_2} - E_{Li_{n_2}@MgAlB_2} + (n_2 - n_1)E_{Li}]/z(n_2 - n_1)e, \quad (11)$$

where $E_{Li_{n_1}@MgAlB_2}$ and $E_{Li_{n_2}@MgAlB_2}$ represent the total energy of the systems when there are $n_1$ and $n_2$ number of lithium atoms adsorbed on the supercell of the $MgAlB_2$ monolayer, respectively. $\mu_{Li}$ is the total energy of a single lithium atom in its stable bcc phase, z is the electronic charge of the Li ion. To determine the stable lithiation sequence, adsorption energies were first evaluated for all energetically favorable Li configurations. Subsequently, the relative formation energies were computed to establish the convex hull and identify the ground-state intermediate phases. The relative formation energy is defined as follows: [101]

$$E_{rf} = E_{Li_m@MgAlB_2} - mE_{Li_{1.0}@MgAlB_2} - (1 - m)E_{MgAlB_2}, \quad (12)$$

where $E_{Li_m@MgAlB_2}$, $E_{Li_{1.0}@MgAlB_2}$, $E_{MgAlB_2}$ are the total energies of the partially lithium adsorbed monolayer at a concentration of 'm', fully lithiated monolayer, and the pristine $MgAlB_2$ monolayers, respectively. Notably, here the concentration 'm' corresponds to 1.0 when the maximum number of lithium atoms, n = 16, are adsorbed on the monolayer, and 'm' corresponds to 0.0625 when there is only one lithium atom, n = 1 is adsorbed. The variation of calculated adsorption energy and formation energy profiles as a function of number of Li atoms/concentration are presented in **Fig. 13.** The convex hull in **Fig. 13(b)** serves as the basis for constructing the voltage profile and evaluating the lithiation thermodynamics of the Janus $MgAlB_2$ monolayer. We had determined the stable intermediate structures to plot the convex hull with the help of the following criteria: [102]

$$\Delta E = E_{Li_{m_i}@MgAlB_2} - \frac{(m_h - m_i)}{(m_h - m_l)} E_{Li_{m_l}@MgAlB_2} - \frac{(m_i - m_l)}{(m_h - m_l)} E_{Li_{m_h}@MgAlB_2}, \quad (13)$$

where $m_i$ denotes an arbitrary Li concentration, $m_l$ and $m_h$ correspond to the lowest and highest Li concentrations are considered, respectively. $E_{Li_{m_i}@MgAlB_2}$, $E_{Li_{m_l}@MgAlB_2}$, and $E_{Li_{m_h}@MgAlB_2}$ are the total energies of the corresponding lithiated configurations. Here, $m_l$ = 0.0625 and $m_h$ = 1. A negative value of $\Delta E$ on the convex hull, i.e., $\Delta E \leq 0$, indicates that the configuration is thermodynamically stable against decomposition into neighboring phases, whereas a positive value, i.e., $\Delta E > 0$, indicates a metastable state. Stable phases with $\Delta E$ lie on the convex hull. Consequently, the structures located on the convex-hull boundary represent the stable state lithiation configurations, while those above the hull correspond to metastable configurations. We used these stable intermediate phases to construct the voltage profile of the Janus $MgAlB_2$ monolayer. The determined OCV profile versus the number of Li atoms is shown in **Fig. 15**. We used the equilibrium voltage between two adjacent stable phases lying on the convex hull to determine the step-like OCV voltage profile. The calculated OCV shows a variation from 2.23 V to 0.18 V throughout the lithiation process. The relatively high voltage at low Li amount originates from the relatively stronger interaction between the first inserted Li atoms and the host monolayer. As we increase the number of Li atoms, the voltage decreases gradually and exhibits several steps-like plateaus, indicating the sequential occupation of energetically favourable adsorption sites and the formation of stable intermediate phases. Importantly, the obtained voltage window lies within the desirable range for anode materials.

The practical viability of an electrode material is closely related to its ability to maintain structural integrity under finite-temperature operating conditions. Therefore, the thermal stability of both pristine and the fully lithiated Janus $MgAlB_2$ monolayers was investigated using ab initio molecular dynamics (AIMD) simulations. The simulations were performed on a 2×2×1 supercell within the NVT ensemble at 500 K using a Nosé-Hoover thermostat, with a time step of 1 fs and a total simulation duration of 10 ps. The evolution of the free energy and representative final configurations obtained from the AIMD trajectories are presented in **Fig. 16**.

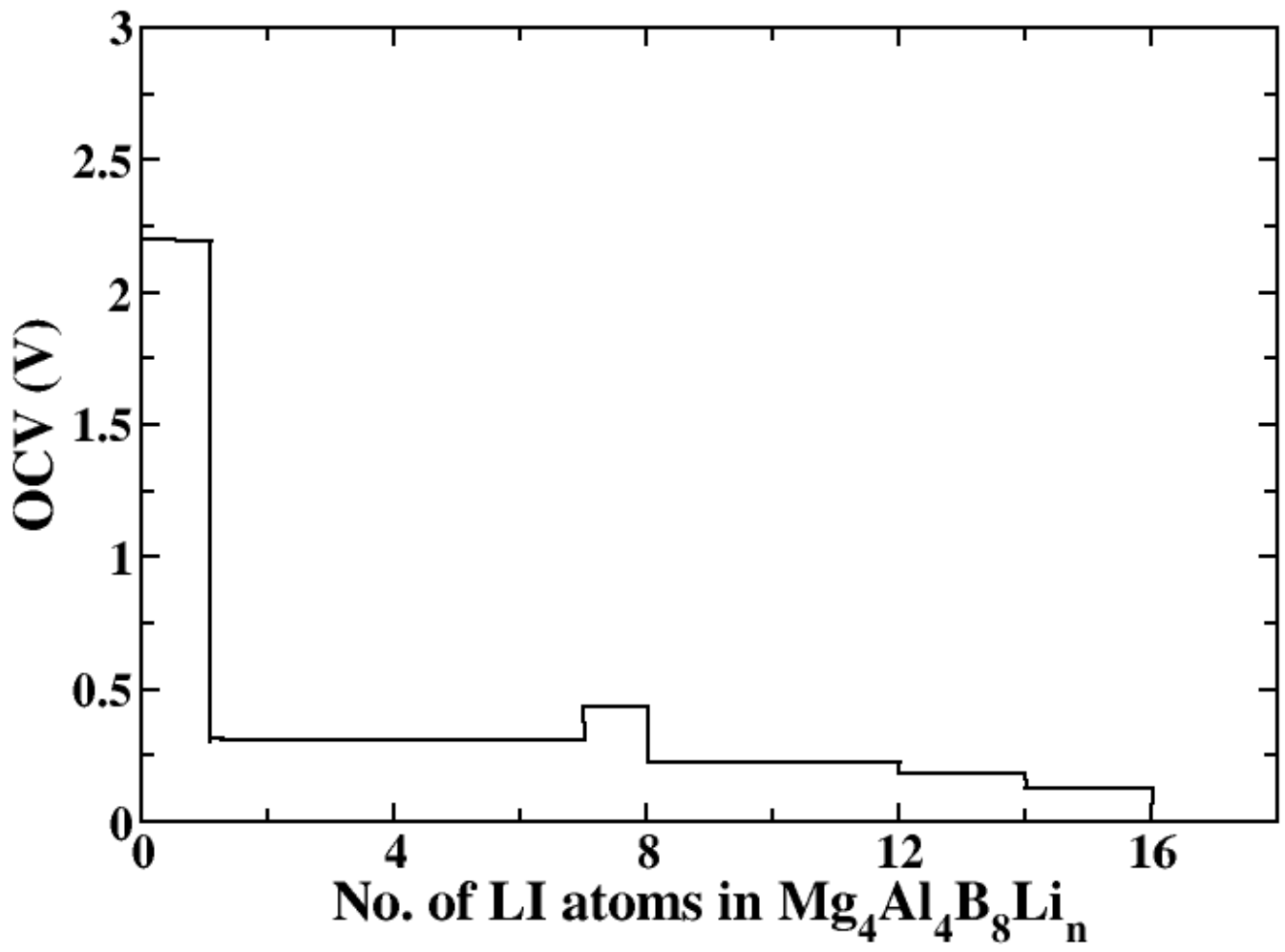

**Figure 15.** Open circuit voltage (OCV) profile against $Li/Li^+$ of $MgAlB_2$.

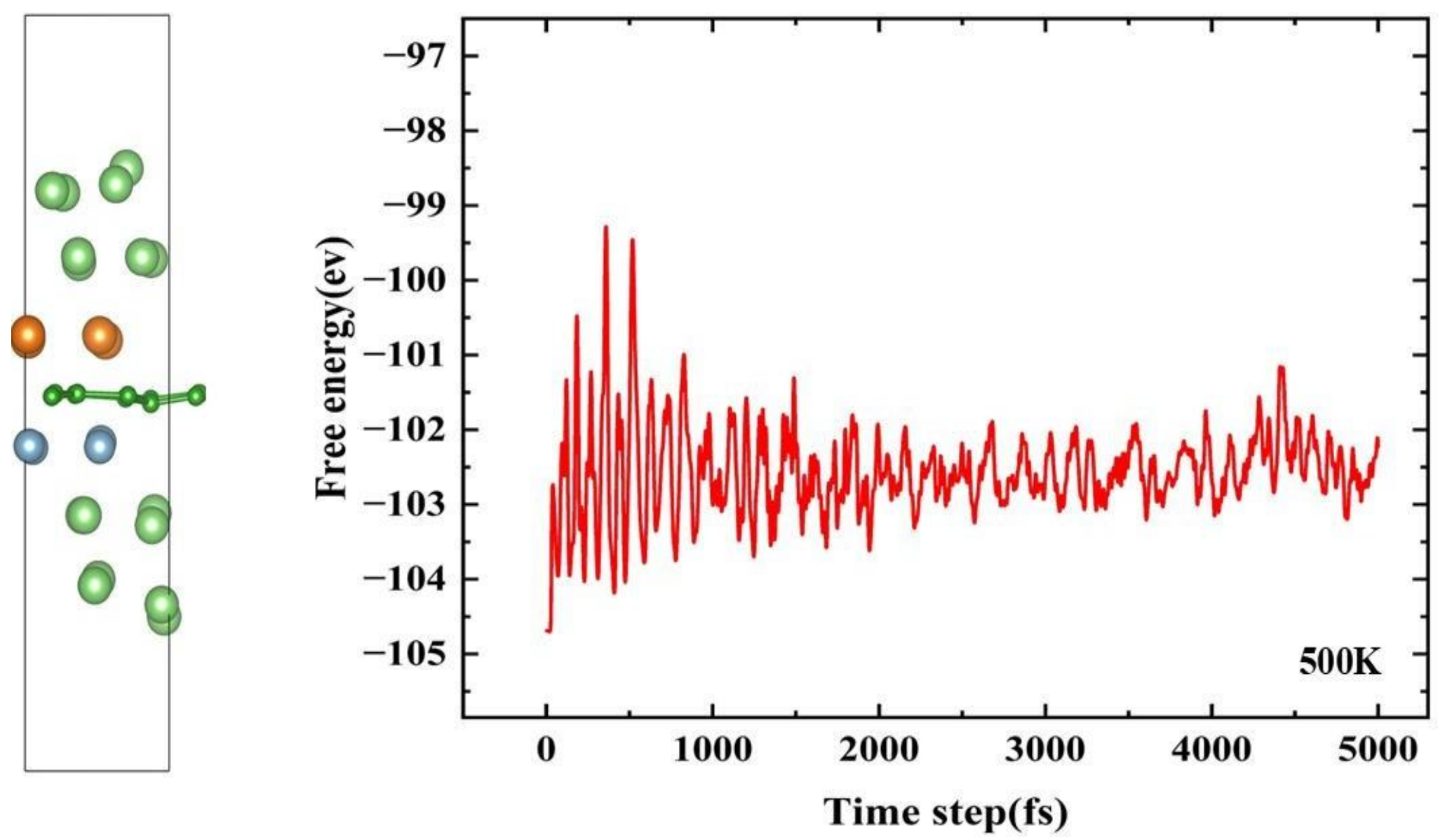

**Fig. 16.** Snapshots of fully lithium adsorbed $MgAlB_2$ monolayer after AIMD simulations for 10 ps with NVT ensemble at 500K.

During the initial stage of the simulation, the free energy exhibits relatively large fluctuations as the system approaches thermal equilibrium. After equilibration, the energy oscillates around a nearly constant average value -102.5 eV to -103 eV without displaying any noticeable long-term drift, indicating that the system remains thermodynamically stable throughout the simulation. Throughout the simulation, both the pristine and fully lithiated structures remain structurally intact without exhibiting any bond breaking, atomic reconstruction, or any intense structural deformation.

Although minor out-of-plane buckling and small displacements of the adsorbed Li atoms from their equilibrium positions are observed, the overall framework of the $MgAlB_2$ monolayer is well preserved. These results suggest that the material can withstand the thermal stresses associated with battery operation while preserving its structural integrity, thereby supporting its suitability as a stable anode material for rechargeable LIBs.

**8. Conclusions**

We proposed a Janus $MgAlB_2$ monolayer, an ordered dual metal MBene composed of main group Mg (group II) and Al (group IIIA), and assessed its viability as a lithium-ion battery anode using first principles simulations. Formation energies, phonon spectra, and ab initio molecular dynamics confirm its structural, dynamical, and thermal stability. The computed elastic constants satisfy mechanical stability criteria, indicating that the lattice can tolerate external perturbations without loss of integrity. Electronic structure analysis shows intrinsic metallic character, which supports efficient charge transport during operation. Al incorporation into parent $Mg_2B_2$ induces charge redistribution and Janus polarity, enabling adsorption of up to three Li layers per surface. As a result, the monolayer achieves a theoretical specific capacity of 1470.24 $mAhg^{-1}$, well above that of $Mg_2B_2$ and $Al_2B_2$. Convex hull calculations identify stable intermediate lithiation states and indicate favourable Li storage thermodynamics across a wide composition range. The Li migration barrier is low (17.1 meV), consistent with rapid ion transport and strong rate capability. In addition, full lithiation causes modest lattice and volume changes, suggesting good structural reversibility and mechanical robustness over cycling. Taken together, high capacity, fast Li diffusion, confirmed thermodynamic stability, and minimal deformation upon lithiation position Janus $MgAlB_2$ as a high-performance anode candidate for lithium-ion batteries.

## Keywords

Janus structure, non-transition metal MBene, Intrinsic polarization, built-in Dipole moment, Ultralow diffusion barrier, Storage capacity, Density functional theory, LIBs, anode material, energy storage applications.

## Conflicts of interest

There are no conflicts of interest.

## Data availability

The data supporting this article have been included as part of the supplementary information (SI). Supplementary information is available.

## Acknowledgements

The author acknowledgement the computational facilities provided by the Department of Physics, Indian Institute of Technology Patna, Bihar, India. PP thanks DST-SERB, Government of India, for ECRA project (ECR/2017/003305).

## References

1. de Oliveira Matias, J.C. and T.C. Devezas, Consumption dynamics of primary-energy sources: The century of alternative energies. Applied Energy, 2007. **84**(7-8): p. 763-770.
2. Chu, S., Y. Cui, and N. Liu, The path towards sustainable energy. Nature Materials, 2017. **16**(1): p. 16-22.
3. Ang, T.-Z., et al., A comprehensive study of renewable energy sources: Classifications, challenges and suggestions. Energy Strategy Reviews, 2022. **43**: p. 100939.
4. Gulagi, A., et al., The role of renewables for rapid transitioning of the power sector across states in India. Nature Communications, 2022. **13**(1): p. 5499.
5. Kumar, A. and P. Parida, Unveiling the potential of a BCN-biphenylene monolayer as a high-performance anode material for alkali metal ion batteries: a first-principles study. Nanoscale, 2024. **16**(27): p. 13131-13147.
6. Kumar, A., P. Samanta, and P. Parida, Electrochemical performance of gold monolayers for lithium-ion batteries: A first-principles study. Journal of Energy Storage, 2026. **163**: p. 122132.
7. Miller, J.R., Valuing reversible energy storage. Science, 2012. **335**(6074): p. 1312-1313.
8. Kim, S., et al., Electrochemically driven mechanical energy harvesting. Nature communications, 2016. **7**(1): p. 10146.
9. Song, Z., et al., Origami lithium-ion batteries. Nature communications, 2014. **5**(1): p. 3140.
10. Samanta, P., A. Kumar, and P. Parida, Two-Dimensional Antiferromagnetic Fe2As2 as an Anode Material for Rechargeable Li-Ion Batteries: A DFT Study. Advanced Theory and Simulations, 2026. **9**(1): p. e01068.
11. Kumar, A. and P. Parida, Theoretical study of δ-5 boron monolayer as an anode material for Li- and non-Li-ion batteries. Journal of Materials Research, 2022. **37**(20): p. 3384-3393.
12. Kumar, A. and P. Parida, Iron-arsenide monolayers as an anode material for lithium-ion batteries: a first-principles study. Physical Chemistry Chemical Physics, 2024. **26**(15): p. 12060-12069.
13. Kato, K., et al., Light-assisted rechargeable lithium batteries: organic molecules for simultaneous energy harvesting and storage. Nano letters, 2021. **21**(2): p. 907-913.
14. Goodenough, J.B. and K.-S. Park, The Li-Ion Rechargeable Battery: A Perspective. Journal of the American Chemical Society, 2013. **135**(4): p. 1167-1176.
15. Larcher, D. and J.M. Tarascon, Towards greener and more sustainable batteries for electrical energy storage. Nature Chemistry, 2015. **7**(1): p. 19-29.
16. Bruce, P.G., B. Scrosati, and J.M. Tarascon, Nanomaterials for rechargeable lithium batteries. Angewandte Chemie International Edition, 2008. **47**(16): p. 2930-2946.
17. Guo, P., H. Song, and X. Chen, Electrochemical performance of graphene nanosheets as anode material for lithium-ion batteries. Electrochemistry Communications, 2009. **11**(6): p. 1320-1324.
18. Al Hassan, M., et al., Emergence of graphene as a promising anode material for rechargeable batteries:

a review. Materials today chemistry, 2019. **11**: p. 225-243.
19. Jiang, T., et al., Battery technologies for grid-scale energy storage. Nature Reviews Clean Technology, 2025. **1**(7): p. 474-492.
20. Liu, Y., et al., Feasibility of Lithium Storage on Graphene and Its Derivatives. J Phys Chem Lett, 2013. **4**(10): p. 1737-42.
21. Zhao, Y., et al., Recent developments and understanding of novel mixed transition-metal oxides as anodes in lithium-ion batteries. Advanced Energy Materials, 2016. **6**(8): p. 1502175.
22. Yang, E., H. Ji, and Y. Jung, Two-dimensional transition metal dichalcogenide monolayers as promising sodium ion battery anodes. The Journal of Physical Chemistry C, 2015. **119**(47): p. 26374-26380.
23. Yun, Q., et al., Layered transition metal dichalcogenide-based nanomaterials for electrochemical energy storage. Advanced Materials, 2020. **32**(1): p. 1903826.
24. Xiao, J., et al., Exfoliated MoS2 nanocomposite as an anode material for lithium ion batteries. Chemistry of Materials, 2010. **22**(16): p. 4522-4524.
25. Wang, K., et al., development status and modification strategies of nano-MoS2-based anode materials. Ionics, 2024. **30**(8): p. 4387-4415.
26. Jing, Y., et al., Metallic VS2 monolayer: a promising 2D anode material for lithium ion batteries. The Journal of Physical Chemistry C, 2013. **117**(48): p. 25409-25413.
27. Deng, J., J.Z. Liu, and N.V. Medhekar, Enhanced lithium adsorption and diffusion on silicene nanoribbons. RSC Advances, 2013. **3**(43): p. 20338-20344.
28. Seyed-Talebi, S.M., I. Kazeminezhad, and J. Beheshtian, Theoretical prediction of silicene as a new candidate for the anode of lithium-ion batteries. Physical Chemistry Chemical Physics, 2015. **17**(44): p. 29689-29696.
29. Regragui, N., et al., Unlocking the potential of a chemically modified blue phosphorene as anode materials for high-performance sodium-ion batteries. Journal of Energy Storage, 2024. **97**: p. 112886.
30. Batmunkh, M., M. Bat-Erdene, and J.G. Shapter, Phosphorene and phosphorene-based materials–prospects for future applications. Advanced Materials, 2016. **28**(39): p. 8586-8617.
31. Zhang, X., et al., Borophene as an extremely high-capacity electrode material for Li-ion and Na-ion batteries. Nanoscale, 2016. **8**(33): p. 15340-15347.
32. Tang, Q., Z. Zhou, and P. Shen, Are MXenes promising anode materials for Li ion batteries? Computational studies on electronic properties and Li storage capability of Ti3C2 and Ti3C2X2 (X= F, OH) monolayer. Journal of the American Chemical Society, 2012. **134**(40): p. 16909-16916.
33. Greaves, M., S. Barg, and M.A. Bissett, MXene-based anodes for metal-ion batteries. Batteries & Supercaps, 2020. **3**(3): p. 214-235.
34. Zhang, B., J. Zhou, and Z. Sun, MBenes: progress, challenges and future. Journal of Materials Chemistry A, 2022. **10**(30): p. 15865-15880.
35. Li, Y., et al., Computational evaluation of ScB and TiB MBenes as promising anode materials for high-performance metal-ion batteries. Physical Review Materials, 2022. **6**(4): p. 045801.
36. Wang, Y., et al., On two-dimensional metal borides (MBenes) as anode materials for metal-ion batteries: A first-principles study. Computational Materials Science, 2024. **233**: p. 112710.
37. Guo, Z., J. Zhou, and Z. Sun, New two-dimensional transition metal borides for Li ion batteries and electrocatalysis. Journal of Materials Chemistry A, 2017. **5**(45): p. 23530-23535.
38. Ramezanzadeh, M., et al., 2D-Transition Metal Borides (MBenes): a comprehensive review of the materials, chemistry, advances, and novel applications. Advanced Composites and Hybrid Materials, 2025. **8**(3): p. 269.
39. Bo, T., et al., Hexagonal Ti 2 B 2 monolayer: a promising anode material offering high rate capability for Li-ion and Na-ion batteries. Physical Chemistry Chemical Physics, 2018. **20**(34): p. 22168-22178.
40. Bhat, A., et al., Prospects challenges and stability of 2D MXenes for clean energy conversion and storage applications. npj 2D Materials and Applications, 2021. **5**(1): p. 61.
41. Jalaludeen, S.A., et al., An Overview of Two-Dimensional Nanomaterial—MXene in Energy Storage and Sensing Application. Plasmonics, 2025. **20**(10): p. 9083-9097.
42. Tripathy, D.B., P. Agarwal, and S. Pradhan, MBenes: emerging two-dimensional metal borides for

multifunctional energy storage and conversion technologies. Journal of Materials Chemistry A, 2026. **14**(11): p. 6217-6241.
43. Rahman, A.U., et al., Unveiling the Structural Stability and Optoelectronic Properties of Pristine and Janus Monolayers. ACS Omega, 2025. **10**(19): p. 19574-19584.
44. Choudhary, T., S. Vinod, and R.K. Biswas, Thermoelectric performance of Janus monolayers embedded in MX2-based superlattices: a computational insight. Physical Chemistry Chemical Physics, 2026. **28**(8): p. 5288-5302.
45. Kumar, A., P. Senapati, and P. Parida, Unravelling the potential of a hybrid borocarbonitride biphenylene 2D network for thermoelectric applications: a first principles study. Nanoscale, 2025. **17**(7): p. 4015-4029.
46. Kumar, A., P. Senapati, and P. Parida, Theoretical insights into the structural, electronic and thermoelectric properties of the inorganic biphenylene monolayer. Physical Chemistry Chemical Physics, 2024. **26**(3): p. 2044-2057.
47. Senapati, P., A. Kumar, and P. Parida, Enhanced thermopower in two-dimensional ruthenium dichalcogenides RuX2 (X = S, Se): a first-principles study. Physica Scripta, 2025. **100**(6): p. 065923.
48. Senapati, P. and P. Parida, Thermoelectric performance of quantum dots embedded in an Aharonov-Bohm ring: a Pauli master equation approach. Scientific Reports, 2025. **15**(1): p. 13232.
49. Xiong, F. and Y. Chen, A first-principles study of Janus monolayer TiSSe and VSSe as anode materials in alkali metal ion batteries. Nanotechnology, 2021. **32**(2): p. 025702.
50. Alfurhud, S. and U. Schwingenschlögl, Engineering of Janus transition metal dichalcogenide bilayers as absorber materials for solar cells. Scientific Reports, 2025. **15**(1): p. 14863.
51. Yan, L., et al., Janus $M\text{SH}$ ($M$ = Sc, Y, V, Nb, Ta) monolayers: Promising candidates for photocatalytic water splitting and out-of-plane piezoelectricity. Physical Review B, 2025. **112**(24): p. 245427.
52. Wang, F., et al., Insight into Janus ${\mathrm{V}}_{2}\mathrm{COS}$ as anode material of high-performance alkali metal ion battery: Diffusion barrier, recyclability, specific capacity, and open-circuit voltage. Physical Review Materials, 2024. **8**(8): p. 085801.
53. Shang, C., et al., Theoretical prediction of Janus MoSSe as a potential anode material for lithium-ion batteries. The Journal of Physical Chemistry C, 2018. **122**(42): p. 23899-23909.
54. Wang, J., et al., Ordered double transition metal MBene: the hexagonal ScTiB2 monolayer as a superior anode material for lithium-ion batteries. Computational Materials Science, 2022. **214**: p. 111736.
55. Anasori, B., et al., Two-Dimensional, Ordered, Double Transition Metals Carbides (MXenes). ACS Nano, 2015. **9**(10): p. 9507-9516.
56. Han, Y., et al., Exploring the potential of MB2 MBene family as promising anodes for Li-ion batteries. RSC Advances, 2024. **14**(16): p. 11112-11120.
57. Sun, Y., et al., Hexagonal Mg2B2 and Ca2B2 monolayers as promising anode materials for Li-ion and Na-ion batteries. Nanoscale, 2024. **16**(33): p. 15699-15712.
58. Ma, S., et al., The superconducting Dirac AlB6 monolayer as an excellent anode material for Li/Na ion batteries. Materials Today Communications, 2024. **39**: p. 109325.
59. Zou, R.-F., et al., Two-dimensional AlB4/Al2B2: high-performance Dirac anode materials for sodium-ion batteries. Physical Chemistry Chemical Physics, 2023. **25**(42): p. 28814-28823.
60. Abedi, S., et al., Prediction of novel two-dimensional Dirac nodal line semimetals in Al2B2 and AlB4 monolayers. Nanoscale, 2022. **14**(31): p. 11270-11283.
61. Kresse, G. and J. Furthmüller, Efficient iterative schemes for ab initio total-energy calculations using a plane-wave basis set. **54**(16): p. 11169.
62. Hafner, J., Ab-initio simulations of materials using VASP: Density-functional theory and beyond. Journal of computational chemistry, 2008. **29**(13): p. 2044-2078.
63. Monkhorst, H.J. and J.D. Pack, Special points for Brillouin-zone integrations. Physical Review B, 1976. **13**(12): p. 5188-5192.
64. Perdew, J.P., K. Burke, and M. Ernzerhof, Generalized Gradient Approximation Made Simple. Physical Review Letters, 1996. **77**(18): p. 3865-3868.

65. Goerigk, L., A comprehensive overview of the DFT-D3 London-dispersion correction. Non-covalent interactions in quantum chemistry and physics, 2017: p. 195-219.
66. Fluckey, S. and W. Vandenberghe, Diamond two-phonon infrared absorption spectrum calculated from first principles using the finite displacement method. Applied Physics Letters, 2024. **124**. 152203.
67. Perdew, J.P., K. Burke, and Y. Wang, Generalized gradient approximation for the exchange-correlation hole of a many-electron system. Physical Review B, 1996. **54**(23): p. 16533-16539.
68. Naik, A.A., et al., A Quantum-Chemical Bonding Database for Solid-State Materials. Scientific Data, 2023. **10**(1): p. 610.
69. Henkelman, G., B.P. Uberuaga, and H. Jónsson, A climbing image nudged elastic band method for finding saddle points and minimum energy paths. The Journal of chemical physics, 2000. **113**(22): p. 9901-9904.
70. Sidler, D. and S. Riniker, Fast Nosé–Hoover thermostat: molecular dynamics in quasi-thermodynamic equilibrium. Physical Chemistry Chemical Physics, 2019. **21**(11): p. 6059-6070.
71. Kumar, A. and P. Parida, Unveiling the potential of a BCN-biphenylene monolayer as a high-performance anode material for alkali metal ion batteries: a first-principles study. Nanoscale, 2024. **16**(27): p. 13131-13147.
72. Jona, F. and P.M. Marcus, Magnesium under pressure: structure and phase transition. Journal of Physics: Condensed Matter, 2003. **15**(45): p. 7727.
73. Witt, W., Absolute Präzisionsbestimmung von Gitterkonstanten an Germanium- und Aluminium-Einkristallen mit Elektroneninterferenzen. Zeitschrift für Naturforschung A, 1967. **22**(1): p. 92-95.
74. Masago, A., K. Shirai, and H. Katayama-Yoshida, Crystal stability of $\ensuremath{\alpha}$- and $\ensuremath{\beta}$-boron. Physical Review B, 2006. **73**(10): p. 104102.
75. Majid, A., et al., First-Principles Investigation of Lithium Titanate Oxide as an Anode Material in Li-, Na-, Mg-, Ca-, and K-Ion Batteries. ACS Omega, 2025. **10**(30): p. 33645-33661.
76. Zhang, H., et al., Dirac State in the FeB2 Monolayer with Graphene-Like Boron Sheet. Nano Letters, 2016. **16**(10): p. 6124-6129.
77. Zhang, Z., et al., Two-Dimensional Tetragonal TiC Monolayer Sheet and Nanoribbons. Journal of the American Chemical Society, 2012. **134**(47): p. 19326-19329.
78. Cahangirov, S., et al., Two- and One-Dimensional Honeycomb Structures of Silicon and Germanium. Physical Review Letters, 2009. **102**(23): p. 236804.
79. Molina-Sánchez, A. and L. Wirtz, Phonons in single-layer and few-layer MoS${}_{2}$ and WS${}_{2}$. Physical Review B, 2011. **84**(15): p. 155413.
80. Muy, S., et al., Tuning mobility and stability of lithium ion conductors based on lattice dynamics. Energy & Environmental Science, 2018. **11**(4): p. 850-859.
81. Mouhat, F. and F.-X. Coudert, Necessary and sufficient elastic stability conditions in various crystal systems. Physical Review B, 2014. **90**(22): p. 224104.
82. Liu, K., et al., Elastic Properties of Chemical-Vapor-Deposited Monolayer MoS2, WS2, and Their Bilayer Heterostructures. Nano Letters, 2014. **14**(9): p. 5097-5103.
83. Abbas, G., et al., Two-dimensional B3P monolayer as a superior anode material for Li and Na ion batteries: a first-principles study. Materials Today Energy, 2020. **17**: p. 100486.
84. Zhang, Y., First principles prediction of two-dimensional tungsten carbide (W2C) monolayer and its Li storage capability. Computational Condensed Matter, 2017. **10**: p. 35-38.
85. Lin, S., et al., Ultrahigh energy density BeN monolayer: A nodal-line semimetal anode for Li-ion batteries. Physical Review Research, 2024. **6**(1): p. 013028.
86. Gao, S., et al., Twin-Graphene: A Promising Anode Material for Lithium-Ion Batteries with Ultrahigh Specific Capacity. The Journal of Physical Chemistry C, 2023. **127**(29): p. 14065-14074.
87. Tritsaris, G.A., et al., Diffusion of Lithium in Bulk Amorphous Silicon: A Theoretical Study. The Journal of Physical Chemistry C, 2012. **116**(42): p. 22212-22216.
88. Persson, K., et al., Thermodynamic and kinetic properties of the Li-graphite system from first-principles calculations. Physical Review B, 2010. **82**(12): p. 125416.
89. He, Q., et al., Computational investigation of 2D 3d/4d hexagonal transition metal borides for metal-ion batteries. Electrochimica Acta, 2021. **384**: p. 138404.

90. Yuan, G., et al., Monolayer Zr2B2: A promising two-dimensional anode material for Li-ion batteries. Applied Surface Science, 2019. **480**: p. 448-453.
91. Wang, S.-F., et al., Theoretical investigation of Ti2B monolayer as powerful anode material for Li/Na batteries with high storage capacity. Applied Surface Science, 2021. **538**: p. 148048.
92. Lv, X., et al., Sc2C as a promising anode material with high mobility and capacity: a first-principles study. ChemPhysChem, 2017. **18**(12): p. 1627-1634.
93. Zhang, X., et al., Mn2C monolayer: A superior anode material offering good conductivity, high storage capacity and ultrafast ion diffusion for Li-ion and Na-ion batteries. Applied Surface Science, 2020. **503**: p. 144091.
94. Wu, Y.-Y., et al., Two-dimensional tetragonal Ti2BN: A novel potential anode material for Li-ion batteries. Applied Surface Science, 2020. **513**: p. 145821.
95. Hu, J., et al., Investigations on Nb 2 C monolayer as promising anode material for Li or non-Li ion batteries from first-principles calculations. RSC advances, 2016. **6**(33): p. 27467-27474.
96. Belasfar, K., et al., First-principles study of BC3 monolayer as anodes for lithium-ion and sodium-ion batteries applications. Journal of Physics and Chemistry of Solids, 2020. **139**: p. 109319.
97. Rajput, K., et al., Ca2C MXene monolayer as a superior anode for metal-ion batteries. 2D Materials, 2021. **8**(3): p. 035015.
98. Sun, Q., et al., Ab initio prediction and characterization of Mo2C monolayer as anodes for lithium-ion and sodium-ion batteries. The journal of physical chemistry letters, 2016. **7**(6): p. 937-943.
99. Pan, H., Electronic properties and lithium storage capacities of two-dimensional transition-metal nitride monolayers. Journal of Materials Chemistry A, 2015. **3**(43): p. 21486-21493.
100. Liu, Y., et al., Feasibility of Lithium Storage on Graphene and Its Derivatives. J Phys Chem Lett, 2013. **4**(10): p. 1737-42.
101. Patel, P., et al., Two-Dimensional α-SiX (X = N, P) Monolayers as Efficient Anode Material for Li-Ion Batteries: A First-Principles Study. ACS Applied Nano Materials, 2023. **6**(3): p. 2103-2115.
102. Abdullahi, Y.Z., et al., Theoretical Screening of Metal Borocarbide Sheets for High-Capacity and High-Rate $\mathrm{Li}$- and $\mathrm{Na}$-Ion Batteries. Physical Review Applied, 2021. **16**(2): p. 024031.

# Supporting Information (SI)

## Janus $MgAlB_2$ MBene: a dipole-engineered anode for ultrafast Li-ion transport and exceptional lithium storage

Pritam Samanta[1], Shashank Pandey[1], Prakash Parida[1*], Amit Kumar Jana[1]
[1]Department of Physics, Indian Institute of Technology Patna, Bihta, Patna, 801106, Bihar, India
*Corresponding authors: pparida@iitp.ac.in

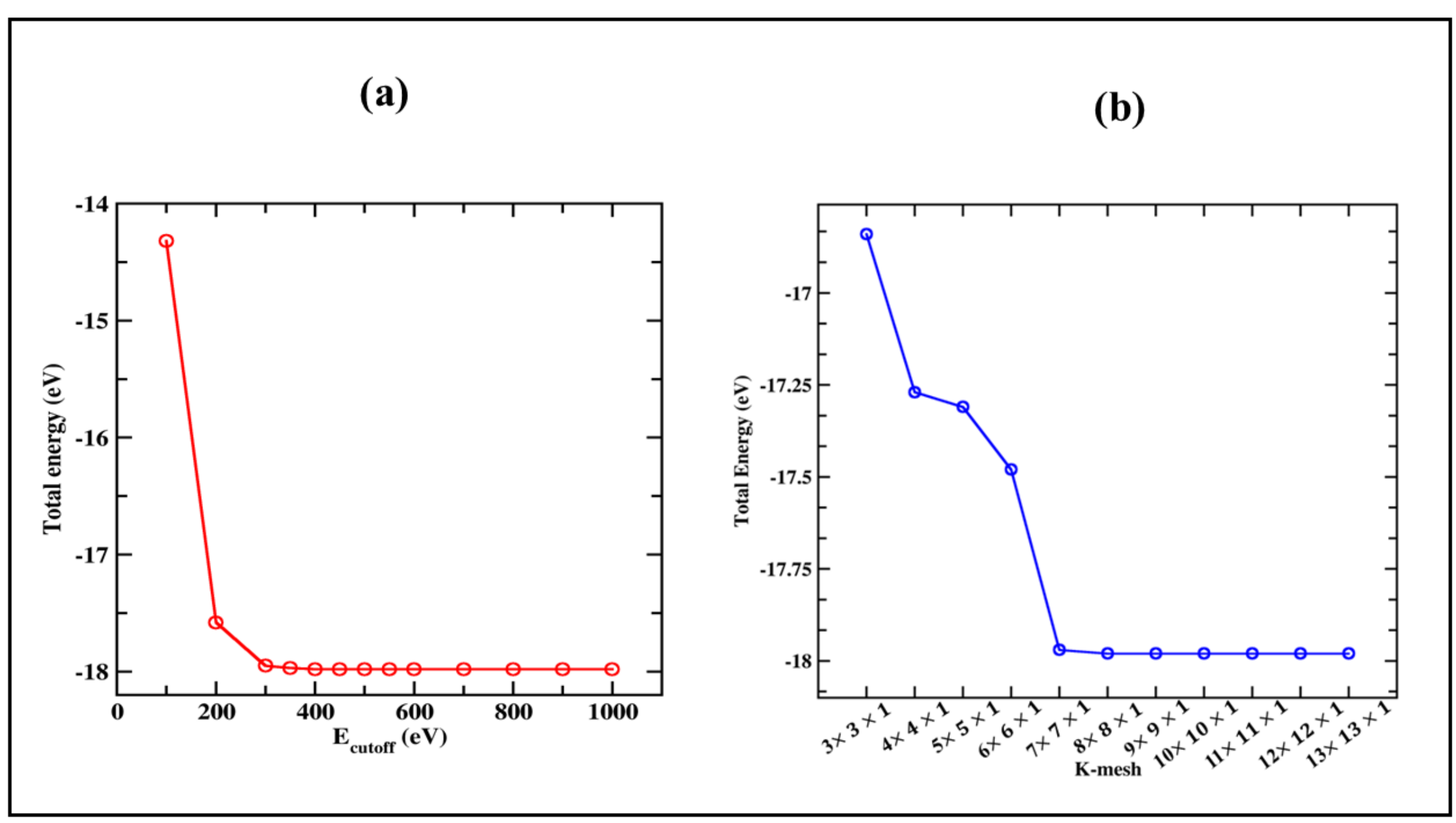

**Fig. S1** (a) Total energy versus plane wave energy cut off for plane wave and (b) the k-mesh diagrams of the $MgAlB_2$ monolayer.

**Table S1** Stable phases of constituent elements of Janus $MgAlB_2$ and Li atoms with energy per atom of the corresponding elements in their bulk stable phase.

| Elements | Stable phase | Space group | Energy per atom (eV/atom) |
|---|---|---|---|
| Mg | Hexagonal close packing(hcp) | $P6_3/mmc$ | -1.23 |
| Al | FCC | $Fm\bar{3}m$ | -3.45 |
| B | Rhombohedral | $R\bar{3}m$ | -6.47 |
| Li | BCC | $Im\bar{3}m$ | -1.92 |

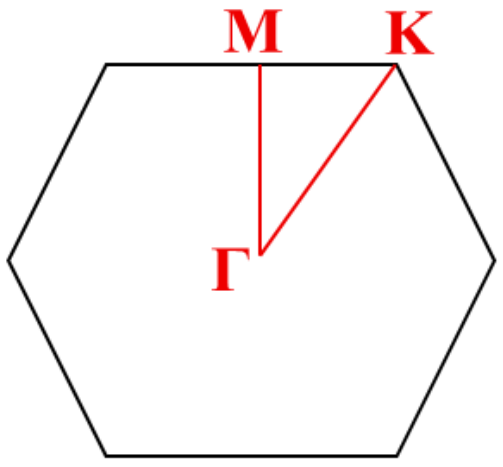

**Fig. S2** The illustration of the Brillouin zone with high-symmetry points.

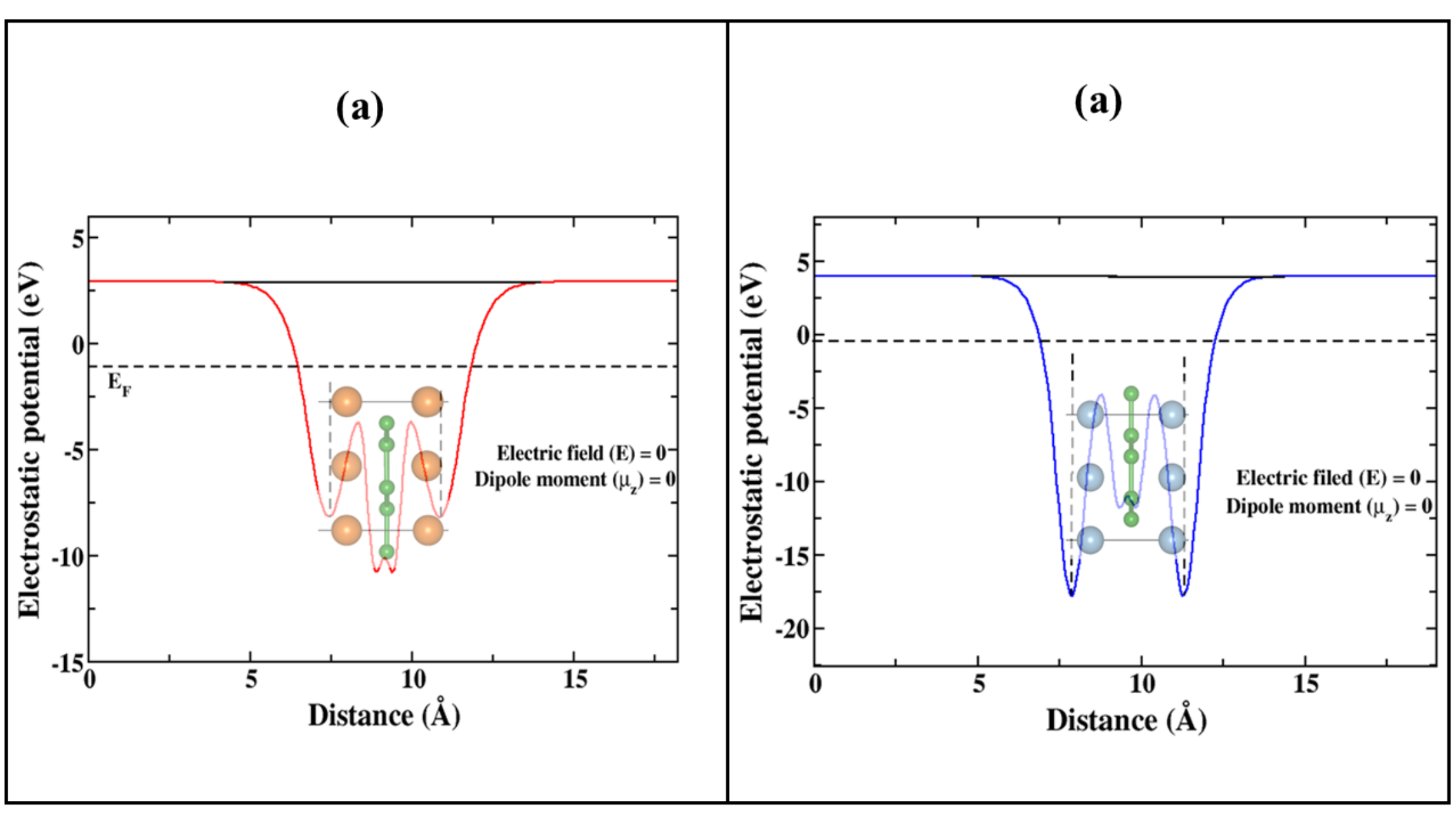

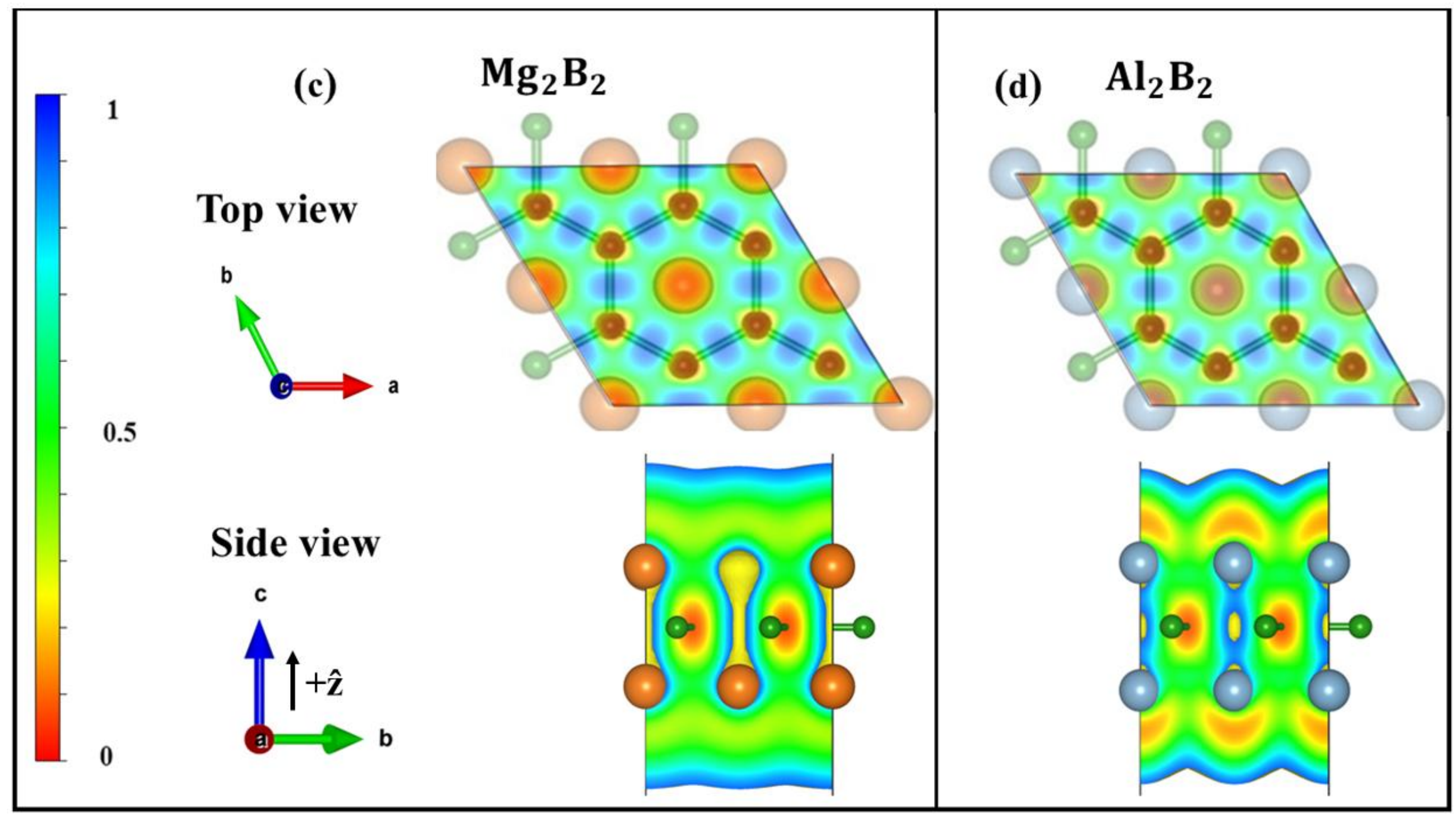

**Fig. S3** Average electrostatic potential of pristine (a) $Mg_2B_2$, (b) $Al_2B_2$ along the z-axis. ELF map of (c) $Mg_2B_2$, (d) $Al_2B_2$.

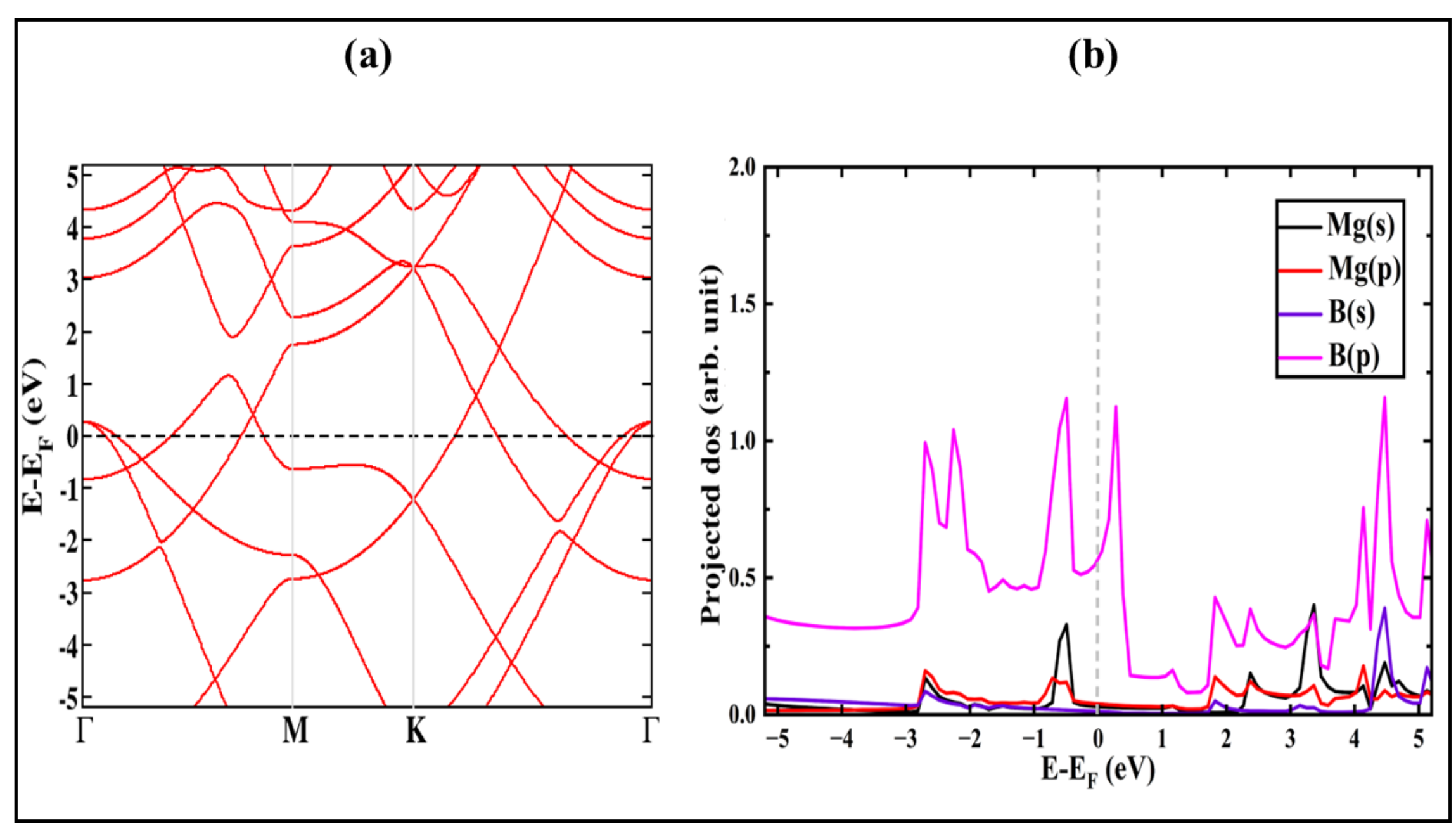

**Fig. S4** (a) Band structure and (b) PDOS of pristine $Mg_2B_2$.

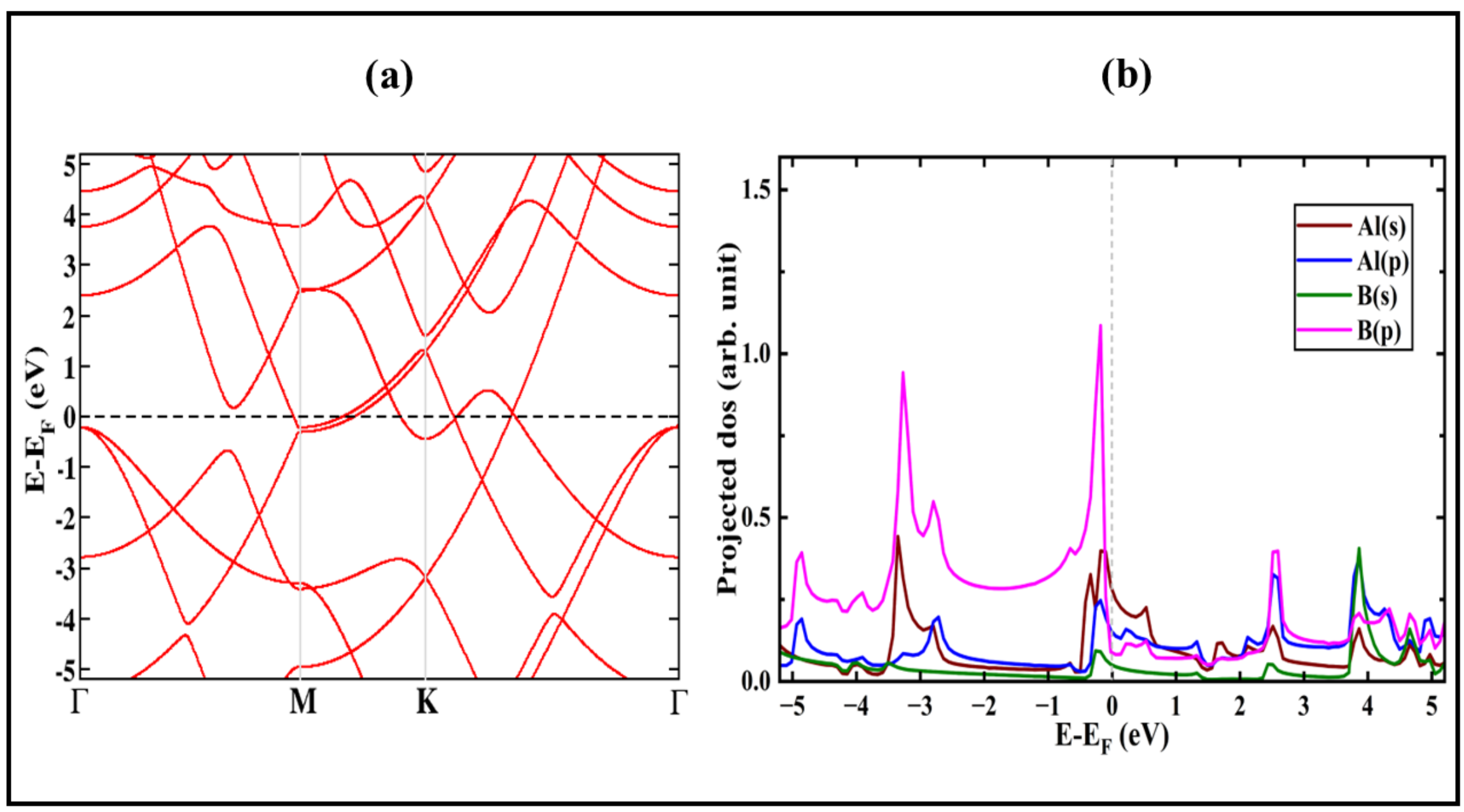

**Fig. S5** (a) Band structure and (b) PDOS of pristine $Al_2B_2$.

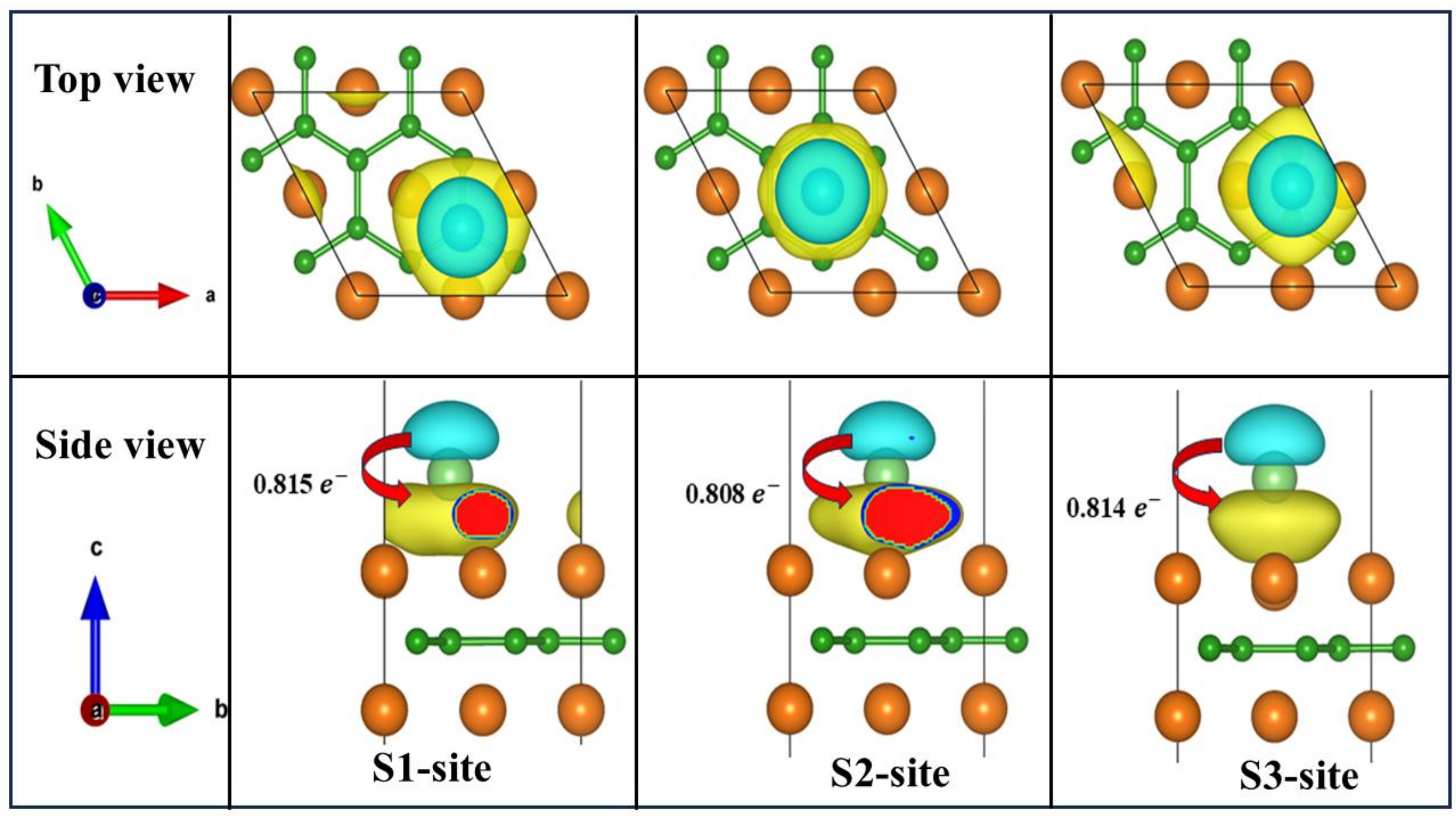

**Fig. S6** Top view & side view of the EDD plots for a single Li atom adsorbed over $Mg_2B_2$ monolayer at three different adsorption sites. In the EDD plots, the cyan and yellow colours represent regions of electron depletion and accumulation, respectively. The iso-surface value is set to 0.01 $eÅ^{-3}$.

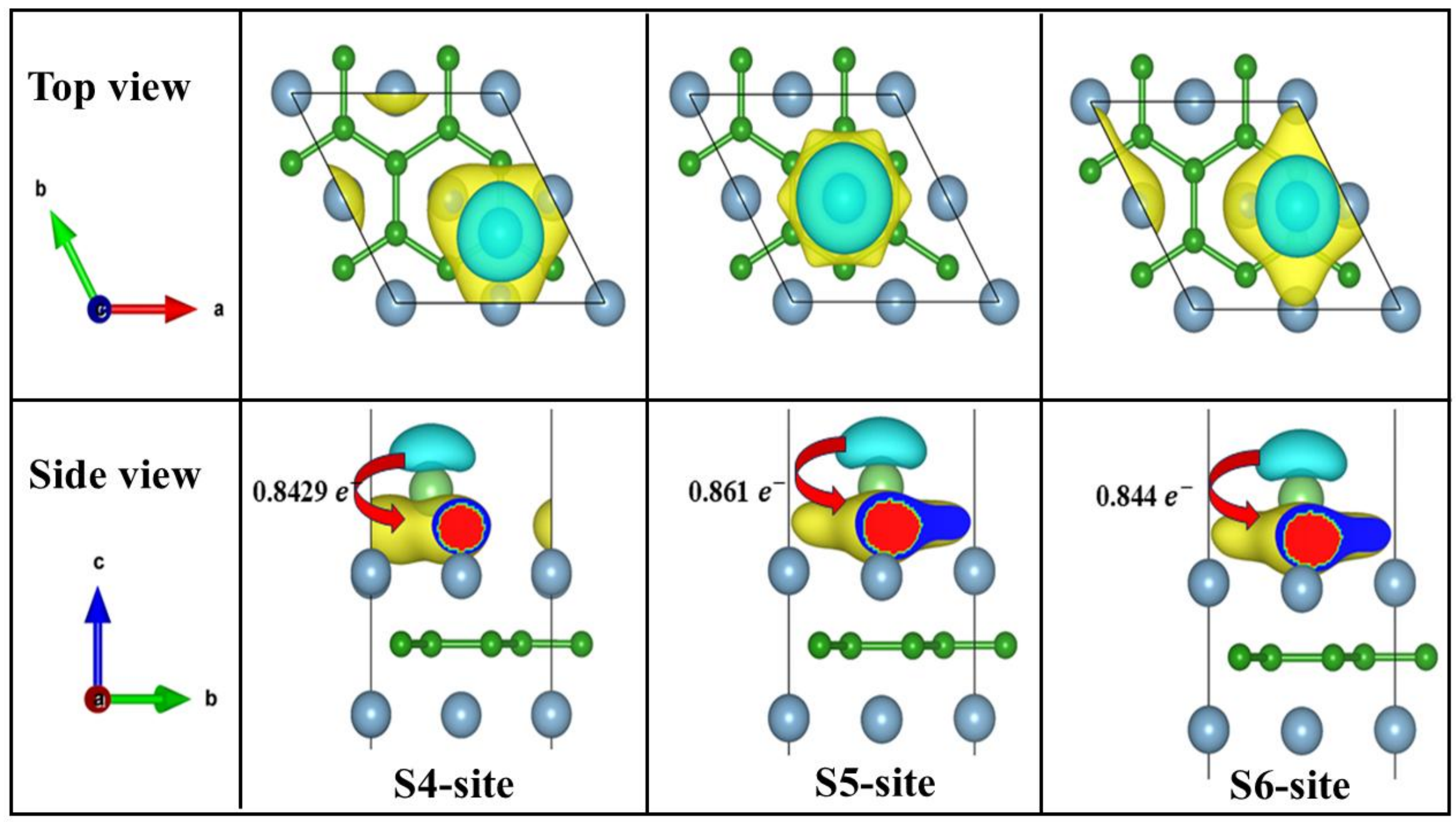

**Fig. S7** Top view & side view of the EDD plots for a single Li atom adsorbed over $Al_2B_2$ monolayer at three different adsorption sites. In the EDD plots, the cyan and yellow colours represent regions of electron depletion and accumulation, respectively. The iso-surface value is set to 0.01 $eÅ^{-3}$.

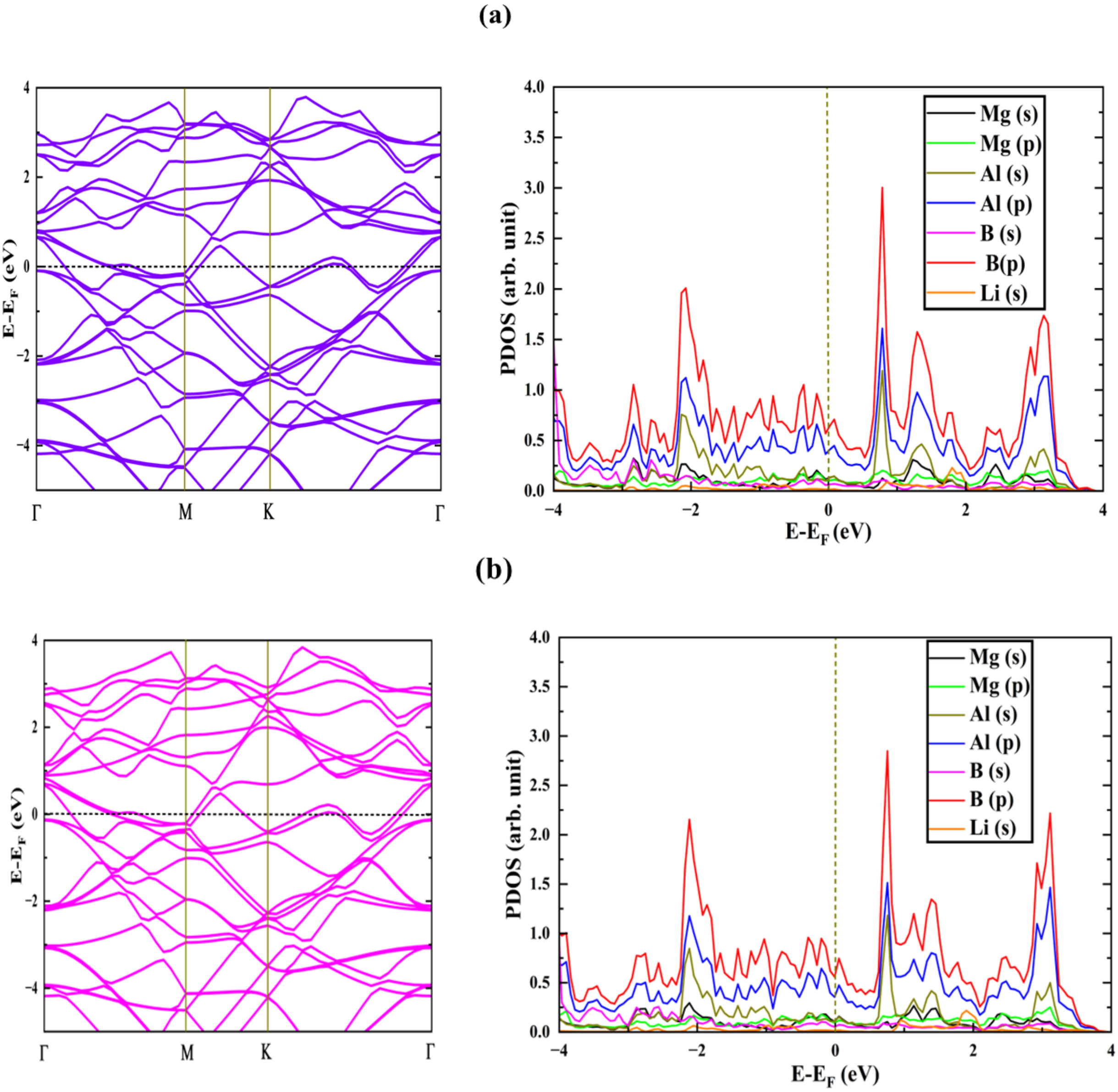

**Fig. S8** Band and PDOS of Li adsorbed at the (a) S2-site and (b) S3-site on the Mg surface of $MgAlB_2$.

(a)

(b)

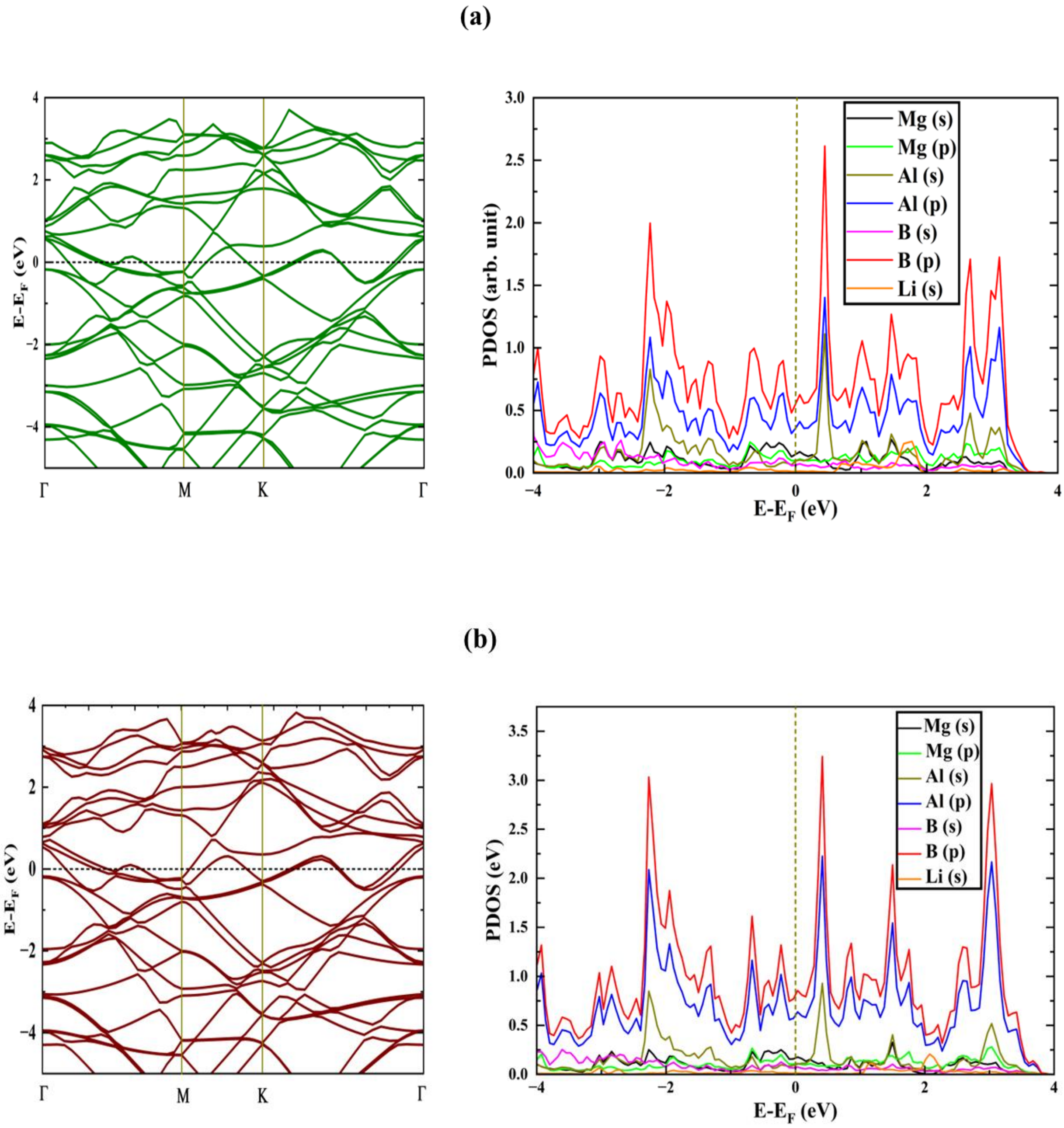

**Fig. S9** Band structure and PDOS of Li adsorbed) at the (a) S5 site and (b) S6 site on the Al surface of Janus $MgAlB_2$.

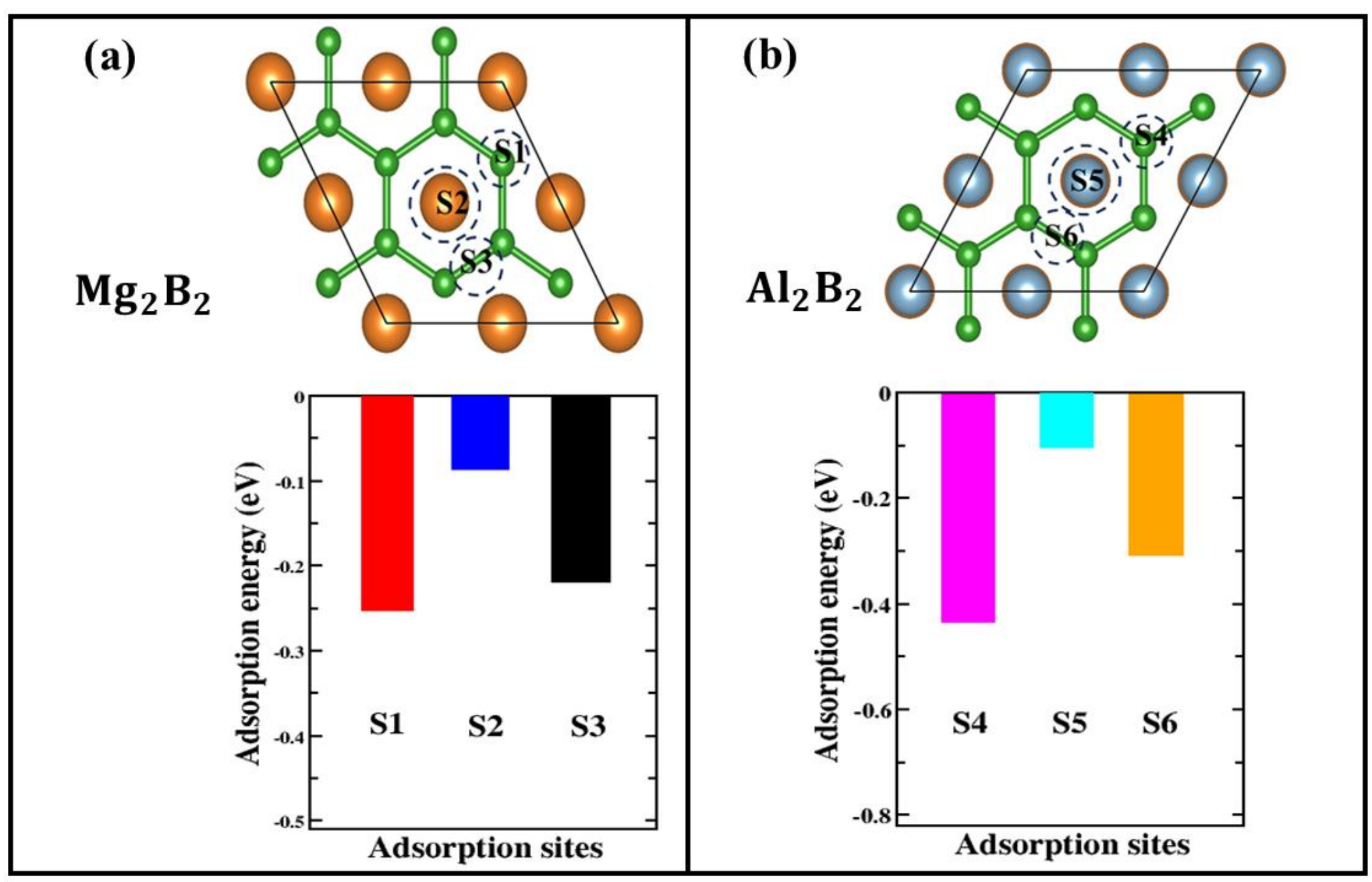

**Fig. S10** Different adsorption sites and histogram representing the adsorption energies at the corresponding adsorption sites on the $2 \times 2 \times 1$ supercell of pristine (a) $Mg_2B_2$ and (b) $Al_2B_2$ monolayer.

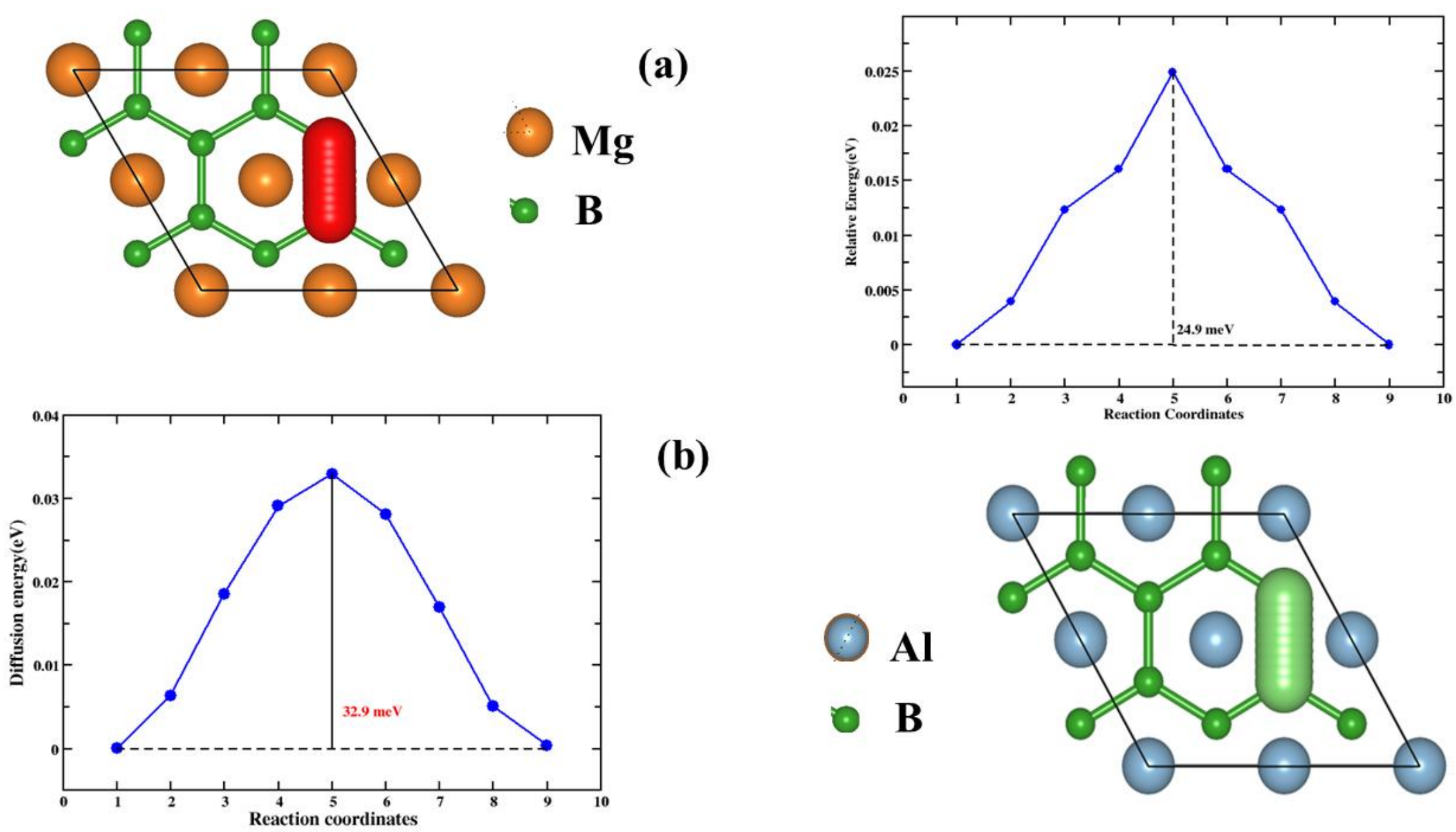

**Fig. S11** Lowest diffusion energy path and the corresponding migration energy barrier profile of (a) $Mg_2B_2$ (S1-S3-S1 pathway) and (b) $Al_2B_2$ (S1-S3-S1 pathway), respectively.

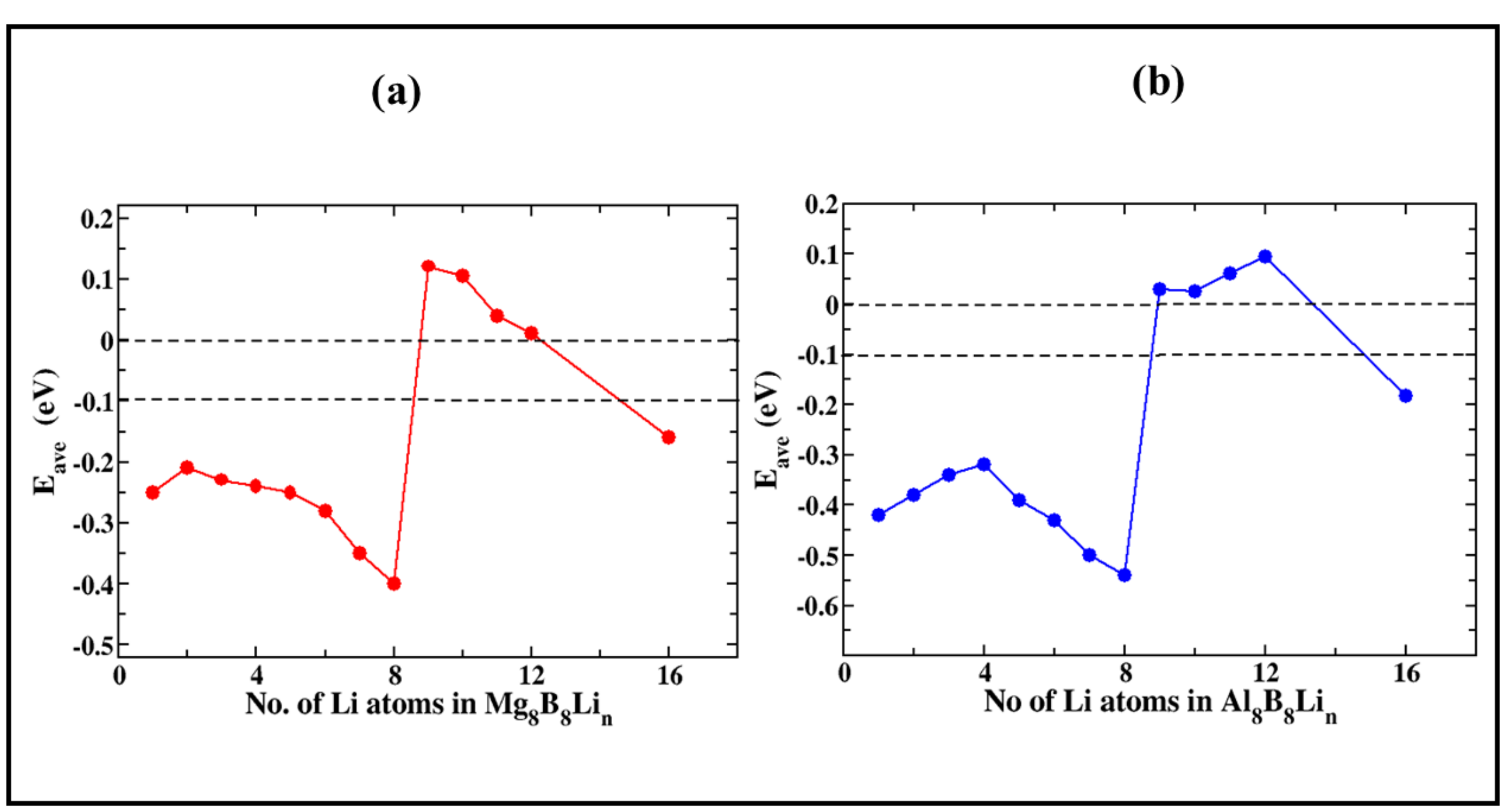

**Fig. S12** Variation of average adsorption energies with number of Li atoms adsorbed on the 2 × 2 × 1 supercell of (a) $Mg_2B_2$ and (b) $Al_2B_2$ monolayer. Here 'n' represents the number of Li atoms adsorbed on the monolayer.

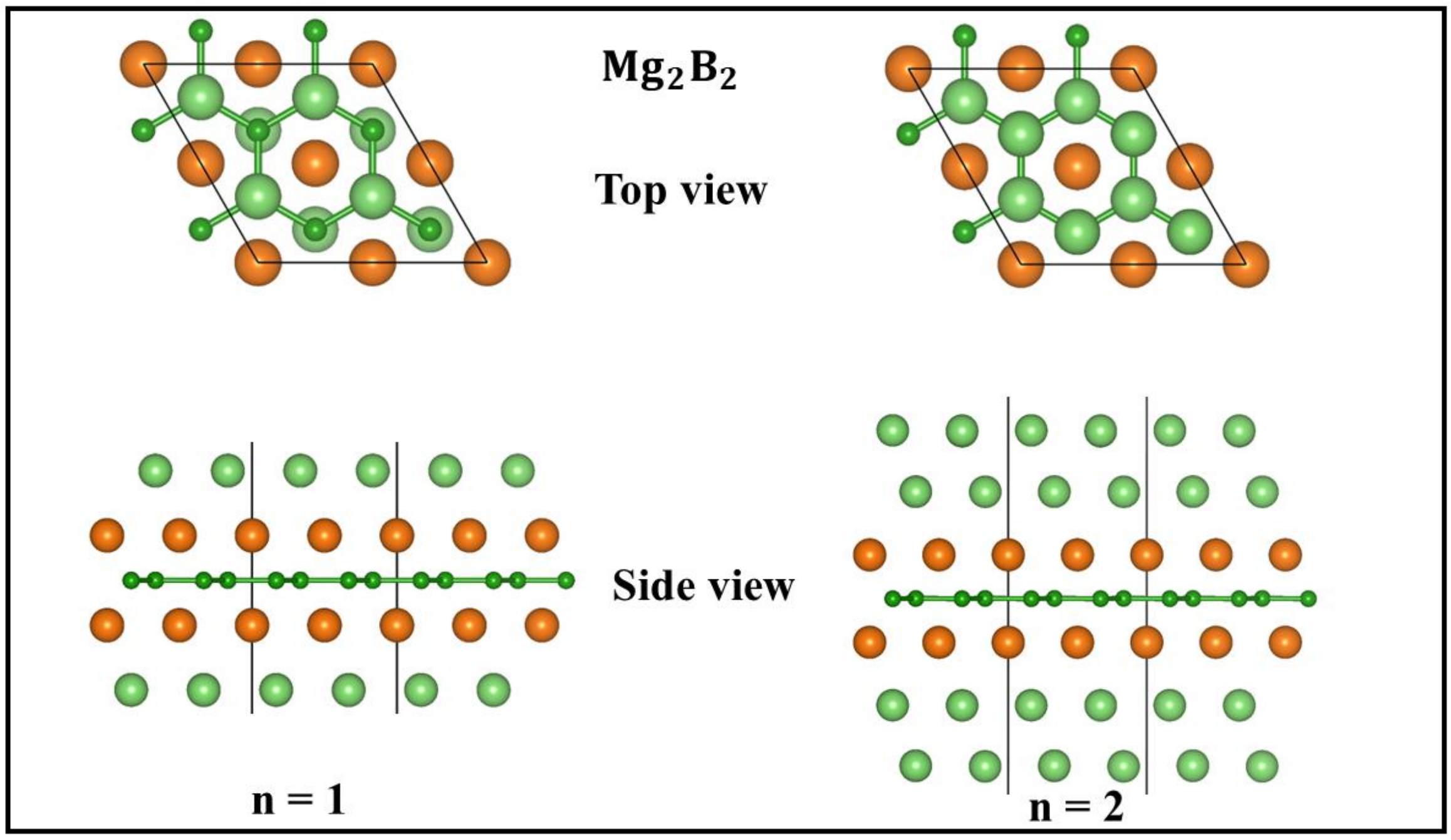

**Fig. S13** The top and side views of optimized $Mg_2B_2$ with maximum Li adsorption. The left and right panels represent the first and second layer, respectively.

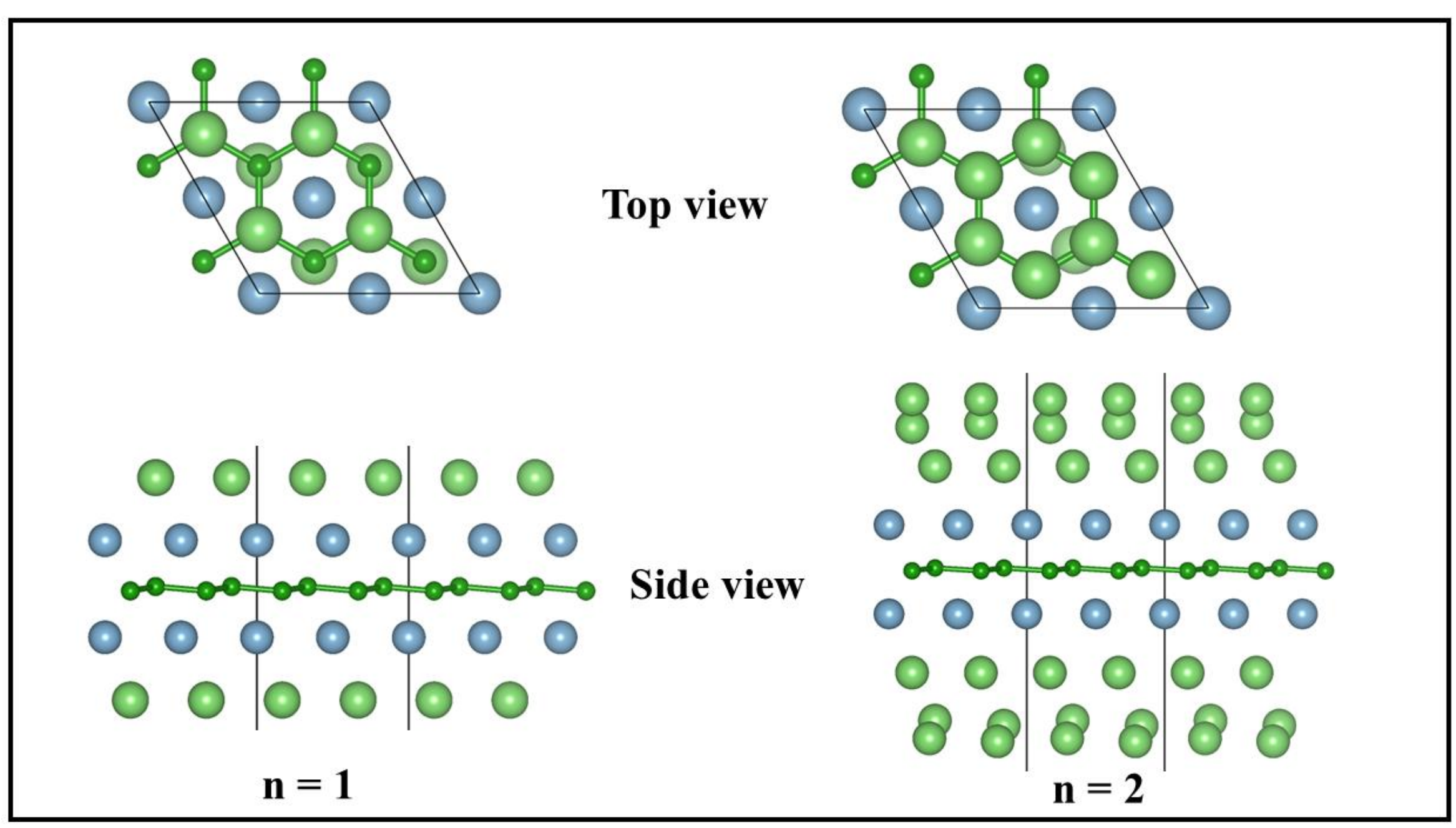

**Fig. S14** The top and side views of optimized $Al_2B_2$ with maximum Li adsorption. The left and right panels represent the first and second layer, respectively.

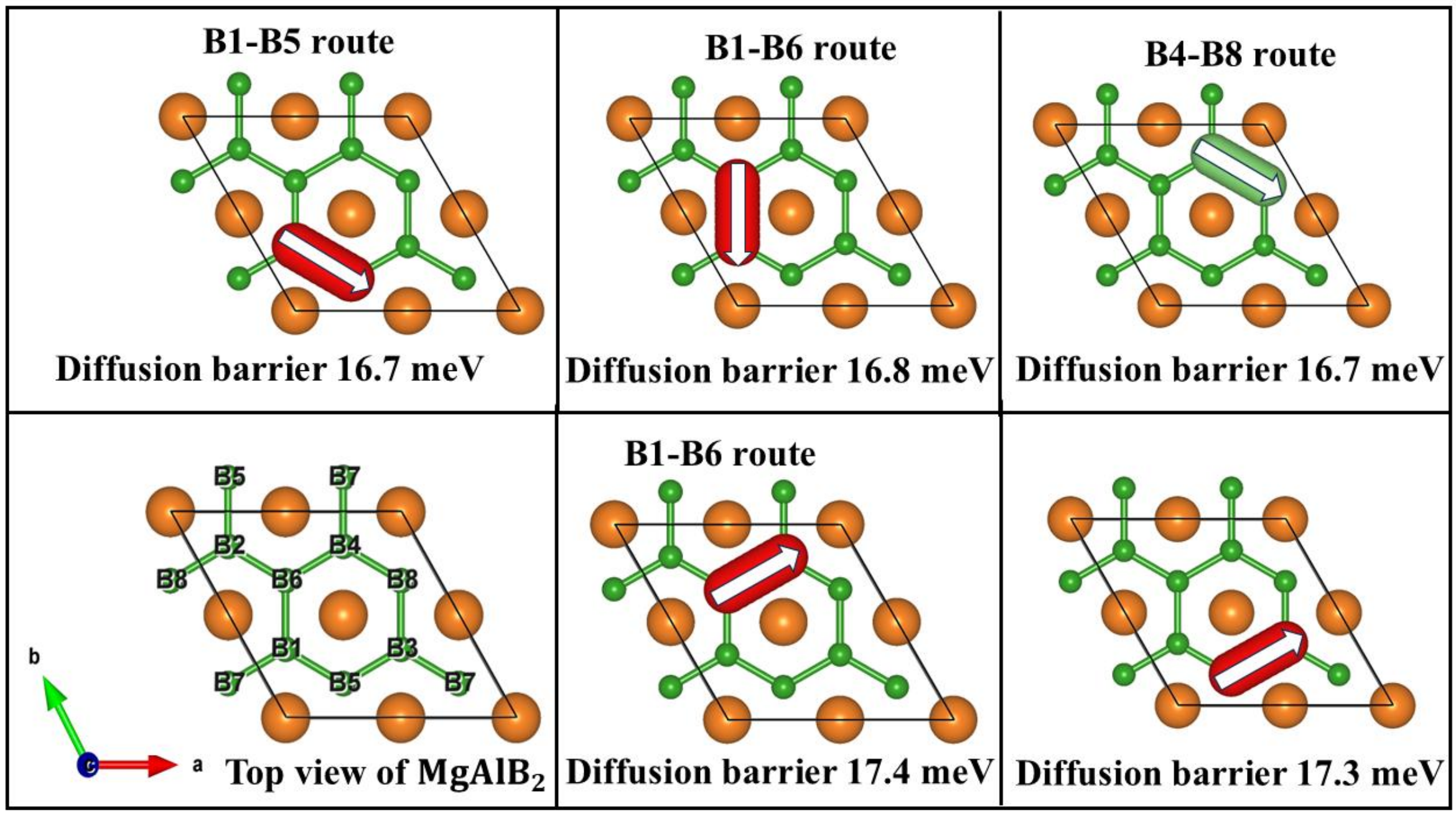

**Fig. S15** Different diffusion paths with corresponding diffusion barriers for S1-S3-S1 root (path1) on the Mg surface on Janus $MgAlB_2$ monolayer.

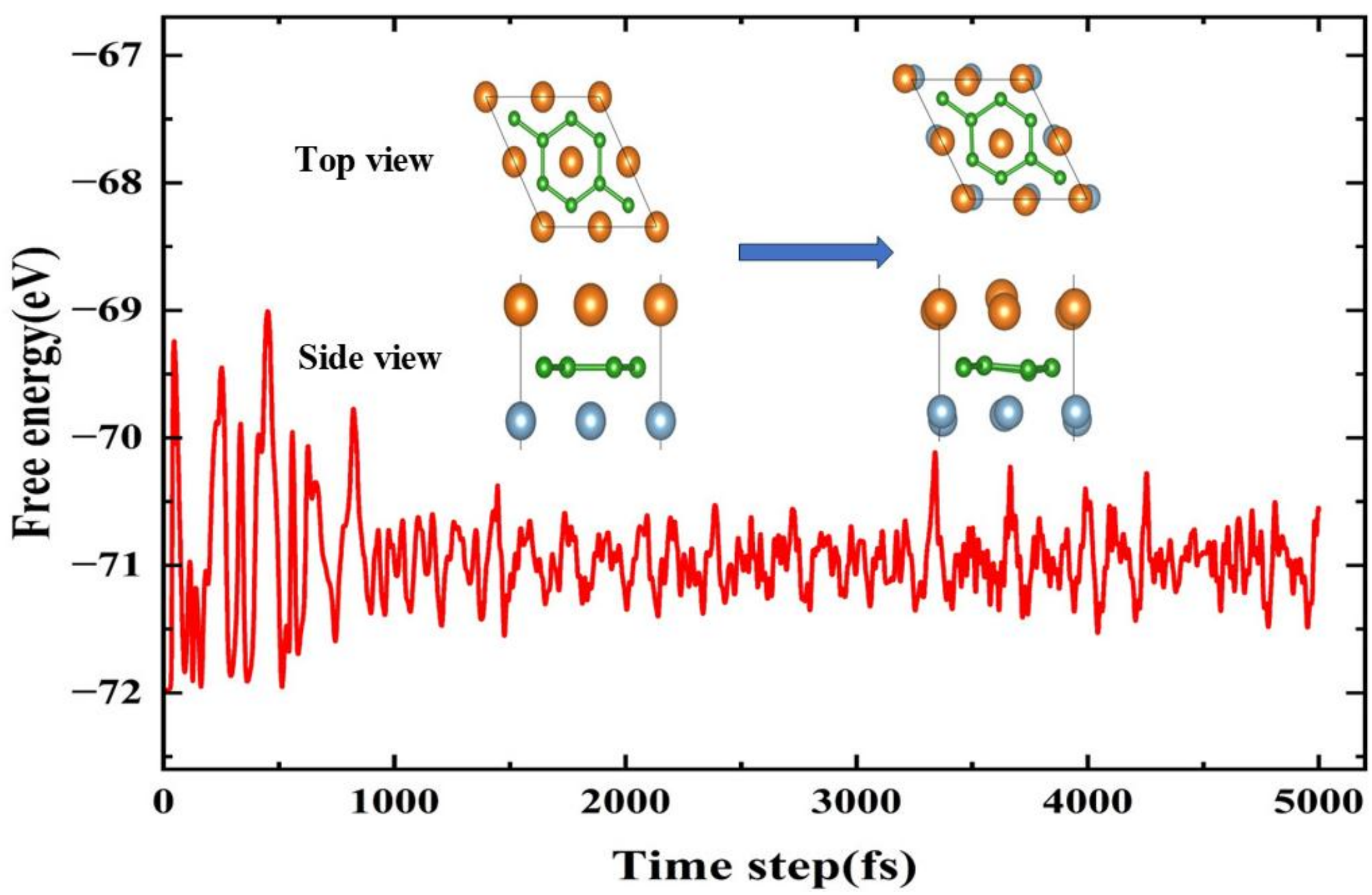

**Fig. 16.** Snapshots of pristine $MgAlB_2$ monolayer after AIMD simulations for 10 ps with NVT ensemble at 500K.